\documentclass[a4paper,11pt]{article}
\usepackage{setspace}
\usepackage{amsmath,amssymb,amsthm}
\usepackage[totalwidth=17cm,totalheight=24cm]{geometry}
\usepackage{empheq}
\usepackage{graphicx}
\newcommand{\ket}[2]{ {\langle #1\,#2\rangle} }
\newcommand{\bra}[2]{ {[#1\,#2]}}
\newcommand{\rmd}{\mathrm{d}}

\newcommand{\im}{{\mathrm i}}

\newcommand{\C}{{\mathbb C}}
\newcommand{\R}{{\mathbb R}}
\newcommand{\Hc}{{\cal H}}

\newcommand{\Tr}[1]{\mathrm{Tr}  \left( {#1}\right)}
\newcommand{\be}{\begin{eqnarray}}
\newcommand{\ee}{\end{eqnarray}}
\newcommand{\nn}{\nonumber}

\def\d{\delta}
\def\a{\alpha}
\def\b{\beta}

\begin{document}

\title{Pure connection formalism for gravity: Feynman rules and the graviton-graviton scattering}
\author{Gianluca Delfino${}^1$, Kirill Krasnov${}^{1}$ and Carlos Scarinci${}^{1,2}$  \\ \\ \it{${}^1$ \, School of Mathematical Sciences, University of Nottingham}\\ \it{University Park, Nottingham, NG7 2RD, UK} \and 
\it{${}^2$ \, Department Mathematik, FAU Erlangen-N\"urnberg}  \\ \it{Cauerstrasse 11, 91058 Erlangen, Germany}}
\date{v2: October 2014}
\maketitle
\begin{abstract}\noindent We continue to develop the pure connection formalism for gravity. We derive the Feynman rules for computing the connection correlation functions, as well as the prescription for obtaining the Minkowski space graviton scattering amplitudes from the latter. The present formalism turns out to be simpler than the metric based one in many aspects. Simplifications result from the fact that the conformal factor of the metric, a source of complications in the usual approach, does not propagate in the connection formulation even off-shell. This simplifies both the linearized theory and the interactions. For comparison, in our approach the complete off-shell cubic GR interaction contains just 3 terms, which should be compared to at least a dozen terms in the metric formalism. We put the technology developed to use and compute the simplest graviton-graviton scattering amplitudes. For GR we reproduce the well-known result. For our other, distinct from GR, interacting theories of massless spin 2 particles we obtain non-zero answers for some parity-violating amplitudes. Thus, in the convention that all particles are incoming, we find that the 4 minus, as well as the 3 minus 1 plus amplitudes are zero (as in GR), but the amplitudes with 4 gravitons of positive helicity, as well as the 3 plus 1 minus amplitudes are different from zero. This serves as a good illustration of the type of parity violation present in these theories. We find that the parity-violating amplitudes are important at high energies, and that a general parity-violating member of our class of theories "likes" one helicity (negative in our conventions) more than the other in the sense that at high energies it tends to convert all present gravitons into those of negative helicity. 
\end{abstract}

\section{Introduction}

In paper \cite{Krasnov:2011pp} one of us showed how $\Lambda\not=0$ General Relativity (GR) can be described in the "pure connection" formalism, in which the only field present in the Lagrangian formulation of the theory is a (complexified) ${\rm SO}(3)$ connection rather than the metric. Paper \cite{Krasnov:2011up} made the first steps towards setting up the perturbation theory in this formalism, analyzing the free theory and obtaining the propagator. In addition, a large class of interacting theories of massless spin 2 particles that are distinct from GR was exhibited, making it particularly clear that GR is just a special member in a large class of gravitational theories with similar properties. Earlier works on the pure connection formalism for gravity (in the $\Lambda=0$ setting) include \cite{Capovilla:1989ac} and \cite{Capovilla:1991kx}. Early works on distinct from GR gravitational theories with two propagating degrees of freedom include \cite{Capovilla:1992ep} and \cite{Bengtsson:1990qg}. In works of the author of \cite{Bengtsson:1990qg} they are referred to as "neighbours of GR". These earlier developments used formulations involving other fields in addition to a connection; we refer the reader to \cite{Krasnov:2009ik} and references therein. 

A recent work \cite{Delfino:2012zy} by the present authors gave a more systematic treatment of the linearized theory in the pure connection setting. One outcome of \cite{Delfino:2012zy} is a realization that the reality conditions satisfied by the connection can only be properly understood for $\Lambda\not=0$, i.e. before the Minkowski limit is taken. This paper also derived the mode decomposition of the connection into creation/annihilation operators. The mode decomposition obtained demonstrates, in particular, that the gauge-theoretic description of gravitons is parity asymmetric, because the two helicities of the graviton are treated quite differently. One can then expect that a generic gravitational theory built using this formalism is parity-violating. This expectation will be confirmed in the present paper by a direct computation of amplitudes of some parity-violating processes.

This is the second paper in a series on the pure connection formalism for gravity. Our main objective here is to derive Feynman rules, as well as spell out the rules for extracting the graviton scattering amplitudes from the connection correlation functions. To illustrate how the formalism can be used for practical computations, we compute the simplest graviton scattering amplitudes, the two-to-two graviton ones. In the next paper from the series we apply the formalism to compute more general, in particular the so-called maximally helicity violating (MHV),  amplitudes with an arbitrary number of external gravitons. The MHV amplitudes turn out to be the same for any member of the family of gravitational theories that we study. An analysis of the loop amplitudes is postponed to later papers in the series. 

One of the conclusions of this paper is that the gauge-theoretic approach \cite{Krasnov:2011up} to GR (and more general gravity theories) works, in the sense that it can be used to reproduce the known GR results. We also compute the helicity-violating scattering amplitudes that are zero in GR, but are present in a general member of our family of theories, thus confirming their parity-violating nature. 

However, we hope to be able to convince the reader in a much stronger statement. Thus, not only the gauge-theoretic description of GR is equivalent (at tree level) to the standard one, but it is actually a significantly more efficient tool for practical calculations than the description based on the Einstein-Hilbert action. In particular, we shall see that the free theory looks considerably simpler in the language of connections, with the gauge-fixed Lagrangian being of the $\phi\Box \phi$ type, and only a set of fields in a single irreducible component of the Lorentz group propagating. This should be contrasted with the situation in the metric-based GR, where both the tracefree and the trace part of the metric perturbation propagate, with different signs in front of the corresponding kinetic terms. Further, we shall see that in the connection formalism the interaction vertices contain much smaller number of terms as compared to the metric-based description. For instance we shall see that the GR cubic interaction vertex contains just 3 terms, of which just a single term is responsible for the graviton-graviton scattering amplitude. The simplicity of this cubic GR vertex figures even more prominently in the next paper on MHV amplitudes. One finds that computations that would be very difficult in the usual metric description are made easy by the use of the gauge theoretic framework. We will see that these significant simplifications are due to the fact that the conformal mode of the graviton, which is gauge in pure gravity, does not propagate in our formalism even off-shell. This simplifies in particular the interactions of the theory, and this in turn allows for computations that are unimaginable in the usual formalism. 

Let us also comment on parity and unitarity in the formalism that we develop. We shall see that parity of GR, while being a symmetry of the theory, is not manifest in our description, as it treats the two graviton helicities differently. Also, as was exhibited in \cite{Delfino:2012zy}, parity in our context is related to the operation of Hermitian conjugation. It is thus directly tied with the issue of the Lagrangian being Hermitian (something that is again not manifest in our approach), and therefore unitarity. In these aspects the formalism developed here is similar to the twistor description of Yang-Mills theory proposed in \cite{Witten:2003nn}, which also makes parity and unitarity not manifest. At the moment of writing this, we do not know whether this is just an accidental similarity or there is a more deep relation between our gauge-theoretic and the twistor approaches. 

Our analysis here will be for a generic point in the space of gravity theories described in \cite{Krasnov:2011up}, with GR results obtained as just a special case of more general formulas. We decided against working out just the GR case because treating a general theory is at almost no expense. For readers only interested in General Relativity we will always make the GR versions of formulas explicit. However, as we shall see, embedding GR into a larger class of theories makes some of its properties more transparent. 

This paper uses spinors, and in particular the spinor helicity basis for the graviton polarizations, quite heavily (in the second half of the paper). We shall see that some aspects of our gauge-theoretic formalism can only be fully appreciated if one works in spinor terms. Relevant facts about spinors are reviewed in the text. We warn the reader that gravity community notations are adopted here, and our 2-component spinors are referred to as unprimed and primed ones (not undotted and dotted). The reason for this choice is our need, at places, to use the doubly-null tetrad representation of the metric and the self-dual 2-forms. These formulas are standard in the gravity literature using spinors, see e.g. \cite{Penrose:1985jw}, and rewriting them in terms of dotted/undotted spinors would make them look awkward (at least to us). We hope that the interested particle physics oriented reader will be able to follow in spite of these choices. 

We would also like to warn the reader that at places this paper becomes quite technical, as we will have to develop from scratch many results that are simply assumed in more standard treatments. This includes a prescription for how the scattering amplitudes can be extracted from the connection correlation functions, as well as a statement of the crossing symmetry. The results we get are not surprising and could have been guessed from the start (modulo some minus signs that are characteristic of our approach treating gravity {\it \`a la} theory of fermions), but we felt we needed to do the exercise to be sure about the internal self-consistency of our approach. A particularly technical part of the paper is the one dealing with the prescription of how to take the Minkowski space limit (our theory begins as a theory of interacting gravitons in de Sitter space). Readers not interested in all these technicalities can simply take the obtained Minkowski space Feynman rules as the starting point, and then follow the part of the paper that computes the graviton scattering amplitudes. For the convenience of the reader we review the Feynman rules in the Appendix, so the paper can well be read in the opposite direction, from the end to the beginning.

The organization of this paper is as follows. We start with a brief review of relevant facts from \cite{Delfino:2012zy}. Then, in section \ref{sec:LSZ} we derive a prescription for how the scattering amplitudes can be obtained from the connection correlation functions. Here we also discuss the tricky issues of taking the Minkowski limit. The outcome of this section is a practical prescription for how to do calculations. We stress that even though the theory starts as being about gravitons in de Sitter space, the final prescription works with Minkowski space quantities, and, in particular, the usual Fourier transform is available.  Then in section \ref{sec:gf} we explain how the gauge-fixing is done, and obtain the propagator. Section \ref{sec:inter} computes the interaction terms, up to quartic interactions. Section \ref{sec:spinor} reviews the necessary spinor technology. This technology is then immediately put to use and the graviton polarization tensors are translated into the spinors language. In section \ref{sec:feyn} we translate everything in spinor terms and state the Feynman rules in the their final, most useful for computational purposes form. Section \ref{sec:scatt} then computes the graviton-graviton scattering amplitudes. We conclude with a discussion.

\section{Brief summary of the previous work}

In this section we write down, for the convenience of the reader, formulas from \cite{Delfino:2012zy} that we need in the present work. 

\subsection{The theory}

The class of gravity theories that we are studying in this work is described by the following action principle \cite{Krasnov:2011up}
\be\label{action}
S[A] = \im \int f(F\wedge F).
\ee
Here $\im=\sqrt{-1}$, and $F^i=d A^i +(1/2)\epsilon^{ijk} A^j\wedge A^k$ is the curvature of a complexified ${\rm SO}(3)$ connection $A^i$. The function $f$, referred to as the defining function of the theory, is a gauge-invariant, homogeneous function of degree one mapping symmetric $3\times 3$ matrices to (complex) numbers. The requirement that $f$ is of degree of homogeneity one makes the quantity $f(F\wedge F)$ in (\ref{action}) a well-defined 4-form that can be integrated over spacetime. As we have already mentioned, general relativity (GR) is only a special member of the class (\ref{action}), and the corresponding defining function is, see \cite{Krasnov:2011pp}
\be\label{GR}
f_{\rm GR}(F\wedge F) =\frac{1}{16\pi G \Lambda} \left( \mathrm{Tr}{ \sqrt{F\wedge F} } \right)^2,
\ee
where $G$ is the usual Newton's constant, and $\Lambda$ is the cosmological constant. 

\subsection{Background}

As is described in details in \cite{Delfino:2012zy}, we fix a background connection and expand (\ref{action}) around this background. An explicit expression for the background connection can be found in \cite{Delfino:2012zy}. Here we will only need the fact that the curvature of this connection satisfies:
\be\label{F-S}
F^i_{\mu\nu} = - M^2 \Sigma^i_{\mu\nu},
\ee
where $M$ is a parameter with dimensions of mass (it gives the unit in terms of which all other dimensionful quantities are later measured), and $\Sigma^i_{\mu\nu}$ are the self-dual 2-forms for the de Sitter background. These 2-forms carry information about the (de Sitter) metric, and all our computations below happen in and use this background metric structure. The parameter $M$ is then just the (inverse of) the radius of curvature of the background de Sitter metric. We shall see that special care will need to be taken when passing to the Minkowski limit $M\to 0$, as many of the intermediate formulas will contain inverse powers of $M$. 

\subsection{Convenient way to write the action}

To obtain variations of (\ref{action}), it is very convenient to rewrite it in a form such that the defining function is applied to a certain $3\times 3$ matrix rather than 4-form valued $3\times 3$ matrix. This is most conveniently achieved by introducing
\be\label{X}
\hat{X}^{ij}=\frac{1}{8\im M^4} \epsilon^{\mu\nu\rho\sigma} F^i_{\mu\nu}F^j_{\rho\sigma}.
\ee
Here $\epsilon^{\mu\nu\rho\sigma}$ is obtained from the volume form $\epsilon_{\mu\nu\rho\sigma}$ of the de Sitter background metric, by raising all of its indices using the metric. The convenience of this choice lies in the fact that on the background (\ref{F-S}) we have $\hat{X}^{ij}=\delta^{ij}$. The integrand in (\ref{action}) can then be written in terms of $\hat{X}^{ij}$
\be\label{action'}
S[A] = -2 M^4 \int d^4 x \sqrt{-g} \, f(\hat{X}),
\ee
where $\sqrt{-g}$ is the square root of the determinant of the de Sitter background metric. Now, when determining the variations of this action, the variations should only be applied to the function $f$ and thus the curvature $F$ on which it non-linearly depends. Thus, the variations of (\ref{action'}) are very easy to compute. 

\subsection{Variations of the matrix $\hat{X}$}

We start by giving the variations of $\hat X$, as a function of the connection, evaluated at the background $\hat{X}^{ij}\,\hat{=}\,\delta^{ij}$. We have:
\be\label{X_variations}
\delta\hat{X}^{ij}\,\hat{=}\,-\frac{1}{M^2}\Sigma^{(i\mu\nu}D_\mu\delta A^{j)}_\nu,
\\\nonumber
\delta^2 \hat{X}^{ij}\,\hat{=}\,\frac{1}{\im M^4}\epsilon^{\mu\nu\rho\sigma}D_\mu\delta A^i_{\nu}D_\rho\delta A^j_\sigma-\frac{1}{M^2}\Sigma^{(i\mu\nu} \epsilon^{j)kl}\delta A_\mu^k   \delta A_\nu^l,
\\\nonumber
\delta^3 \hat{X}^{ij}\,\hat{=}\,\frac{3}{\im M^4}\epsilon^{\mu\nu\rho\sigma}D_\mu\delta A_\nu^{(i}\epsilon^{j)kl}\delta A_\rho^k\delta A^l_\sigma.
\ee
Finally, the fourth variation is zero $\delta^4 \hat{X}^{ij}=0$ even away from the background. In all expressions above $D_\mu$ is the covariant derivative with respect to the background connection. Thus, it is important to keep in mind that $D$'s do not commute:
\be\label{comm-DD}
2 D_{[\mu} D_{\nu]} V^i = \epsilon^{ijk} F_{\mu\nu}^j V^k,
\ee
for an arbitrary Lie algebra valued function $V^i$. Here $F_{\mu\nu}^i$ is the background curvature (\ref{F-S}). Thus, the commutator (\ref{comm-DD}) is of the order $M^2$. This has to be kept in mind when (in the limit $M\to 0$) replacing the covariant derivatives $D$ with the usual partial derivatives.

\subsection{Variations of the action}

The variations of the action (\ref{action'}) are now easy to compute. We have 
\be\label{variations-action-general}
\delta S\,\hat{=}\,-2M^4\int f^{(1)}_{ij}\delta\hat X^{ij},\qquad 
\delta^2 S\,\hat{=}\,-2M^4\int \left[f^{(2)}_{ijkl}\delta\hat X^{ij}\delta\hat X^{kl}+f^{(1)}_{ij}\delta^2\hat X^{ij}\right],
\\ \nonumber
\delta^3 S\,\hat{=}\,-2M^4\int \left[f^{(3)}_{ijklmn}\delta\hat X^{ij}\delta\hat X^{kl}\delta\hat X^{mn}+3f^{(2)}_{ijkl}\delta^2\hat X^{ij}\delta\hat X^{kl}+f^{(1)}_{ij}\delta^3\hat X^{ij}\right],
\\ \nonumber
\delta^4 S\,\hat{=}\,-2M^4\int \left[f^{(4)}_{ijklmnpq}\delta\hat X^{ij}\delta\hat X^{kl}\delta\hat X^{mn}\delta\hat X^{pq}+6f^{(3)}_{ijklmn}\delta^2\hat X^{ij}\delta\hat X^{kl}\delta\hat X^{mn}\right.
\\ \nonumber
\left.+4f^{(2)}_{ijkl}\delta^3\hat X^{ij}\delta\hat X^{kl}+3f^{(2)}_{ijkl}\delta^2\hat X^{ij}\delta^2\hat X^{kl}\right].
\ee
Here we have omitted the integration measure $\sqrt{-g} \, d^4x$ for brevity. The quantities $f^{(n)}_{ij\ldots}$ are partial derivatives of the defining function with respect to its matrix-valued argument. As is explained in \cite{Delfino:2012zy}, when evaluated at the background $\hat{X}^{ij}=\delta^{ij}$, these derivatives can be parameterized as
\be\nonumber
 f^{(1)}_{ij}=  \frac{f(\delta)}{3} \delta_{ij},  \\ \label{f2}
f^{(2)}_{ijkl} = -\frac{g^{(2)}}{2} P_{ijkl}, \\ \label{f3}
f^{(3)}_{ijklmn} =g^{(3)} \sum_{\text perm} \frac{1}{3!} P_{ijab}P_{klbc}P_{mnca}+\frac{g^{(2)}}{6}\left(\delta_{ij} P_{klmn}+\delta_{kl} P_{ijmn}+\delta_{mn} P_{ijkl}\right), \\ \label{f4}
 f^{(4)}_{ijklmnpq}=- \tilde{g}_1^{(4)} \sum_{\text perm} \frac{1}{4!} P_{ijab}P_{klbc}P_{mncd}P_{pqda} + \tilde{g}_2^{(4)} \sum_{\text perm} \frac{1}{3} P_{ij kl} P_{mn pq} + \ldots,
\ee
where the terms denoted by dots contain $\delta_{ij}$ in at least one of the "channels" and are not going to be important for us since they give rise to a contraction that is zero in view of our gauge-fixing condition, see below. Note, however, that these terms would be of importance for vertices of valency higher than 4, but we do not consider those in this work. Here
 \be\label{P}
 P_{ijkl}:=\frac{1}{2}(\delta_{ik}\delta_{jl} +\delta_{il}\delta_{jk}) - \frac{1}{3}\delta_{ij}\delta_{kl}
 \ee
 is the projector on symemtric tracefree matrices. Thus, a general theory from the class (\ref{action}) is parameterized by an infinite number of (dimensionless) independent coupling constants $g^{(2)}, g^{(3)}, \ldots$, as well as the value $f(\delta)$ of the defining function evaluated at the background. For GR we get
\be\label{g-GR}
f_{\rm GR}(\delta)=\frac{3M_p^2}{M^2},\qquad g^{(2)}_{\rm GR}=\frac{M_p^2}{M^2}, \qquad g_{\rm GR}^{(3)}= \frac{3M_p^2}{4M^2}, \qquad \tilde{g}_{\rm GR\, 1}^{(4)} = \frac{15M_p^2}{8M^2}, \qquad \tilde{g}^{(4)}_{\rm GR\, 2}= \frac{M_p^2}{8M^2}.
 \ee
Here we have defined $M_p^2=1/16\pi G$ and chose the scale parameter $M$ to be related to the cosmological constant as $M^2=\Lambda/3$. 

As we realized after \cite{Delfino:2012zy} was completed, the two terms appearing in (\ref{f4}) are actually multiples of each other, so that there is just a single independent coupling constant at the fourth order. Explicitly, the sum over permutations in the first term in (\ref{f4}) is 
\be\label{f4-1}
\frac{1}{3} \left( P_{ijab}P_{klbc}P_{mncd}P_{pqda}+ P_{ijab}P_{klbc}P_{pqcd}P_{mnda}+ P_{ijab}P_{mnbc}P_{klcd}P_{pqda}\right),
\ee
all other permutations just giving a multiplicative factor reducing $1/4!$ to $1/3$. The sum over permutations in the second term is
\be\label{f4-2}
\frac{1}{3}\left( P_{ij kl} P_{mn pq}+ P_{ij mn} P_{kl pq}+P_{ij pq} P_{kl mn}\right).
\ee
It can be checked by an explicit computation that these two quantities are multiples of each other. To compute the coefficient we contract both expressions in the index pairs $ij$ and $kl$. We have
\be
P_{ij mn} P_{ij pq} = P_{mn pq}, \qquad P_{ij kl} \delta^{ik} = \frac{5}{3} \delta_{jl}, \qquad P_{ij kl} \delta^{ik} \delta^{jl} = 5.
\ee
The contraction of (\ref{f4-1}) gives $(7/6) P_{mn pq}$. The contraction of (\ref{f4-2}) gives $(7/3) P_{mnpq}$. We learn that two times (\ref{f4-1}) equals (\ref{f4-2}). This allows us to write (\ref{f4}) as
\be\label{f4*}
f^{(4)}_{ijklmnpq}=- g^{(4)} \frac{1}{3}\left( P_{ij kl} P_{mn pq}+ P_{ij mn} P_{kl pq}+P_{ij pq} P_{kl mn}\right) + \ldots,
\ee
where for GR we get
\be
g^{(4)}_{\rm GR} = \frac{13 M_p^2}{16 M^2}.
\ee
This is our final expression for the matrix of fourth derivatives of the defining function, to be used below in the computation of the 4-vertex.

\subsection{Linearized action}

The linearized action is obtained as half of the second variation of the action (\ref{action'}). We also define the new, canonically normalized field $a_\mu^i = (\sqrt{g^{(2)}}/\im) \delta A_\mu^i$. The linearized Lagrangian then reads
\be\label{free}
{\cal L}^{(2)}=-\frac{1}{2}P_{ijkl} (\Sigma^{i\mu\nu}D_\mu a^{j}_\nu) (\Sigma^{k\rho \sigma}D_\rho a^{l}_\sigma).
\ee
We emphasize that this is a Lagrangian for gravitons in de Sitter space. It should be contrasted with a significantly more complicated linearized Lagrangian in the metric-based description of GR. As is explained in \cite{Krasnov:2012pd}, see also below, this Lagrangian assumes the most transparent form when written in terms of spinors. It is then quite a direct generalization of the linearized Lagrangian of Yang-Mills theory. It is worth stressing that the linearized theory is the same for any member of the class (\ref{action}). 

\subsection{Mode expansion}

Once rewritten via the space plus time decomposition, the Lagrangian (\ref{free}) describes two propagating polarizations of the graviton. Only the symmetric tracefree transverse part of the spatial connection $a^i_j$ propagates, and, taking into account the connection reality conditions, one obtains the following rather non-trivial mode decomposition for $a^{ij}$ 
\be\label{a-mode-dec}
a_{ij}(t,\vec{x}) = \int \frac{d^3k}{(2\pi)^3 2\omega_k} \Big[ \varepsilon_{ij}^-(k) u_k(x) a_k^-  + \varepsilon_{ij}^+ (k) v_k^*(x) (a_k^-)^\dagger
-\varepsilon_{ij}^+(k)  v_k(x) a_k^+ - \varepsilon_{ij}^-(k) u_k^*(x) (a_k^+)^\dagger \Big],
\ee
where we have introduced the following polarization tensors
\be\label{polars}
\varepsilon_{ij}^-(k) = \frac{M}{\sqrt{2}\omega_k} m_i(k) m_j(k), \qquad
\varepsilon_{ij}^+(k) = \frac{\sqrt{2}\omega_k}{M} \bar{m}_i(k) \bar{m}_j(k),
\ee
and the following modes
\be\label{modes}
u_k(x) = \frac{\Hc}{M} e^{-\im\omega_k t+\im\vec{k}\vec{x}}, \qquad
v_k(x) = \frac{M}{\Hc} e^{-\im\omega_k t+\im\vec{k}\vec{x}} \left( 1 + \frac{\im \Hc}{\omega_k} - \frac{\Hc^2}{2\omega_k^2} \right).
\ee
Note that the polarization tensors $\varepsilon^\pm_{ij}(k)$ and the modes $u_k(x), v_k(x)$ are {\it not} complex conjugates of each other, and so the connection is not Hermitian. Instead, it satisfies a reality condition that relates the Hermitian conjugation of the connection to its second derivative, see \cite{Delfino:2012zy}.

Here we work in the de Sitter background with $ds^2 = c^2(t) \left( -dt^2 + \sum_i (dx^i)^2\right)$, and $c(t)=\Hc/M$, where the Hubble parameter is given by
\be
\Hc = \frac{M}{1-Mt}.
\ee
The rather complicated-looking time dependent functions multiplying the usual plane waves in (\ref{modes}) have to do with the fact that we are in the time-dependent de Sitter background. We have normalized the mode functions in such a way that they have a well-defined limit as $M\to 0$, and in this limit they become the usual plane waves. Later we will give a prescription for how the graviton scattering calculations can effectively be done in Minkowski space. The quantity $\omega_k$ in (\ref{a-mode-dec}), (\ref{polars}) is $\omega_k = |\vec{k}|$, and the vectors $m^i,\bar{m}^i$ in (\ref{polars}) are $\vec{k}$-dependent. Suppressing this dependence to have compact formulas, these vectors satisfy
\be\label{ms}
\im \epsilon^{ijk} z_j m_k=m_i, \qquad \im \epsilon^{ijk} z_j \bar{m}_k=-\bar{m}_i, \qquad \im \epsilon^{ijk}m_j \bar{m}_k=z_i,
\ee
and $\vec{z}=\vec{k}/|\vec{k}|$ is the unit vector in the direction of the spatial momentum $\vec{k}$. Finally $(a_k^\pm)^\dagger, a_k^\pm$ are the creation-annihilation operators for the two helicities, which satisfy the canonical commutational relations
\be
[a_k^\pm, (a_k^\pm)^\dagger]= (2\pi)^3 2\omega_k \delta^3(k-p).
\ee

\section{LSZ reduction and the Minkowski limit}
\label{sec:LSZ}

In this section we describe how graviton scattering amplitudes can be derived from the connection correlation functions. We will also give a detailed prescription of how the Minkowski spacetime amplitudes are extracted. We shall see that, if one is only interested in the Minkowski space graviton scattering amplitudes, then all calculations can effectively be done in Minkowski space, where the usual Fourier transform (including in the time direction) is available. However, setting up the corresponding formalism requires some care, because of the blowing up factors of $1/M$ in the interaction vertices, see below. The content of this section is new. 

\subsection{Creation-annihilation operators}

To obtain a version of the LSZ reduction for our theory, we need expressions for the graviton creation-annihilation operators in terms of the field operator. To obtain these, let us first give expressions for the Fourier transform of the connection field operator. From (\ref{a-mode-dec}) we get:
\be
\int d^3 x \, e^{-\im\vec{k}\vec{x}} \, a^{ij}(t,\vec{x}) = \frac{1}{2\omega_k} \Big[ m^i m^j a_k^- \frac{\Hc}{\sqrt{2}\omega_k} e^{-\im\omega_k t} + m^i m^j (a_{-k}^-)^\dagger \frac{\sqrt{2}\omega_k}{\Hc} e^{\im\omega_k t} \left( 1 - \frac{\im \Hc}{\omega_k} - \frac{\Hc^2}{2\omega_k^2} \right) \\ \nonumber
-\bar{m}^i\bar{m}^j a_k^+ \frac{\sqrt{2}\omega_k}{\Hc} e^{-\im\omega_k t} \left( 1 + \frac{\im \Hc}{\omega_k} - \frac{\Hc^2}{2\omega_k^2} \right) - \bar{m}^i \bar{m}^j (a_{-k}^+)^\dagger \frac{\Hc}{\sqrt{2}\omega_k} e^{\im\omega_k t} \Big],
\ee
We have used $m^i(-k) = \bar{m}^i(k)$ in the second and fourth terms. For compactness, the $k$ dependence of the null vectors $m^i,\bar{m}^i$ is suppressed in the above formula, and it is assumed that they are all evaluated at the 3-vector $k$. We can now take the projection of the $m^i m^j$ or $\bar{m}^i\bar{m}^j$ terms, and then device an appropriate linear combination of the connection and its first time derivative to extract the creation-annihilation operators. We get
\be
a^-_k = \im \varepsilon_{ij}^+(k) \int d^3x\, v_k^*(x) \overleftrightarrow{\partial_t} a^{ij}, \qquad
a^+_k = -\im \varepsilon_{ij}^-(k) \int d^3x\, u_k^*(x) \overleftrightarrow{\partial_t} a^{ij}, \\ \nonumber
(a^-_k)^\dagger = -\im \varepsilon_{ij}^-(k) \int d^3x\, u_k(x) \overleftrightarrow{\partial_t} a^{ij}, \qquad
(a^+_k)^\dagger = \im \varepsilon_{ij}^+(k) \int d^3x\, v_k(x) \overleftrightarrow{\partial_t} a^{ij},
\ee
where, as usual $f \overleftrightarrow{\partial_t} g = f\partial_t g- g \partial_t f$, and $u_k(x), v_k(x)$ are the modes given by (\ref{modes}). Importantly, all the creation-annihilation operators are expressed solely in terms of the field $a^{ij}$, and the complex conjugate field never appears. Thus, it is quite non-trivial to see that e.g. $(a^-_k)^\dagger$ is the Hermitian conjugate of $a_k^-$. This would involve using the reality condition for the field operator $a^{ij}$. 

As usual in the proof of the LSZ reduction formulas, see e.g. \cite{Srednicki:2007qs} Chapter 5, we now take time integrals of the time derivative of the creation-annihilation operators. These are zero in free theory, but the corresponding expressions are used in an interacting theory to extract the scattering amplitudes. So, we have
\be\nonumber
a^-_k(\infty)-a^-_k(-\infty) = \int dt \, \partial_t a^-_k = \im \varepsilon_{ij}^+(k) \int d^4x\, v_k^*(x) \bar{D}D a^{ij},
\\ \nonumber
a^+_k(\infty)-a^+_k(-\infty) = \int dt \, \partial_t a^+_k = -\im \varepsilon_{ij}^-(k) \int d^4x\, u_k^*(x) \bar{D}D a^{ij}, \\ \nonumber
(a^-_k)^\dagger(\infty)-(a^-_k)^\dagger(-\infty) = \int dt \, \partial_t (a^-_k)^\dagger = -\im \varepsilon_{ij}^-(k) \int d^4x\, u_k(x) \bar{D}D a^{ij}, \\ \nonumber
(a^+_k)^\dagger(\infty)-(a^+_k)^\dagger(-\infty) = \int dt \, \partial_t (a^+_k)^\dagger = \im \varepsilon_{ij}^+(k)\int d^4x\, v_k(x) \bar{D}D a^{ij},
\ee
where the readers are referred to \cite{Delfino:2012zy} for the definitions of operators $D,\bar{D}$. Note that on connection satisfying its free theory field equation $\bar{D}D a=0$ all these quantities are zero. We can now use these expressions to state the rules for extracting the graviton scattering amplitudes from the interacting theory connection correlation functions. 

\subsection{LSZ reduction}

Quantum field theory in de Sitter space is an intricate subject with many subtleties. Because the background is time-dependent, one may argue that even the very in-out S-matrix is no longer defined, see e.g. \cite{Marolf:2012kh} for a recent nice description of the difficulties that arise (and the possible ways to handle them). However, since in this paper we are only interested in extracting the Minkowski limit results from our formalism, we can ignore all the subtleties and proceed in an exact analogy to what one does in Minkowski space. 

We thus insert a set of graviton creation operators in the far past, and then a set of annihilation operators in the far future to form a graviton scattering amplitude
\be
\langle a^-_{k_-}(\infty) \ldots a^+_{k_+}(\infty) \ldots  | (a^-_{p_-})^\dagger(-\infty) \ldots  (a^+_{p_+})^\dagger(-\infty) \ldots \rangle \\ \nonumber
\equiv \langle k_- \ldots k_+ \ldots | p_- \ldots  p_+ \ldots \rangle.
\ee
Here $k_- \ldots$ and $k_+ \ldots$ are the set of $n$ negative and $m$ positive helicity outgoing graviton momenta, and $p_- \ldots p_+\ldots$ are the incoming momenta of $n'$ negative and $m'$ positive helicity gravitons. We now add the time ordering, and then express the annihilation operators in the future in terms of those in the past, and creation operators in the past in terms of those in the future via the formulas obtained in the previous subsection. This results in the following formula for the scattering amplitude
\be\label{LSZ}
\langle k_- \ldots  k_+ \ldots  | p_- \ldots p_+ \ldots \rangle  = 
\im^{n+n'-m-m'}  
\int d^4 x_- \,\varepsilon^+_{i j} (k_-) v^*_{k_-}(x_-) \bar{D} D \ldots   \\ \nonumber
\int d^4 x_+ \,\varepsilon^-_{k l} (k_+) u^*_{k_+}(x_+) \bar{D} D \ldots 
\int  d^4y_- \,\varepsilon^-_{m n} (p_-) u_{p_-}(y_-)\bar{D} D \ldots  
\int  d^4y_+ \,\varepsilon^+_{r s} (p_+)v_{p_+}(y_+)\bar{D} D  \ldots \\ \nonumber
\langle T a^{i j}(x_-) \ldots a^{k l}(x_+)\ldots a^{m n}(y_-) \ldots a^{r s}(y_+)\rangle.
\ee
The time-ordered correlation functions are then obtained from the functional integral via the usual perturbative expansion. 

We note some unusual features of the formula (\ref{LSZ}). A similar formula can be written for extracting the amplitudes from the metric correlators. In this case, however, because the field equation satisfied by the metric perturbation is real, there is only one solution for each sign of the frequency. In other words, only one type of mode would appear in (\ref{LSZ}), together with its complex conjugate. In the case of the connection the field equation $\bar{D}D a=0$ is complex. This has the effect that for a given sign of the frequency, e.g. positive, there are two linearly independent solutions that we denoted by $u_k(x), v_k(x)$. This is of course just a manifestation of the parity asymmetry of our formalism, because one of the modes is used for the negative helicity and the other for the positive. 

\subsection{Minkowski limit: general discussion}

Now we would like to understand how the Minkowski spacetime graviton scattering amplitudes can be extracted from the general formula (\ref{LSZ}). The standard procedure for taking such a limit is to keep the past and future time limits in the LSZ formula finite, take the $M\to 0$ limit, and then take the time limits to infinity.\footnote{KK is grateful to Richard Woodard for a discussion on this point.} This procedure, however, requires performing all computations in the position space, which is doable, but has the drawback of the spinor helicity formalism being not available to aid the computation. For this reason we will follow an alternative route that will eventually allow us to use the usual Fourier transform and the momentum spinors. 

To set the stage, let us first discuss in more details how to take the Minkowski limit in the metric formalism. In this case one can also write a version of LSZ formula (\ref{LSZ}), and the amplitudes are obtained from correlation functions of the metric perturbation. For definiteness let us  consider just one Feynman diagram contributing to some scattering amplitude. In this diagram, every graviton (on internal or external line) is characterized by its energy $\omega_k$, and we would like to concentrate on processes for which for all gravitons $\omega_k \gg M$. A systematic way to do this is to expand all building blocks of this Feynman diagram (i.e. external wave functions, propagators and vertices) in powers of $M/\omega_k$. The leading order term in this expansion is the desired Minkowski limit. 

Then, instead of first computing the result in de Sitter space and then taking a limit, one can reduce the problem to a computation in Minkowski space. To see this, we need the above mentioned rule that the time intervals are taken to be finite, the $M\to 0$ limit is computed and then the time interval is sent to infinity. Indeed, consider the time dependence of the external wave-functions and the propagators. For $\omega_k \gg M$ this can be separated into the "fast" time dependence coming from the exponents $e^{\im \omega_k t}$ and "slow" coming from the factors of $\Hc$. The covariant derivatives present in the vertices act on the external wave functions and the propagators, and the result is sensitive both to the fast and slow dependence. Let us assume that all the time derivatives have been evaluated. Then, keeping in mind that we take the limit $M\to 0$ keeping all time intervals finite, we can switch off the "slow" time dependence by setting $Mt\to 0$ everywhere. In physical terms, this corresponds to assuming that the duration of any process is much shorter than the de Sitter time scale. With the slow time dependence switched off, the space and time dependence of all the quantities is as in flat space, and the Fourier transform becomes available. Effectively, the above discussion implies that we can just do the computation in Minkowski space, using the Minkowski limits of the vertices, propagators and the external wave-functions.  

The situation in our formalism is not so simple. The source of complication, absent in the metric formulation is that our interaction vertices have powers of $1/M$ in front of them, see below. There are similar factors of $M$ and $1/M$ in the polarization tensors. Thus, in computing Feynman diagrams and applying (\ref{LSZ}) we face the problem that we can only take the $M\to 0$ limit of the whole scattering amplitude, but not of the pieces that it is built from, as some of these pieces diverge in the limit, while some tend to zero. This appears to prevent us from doing the computation in Minkowski space. 

We can, however, circumvent this problem by taking the limit in two steps. First, we can compute all time derivatives (coming from the vertices) present in a given Feynman diagram, and then switch off the slow time dependence. At this stage all the quantities are still $M$-dependent, only the time dependence became the fast time dependence of the exponentials $e^{\im \omega_k t}$. We can then expand all the quantities, i.e. the external wave-functions and propagators (possibly with derivatives applied to them) in powers of $M/\omega_k$. At this stage, while $M$-dependence is still there, we can already do computations in Minkowski space. In the second step, after the Feynman diagram is evaluated, one keeps only the leading, zeroth order in $M$ term in the result. 

An apparent difficulty with this prescription is that it seems hard to decide how many orders of the expansion in powers of $M/\omega_k$ we need to keep for each quantity. Indeed, in principle there could be powers of $1/M$ coming from the (positive) polarization tensors and the vertices cancelling the powers of $M$ coming from $(M/\omega_k)^n$, and producing a finite result. This difficulty is resolved by the following consideration. Let us start by considering the leading order terms in the expansion of all the quantities. Let us call the result of the computation of the Feynman diagram where only the leading orders are kept its {\it leading order part}. The first assumption we need to make is that for physically relevant diagrams (i.e. those arising in computations of scattering amplitudes) it is never the case that the leading order part blows up in the $M\to 0$ limit. We will see that this assumption is true for all the amplitudes considered in this paper. It is possible that one can prove this in full generality, but we will not attempt it here. However, see some further discussion about this point below. 

Let us now consider some amplitude with the leading part surviving in the limit $M\to 0$. For such an amplitude it is clear that the corrections obtained by keeping the subleading terms in the expansion of all its building blocks will be vanishing in the $M\to 0$ limit. This establishes that it is sufficient to consider only the leading parts in the expansion of the (derivatives of the) wave functions and the propagators. 

Let us now come back to our assumption that the leading order part is never singular in the $M\to 0$ limit. What we will see below when we do the computations is that the potentially singular leading order parts of the amplitudes, i.e. containing a negative power of $M$ in front of them, are always zero for reasons of spinor contractions. In other words, these amplitudes are such that the spinors coming from the helicity spinors inserted on the external lines contract to zero. It could, in principle happen that there are {\it subleading} in $M/\omega_k$ terms for which the spinor contractions do not give a null result, and this conspires to give a non-zero amplitude in the $M\to 0$ limit. If this was the case, it would not be sufficient to just work with the leading parts of all the building blocks of the diagrams, which would make the analysis significantly more complicated. While we cannot give a general proof that this does not happen, the reasons for why the spinor contractions give a null result for potentially "dangerous" amplitudes are always very general, and appear to hold independently of any expansion in powers of $M/\omega_k$. We thus assume that the "dangerous" diagrams that have zero leading parts for reasons of spinor contractions are zero precisely, i.e. even before any expansion in $M/\omega_k$ is performed. This allows us to work with only the leading parts of all the objects. As we have said, to us this assumption appears to be well motivated by the details of the computations that are made below. However, it would be comforting to either perform explicit checks (by doing computations in de Sitter space), or find another direct argument proving this. At the moment, all our analysis is based on this assumption. Our results, and in particular the fact that the GR amplitudes are correctly reproduced, seem to justify the assumption, but more direct arguments may be necessary to establish this beyond doubt. 

\subsection{Minkowski limit: prescription for the leading order parts}

Let us see how all this works in practice. From the above discussion we know that we need to keep the leading order of the external wave-functions and propagators, as well as their derivatives (as coming from the vertices), when everything is expanded in powers of $M/\omega_k$. For propagators this is easy, as the leading order of the propagator itself is just its Minkowski limit, and the leading order of a covariant derivative of the propagator is just the usual partial derivative of the Minkowski limit. In physical terms, this can be rephrased by saying that in Minkowski limit all internal lines, as well as the derivatives acting on them, are unaware of the fact that they are actually objects in de Sitter space. Thus, we only need to worry about the leading order parts of the external wave-functions and their covariant derivatives. 

The covariant derivative that is applied to an external graviton wave-function appears in our interaction vertices only in the combination $D_{[\mu} a_{\nu]}^i$. This Lie algebra-valued two-form can be further decomposed into its self-dual and anti-self-dual parts as in (\ref{Da}) below. From the Hamiltonian analysis in \cite{Delfino:2012zy}, see formula (87) of this reference, we know that the self-dual part of $D_{[\mu} a_{\nu]}^i$ is essentially given by the action of the operator $D$ on the spin 2 component $a^{ij}$ of the spatial connection. This is modulo the term involving the derivative of the temporal component of the connection $a_0^i$ (shifted by $c^i$, see (87) of \cite{Delfino:2012zy}), which is set to zero in the Hamiltonian treatment. Similarly, it is clear that the anti-self-dual part of $D_{[\mu} a_{\nu]}^i$ is obtained by applying the derivative operator $\bar{D}$ to $a^{ij}$, see formula (98) of \cite{Delfino:2012zy} for the definition of both operators. Thus, we have to consider the action of $D,\bar{D}$ on all the wave-functions that appear in (\ref{LSZ}), i.e. on $\epsilon^-(k) u_k(x)$ and $\epsilon^-(k) u^*_k(x)$, as well as on $\epsilon^+(k) v_k(x)$ and $\epsilon^+(k) v^*_k(x)$. Moreover, as we discussed above, we are only interested in the leading-order behavior of these derivatives in the limit $M\to 0$. We get
\be\nonumber
D \epsilon^-(k) u_k = D \epsilon^-(k) u^*_k =0, \\ \nonumber \qquad D \epsilon^+(k) v_k \to -2\omega_k \epsilon^+(k)  e^{-\im\omega_k t + \im\vec{k}\vec{x}} , \qquad D \epsilon^+(k) v^*_k \to 2\omega_k \epsilon^+(k)  e^{\im\omega_k t - \im\vec{k}\vec{x}} .
\ee
The action of the $\bar{D}$ operator has been worked out in (135) of \cite{Delfino:2012zy}. In the limit $M\to 0$ we can write
\be\nonumber
\bar{D} \epsilon^-(k) u_k \to 2\omega_k e^{-\im\omega_k t + \im\vec{k}\vec{x}} \epsilon^-(k) , \qquad
\bar{D} \epsilon^-(k) u^*_k \to - 2\omega_k e^{\im\omega_k t - \im\vec{k}\vec{x}} \epsilon^-(k) , \\ \nonumber
\bar{D} \epsilon^+(k) v_k \to - \frac{M^2}{\omega_k} e^{-\im\omega_k t + \im\vec{k}\vec{x}} \epsilon^+(k) , \qquad
\bar{D} \epsilon^+(k) v^*_k \to \frac{M^2}{\omega_k} e^{\im\omega_k t - \im\vec{k}\vec{x}} \epsilon^+(k) .
\ee
The idea now is to device some Minkowski spacetime wave-functions that give exactly the same leading order results when the operators $\lim_{M\to 0} D = -\im \partial_t + \epsilon\partial, \lim_{M\to 0} \bar{D} = \im \partial_t + \epsilon\partial$ are applied. We also note that the limiting case operators are just the complex conjugates of each other. 

For the modes involving $u_k$ and its complex conjugate the answer is obvious --- one should just take $\lim_{M\to 0} u_k$ as the corresponding wave-function. Thus, we set
\be\label{u-M}
u_k^{M}(x) := e^{-\im\omega_k t + \im\vec{k}\vec{x}},
\ee
and use this wave-function (and its complex conjugate) instead of $u_k(x)$ and $u_k^*(x)$ every time it appears in the LSZ formula (\ref{LSZ}). The operators that act on $u_k^M(x)$ (and its complex conjugate) are the Minkowski limit ones $-\im \partial_t + \epsilon\partial$ and $\im \partial_t + \epsilon\partial$.

For the $v_k$ mode the situation is more non-trivial. If we choose it to be just the ordinary plane wave we will correctly reproduce the leading order of the action of the $D$ operator. However, the $\lim_{M\to 0} \bar{D} = \im \partial_t + \epsilon\partial$ operator will give zero. The reason why one gets a non-zero answer when acting on the full wave-function $v_k(x)$ is that this has a non-trivial time-dependent factor multiplying $e^{-\im\omega_k t + \im\vec{k}\vec{x}}$. The idea is then to change the frequency $\omega_k$ in the plane wave to model this non-trivial time-dependence factor. This is achieved by the following plane wave
\be\label{v-M}
v^{M}_k(x):=e^{-\im\tilde{\omega}_k t + \im\vec{k}\vec{x}},
\ee
where
\be\label{omega-mass}
\tilde{\omega}_k:= \sqrt{\omega_k^2- 2M^2}=\omega_k \left( 1- \frac{M^2}{\omega_k^2} +O\left(\frac{M^4}{\omega_k^4}\right) \right).
\ee
Indeed, modulo the higher order corrections this choice of the wave-function gives precisely the required
\be
(\im\partial_t +\epsilon\partial) \varepsilon^+(k) v_k^M(x) = - \frac{M^2}{\omega_k} \varepsilon^+(k) v_k^M(x),\quad (\im\partial_t +\epsilon\partial) \varepsilon^+(k) v_k^{*M}(x) = \frac{M^2}{\omega_k} \varepsilon^+(k) v_k^{*M}(x).
\ee
At the same time, the leading order term in the result of the action of $-\im\partial_t +\epsilon\partial$ on this mode is unchanged by this shift of the frequency. 

Thus, the choice (\ref{u-M}), (\ref{v-M}) satisfies the requirement that when acted upon by the operators $-\im \partial_t + \epsilon\partial$ and $\im \partial_t + \epsilon\partial$ one reproduces the leading behavior of the derivatives of the wave-functions with the full time-dependence. Therefore, if we are interested in the leading order parts of up to the first derivatives of the wave-functions, it is sufficient to replace the full wave-functions by  (\ref{u-M}), (\ref{v-M}), and the operators $D,\bar{D}$ by the corresponding $M\to 0$ limit operators. 

\subsection{The prescription for Minkowski limit amplitudes}

All in all, we see that the computation of the Minkowski graviton amplitudes can be reduced to computations with the Minkowski $1/k^2$ propagator (\ref{prop}), the vertices obtained by replacing the covariant derivatives with partial ones everywhere, with the polarization tensors (\ref{polars}), and with the wave-functions (\ref{u-M}), (\ref{v-M}). The rule for the wave-functions is that the on-shell condition for the negative helicity graviton is the usual one $\omega=\pm\omega_k$, while the positive helicity graviton should be taken to have a small mass $\omega=\pm\tilde{\omega}_k$, with $\tilde{\omega}_k$ given by (\ref{omega-mass}). In doing the computation one keeps all the $M$-dependent prefactors, and at the end takes the limit $M\to 0$ (if this exists). 

Let us write the above prescription in terms of a formula. Given that our Minkowski space wave-functions (\ref{u-M}) and (\ref{v-M}) are just the standard plane waves, we can immediately write down the momentum space LSZ prescription. Indeed, we note that the position space integrals give us the Fourier transforms of the time-ordered correlation function. As usual we will be assigning an arrow to each line in the Feynman diagram, with the arrows on incoming lines pointing towards the diagram, and the arrows on outgoing lines away from it. Then as usual the 4-momentum of each particle is to be understood as in the direction of the arrows. This prescription takes care of the factors of $u_k^M, v_k^M$ and their complex conjugates. The factors of $\bar{D}D$ will then amputate the propagators on the external lines. This will also absorb the overall factors of $\im$ in the formula (\ref{LSZ}). However, with our conventions for the mode decomposition there are some signs that are left over. Indeed, we have $\bar{D}D= \partial_t^2 - \Delta = -\partial^\mu\partial_\mu = - \Box$. At the same time we have $-\Box e^{\im k x}= k^2 e^{\im k x}$, and so the operator $\bar{D}D$ will cancel $k^2$ from each external line. However, in our conventions the propagator is $1/\im k^2$, and so we will have a prefactor of $(-1)^{m+m'}$ left, where $m,m'$ are the numbers of incoming and outgoing {\it positive} helicity gravitons. This nicely follows the pattern that positive helicity gravitons are a source of headache in the formalism. However these minus signs in the amplitudes are convention dependent and are not of any physical significance. We finally have:
\be\label{LSZ-mom}
\langle k_- \ldots  k_+ \ldots  | p_- \ldots p_+ \ldots \rangle  = (-1)^{m+m'} (2\pi)^4 \delta^4\left(\sum k - \sum p\right)  \\ \nonumber
\varepsilon^+_{i j} (k_-) \varepsilon^-_{k l} (k_+) \varepsilon^-_{m n} (p_-) \varepsilon^+_{r s} (p_+)\, \langle T a^{i j}(k_-) \ldots a^{k l}(k_+)\ldots a^{m n}(p_-) \ldots a^{r s}(p_+)\rangle_{\rm amp},
\ee
where the momentum space amplitude is amputated from its external line propagators. We should add to this formula the prescription that the positive helicity incoming particle 4-momentum, as well as the 4-momentum of the negative helicity outgoing particle is slightly massive, see (\ref{omega-mass}), while the 4-momenta of the incoming negative and outgoing positive helicity are massless $\omega=\omega_k=|k|$. 

The above prescription is guaranteed to reproduce the correct leading order $M$-dependence. Thus, it is only consistent if the answers one gets turn out to have non-zero limit as $M\to 0$, i.e. if all the factors of $M$ cancel out from the end result. Below we will see that this is the case for the graviton-graviton amplitudes.

\subsection{Crossing symmetry}

We can now ask about an analog of the field theory crossing symmetry relation for our scattering amplitudes.  We shall discuss the crossing symmetry in the Minkowski space limit only. We recall that the usual QFT crossing symmetry arises if one takes an incoming particle of momentum $\vec{k}$ and energy $\omega_k$, and analytically continues the amplitude to energy $-\omega_k$. The amplitude can then be interpreted as that of an outgoing anti-particle of momentum $-\vec{k}$. 

Let us see what happens in our case. Since the field is electrically neutral, our particles are their own anti-particles, so we should only expect the helicity to change if we flip one graviton from the initial to the final state. That this is indeed the case is seen from our formula (\ref{LSZ-mom}). Indeed, to make an outgoing particle incoming one should just flip the direction of the arrow on Feynman diagram external line corresponding to that particle. This will correctly continue $\omega_k\to -\omega_k$ and $\tilde{\omega}_k\to -\tilde{\omega}_k$, as well as change the sign of the corresponding 3-momentum. Note that we do not touch the helicity states. It is then clear that if we apply the crossing symmetry to a negative helicity outgoing graviton, we produce a positive helicity incoming one, and similarly if we make a positive helicity outgoing graviton to be incoming, we will get a negative helicity one. This is clear from the fact that we are projecting on the negative/positive polarization tensors for the incoming negative/positive helicity gravitons, but do the reverse projection for the outgoing ones. Since by flipping one graviton from outgoing to incoming state we always change the total number of positive helicity gravitons in the amplitude (negative becomes positive, positive becomes negative, so there is always a change in the number of total positive helicity particles), then any crossing symmetry flip always introduces a minus sign coming from $(-1)^{m+m'}$ prefactor. Such a minus sign is appropriate for fermions, and here we see it occurring for purely bosonic particles, which again signifies the analogy between our treatment and that of fermions. 

Thus, we see that the crossing symmetry operates within our formalism, and we can from now on restrict our attention to all gravitons being e.g. incoming. Realistic scattering amplitudes can then be obtained from these by applying the crossing symmetry relations. 

\section{Gauge-fixing and the propagator}
\label{sec:gf}

We will now derive the Feynman rules, starting with the propagator (that was already referred to in the previous section), and then finishing with the interaction vertices. The propagator in the pure connection formalism was obtained in \cite{Krasnov:2011up}. Here we repeat the analysis, by a slightly simpler method, and add some details such as the gauge-fixed linearized action before the Minkowski limit is taken. We also derive the ghost sector Feynman rules. We first work in de Sitter space and then take the Minkowski limit, as is explained in the previous section. 

\subsection{Diffeomorphisms}

As is discussed in more details in \cite{Krasnov:2011up}, \cite{Delfino:2012zy}, at the linearized level diffeomorphisms act as $\delta_\xi a_\mu^i = \xi^\alpha \Sigma_{\alpha\mu}^i$. It is not hard to see what this action is by decomposing the field $a_\mu^i$ into its irreducible components with respect to the action of the Lorentz group. Thus, let us introduce the following projectors
\be\label{proj-31}
P^{(3,1)}_{\mu i| \nu j} := \frac{2}{3} \left( \delta_{ij} g_{\mu\nu} - \frac{1}{2} \epsilon_{ijk} \Sigma^k_{\mu\nu} \right), \qquad P^{(1,1)}_{\mu i| \nu j} := \frac{1}{3} \left( \delta_{ij} g_{\mu\nu}+ \epsilon_{ijk} \Sigma^k_{\mu\nu} \right).
\ee
Both act on pairs $\mu i$ of a spacetime index and an internal one. The projector $P^{(3,1)}$ is on the irreducible component $S_+^3\otimes S_-$, and $P^{(1,1)}$ is on $S_+\otimes S_-$ in the spinor representation $S_+^2 \otimes S_+\otimes S_-$ that the pair $\mu i$ lives in. Here $S_+, S_-$ are two 2-dimensional fundamental representations of the Lorentz group (i.e. the representation realized by unprimed and by primed 2-component spinors). The diffeomorphisms are then simply shifts of the $(1,1)$ component. They can be completely gauge-fixed by requiring
\be\label{gf-cond}
a_\mu^i=\epsilon^{ijk} \Sigma^j_{\mu}{}^\nu a_\nu^k,
\ee
or
\be\label{diff-gf}
P^{(1,1)}_{\mu i | \nu j} a^{\nu j}=0.
\ee
It is important to stress that the gauge-fixing condition for the diffeomorphisms does not contain derivatives (as is appropriate for the transformation that is merely a shift of the field in some direction in the field space). 

We could now add the square of this gauge-fixing condition to the action (with some gauge-fixing parameter) and make the corresponding components of the connection propagating. However, this would have the effect that some components of the connection have the $1/k^2$ propagator, while the pure diffeomorphism gauge modes have a mode-independent, algebraic propagator. This would require dealing with the two components separately, which would make the formalism very cumbersome. To avoid this, we fix the gauge (\ref{diff-gf}) sharply, i.e. work in the corresponding Landau gauge. We shall later see that the gauge-fixing condition (\ref{diff-gf}) is particularly transparent when expressed in spinor terms. It will simply state that everything but the completely symmetric (in spinor indices) component of the connection is zero. This condition will be then easy to impose and it will simplify the computations significantly.

On the other hand, the remaining gauge freedom, namely the usual ${\rm SO}(3)$ transformations, will be gauge-fixed (in the next subsection) as in Yang-Mills theory, i.e. by adding a $D^\mu a_\mu^i$ term squared to the Lagrangian. It is interesting to note that the way the gauge is fixed in our pure connection approach is opposite to that used in e.g. the first order formulation of general relativity that possesses both the internal as well as diffeomorphism gauge symmetry, as in our case. In the first-order formulation the gauge symmetry corresponding to ${\rm SO}(1,3)$ local gauge transformations is fixed by requiring the tetrad perturbation $h^{IJ}, I,J=0,1,2,3$ to be symmetric, i.e. by projecting away some irreducible component of the field with respect to the action of the Lorentz group. The diffeomorphisms are then fixed in the usual derivative way, using the de Donder gauge. What happens in our formalism is precisely the opposite. The diffeomorphisms are fixed in a non-derivative way by projecting away some irreducible component of the field. The gauge rotations are then fixed by adding to the action the square of a term containing the derivative of the field. 

\subsection{Gauge-fixing the gauge rotations}

We now add to the Lagrangian the gauge-fixing term
\be\label{so-gf}
{\cal L}_{\rm gf} = - \frac{\alpha}{2} \left( D^\mu P^{(3,1)}_{\mu i| \nu j} a^{\nu j} \right)^2,
\ee
where the projector is inserted to make the gauge-fixing condition diffeomorphism-invariant. As in \cite{Krasnov:2011up}, we could have then done the transformations in full generality, without imposing the gauge-fixing (\ref{diff-gf}). We would find that (for a choice of $\alpha$ to be given below) the gauge-fixed linearized Lagrangian (modulo the background curvature term) is just a multiple of $a  D^\mu D_\mu P^{(3,1)} a$. However, we can simplify the computation significantly by imposing the gauge condition (\ref{diff-gf}) from the very beginning. 

The simpler computation is as follows. First, we note that the gauge-fixed connections satisfy $\Sigma^{\mu\nu i} a_\nu^i =0$. We can then ignore the $\delta_{ij}\delta_{kl}$ term in the projector $P_{ijkl}$, and write the Lagrangian (\ref{free}) as
\be\label{L-1}
{\cal L}^{(2)}=-\frac{1}{2}\delta_{ik} \delta_{jl}  (\Sigma^{(i\mu\nu}D_\mu a^{j)}_\nu) (\Sigma^{(k\rho \sigma}D_\rho a^{l)}_\sigma).
\ee
We now use the gauge-fixing condition (\ref{diff-gf}) to obtain the following identity
\be\label{ij-ji-ident}
\Sigma^{i\mu\nu} D_\mu a_\nu^j - \Sigma^{j\mu\nu} D_\mu a_\nu^i=  \epsilon^{ijk} D^\mu a_{\mu}^k.
\ee
In other words,
\be\label{SD-symm}
\Sigma^{(i\mu\nu}D_\mu a^{j)}_\nu = \Sigma^{i\mu\nu}D_\mu a^{j}_\nu - \frac{1}{2} \epsilon^{ijk} D^\mu a_{\mu}^k.
\ee
In the derivation of (\ref{ij-ji-ident}) we have secretly extended the covariant derivative with respect to the background connection to a derivative that also acts on the spacetime indices, so that $\Sigma^{\mu\nu i}$ can be taken trough the covariant derivative. Thus, from now on there is also the usual Christoffel symbol inside $D_\mu$. 

We now substitute (\ref{SD-symm}) into the Lagrangian (\ref{L-1}), and use the gauge-fixing condition once more to convert the result into a sum of just two terms:
\be
{\cal L}^{(2)} = -2 P^{+ \mu\nu\rho\sigma} \delta_{ij}  D_\mu a_\nu^i D_\rho a_\sigma^j +\frac{1}{4} \delta_{ab} (D^\mu a_{\mu}^i) (D^\nu a_{\nu}^j).
\ee
Here we have used the fact that
\be
\Sigma^{\mu\nu i} \Sigma^{\rho\sigma i} = 4 P^{+ \mu\nu\rho\sigma} = g^{\mu\rho} g^{\nu\sigma} - g^{\mu\sigma}g^{\nu\rho} - \im \epsilon^{\mu\nu\rho\sigma},
\ee
which is just a multiple of the self-dual projector. We now integrate by parts in the first term, and then represent the self-dual projector as
\be
4 P^{+ \mu\nu\rho\sigma}  = g^{\mu\rho} g^{\nu\sigma} - g^{\mu\nu} g^{\rho\sigma} + 4 P^{- \mu\rho\nu\sigma},
\ee
where $P^{-}$ is the anti-self-dual projector. Then in terms that are anti-symmetric in the covariant derivatives, we express the commutator\footnote{Our conventions are $\nabla_\mu V_\nu := \partial_\mu V_\nu - \Gamma_{\mu\nu}^\rho V_\rho$ for the covariant derivative and $2\nabla_{[\mu} \nabla_{\nu]} V_\rho = - R^\alpha{}_{\rho\mu\nu} V_\alpha = R_{\mu\nu\rho}{}^\alpha V_\alpha$ for the curvature.} of two covariant derivatives via the curvature tensors. We have
\be\label{DD-a}
D_{[\mu} D_{\rho]} a_\sigma^i = \frac{1}{2} R_{\mu\rho\sigma}{}^\alpha a_\alpha^i + \frac{1}{2} \epsilon^{ijk} F^j_{\mu\rho} a_\sigma^k,
\ee
where $R_{\mu\rho\sigma}{}^\alpha$ is the Riemann curvature. We can now use the fact that the last term here is a purely self-dual quantity, and thus drop this part as it will be multiplied by $P^{-\mu\rho\nu\sigma}$. We get:
\be
4 P^{-\mu\rho\nu\sigma} D_{[\mu} D_{\rho]} a_\sigma^i  = \frac{1}{2} (g^{\mu\nu} g^{\rho\sigma} - g^{\mu\sigma} g^{\rho\nu} + \im \epsilon^{\mu\rho\nu\sigma}) R_{\mu\rho\sigma}{}^\alpha a_\alpha^i = -R_{\nu}{}^\alpha a_\alpha^i,
\ee
where $R_{\nu}{}^\alpha:= g^{\sigma\rho} R_{\sigma\nu\rho}{}^\alpha$ is the Ricci tensor. We finally get
\be
{\cal L}^{(2)}  = - \frac{1}{2} \delta_{ij}  g^{\rho\sigma} D^\mu a_\rho^i D_\mu a_\sigma^j + \frac{3}{4} \delta_{ij} (D^\mu a_{\mu}^i) (D^\nu a_{\nu}^j)-\frac{1}{2}\delta_{ij} R^{\mu\nu} a_\mu^i a_\nu^j .
\ee
It is now clear that the choice $\alpha=3/2$ gives
\be\label{L-g-fixed}
{\cal L}^{(2)} +{\cal L}_{\rm gf} = - \frac{1}{2} \delta_{ij}  g^{\rho\sigma} D^\mu a_\rho^i  D_\mu a_\sigma^j -\frac{1}{2}\delta_{ij} R^{\mu\nu} a_\mu^i a_\nu^j,
\ee
which is just the scalar field Lagrangian for every component of $a_\mu^i$ (projected onto the $(3,1)$ representation by the gauge-fixing condition (\ref{diff-gf})), plus a curvature term. 

We can further rewrite (\ref{L-g-fixed}) by using the fact that the background metric is Einstein with positive scalar curvarture:
\be
R_{\mu\nu} = 3M^2 g_{\mu\nu}.
\ee
This gives for the gauge-fixed Lagrangian (after integrating by parts)
\be\label{L-g-fixed*}
{\cal L}^{(2)} +{\cal L}_{\rm gf} = \frac{1}{2} a^{\mu i}  \left( D^2- 3M^2\right)  a_\mu^i,
\ee
where $D^2=D^\mu D_\mu$. 

What is significant about the gauge-fixed Lagrangian (\ref{L-g-fixed*}) is that all modes appear in it with the same sign in front of their kinetic term, unlike in the metric-based description that exhibits the conformal factor problem. This is also the reason why the Lagrangian (\ref{L-g-fixed*}) gives rise to simpler Feynman rules than in the metric case. Indeed, in our case all the fields are treated uniformly, while in the metric case one often meets (e.g. in the vertices) the tracefree part of $h_{\mu\nu}$ and its trace separately, which makes computations more involved. 

\subsection{Minkowski space propagator}

The quadratic form in the gauge-fixed Lagrangian (\ref{L-g-fixed}) can be inverted in full generality, using the relevant de Sitter space modes to construct the associated Green's function. However, as we have already stated above, effectively we are doing all our graviton scattering calculations in Minkowski space. Then the Lagrangian (\ref{L-g-fixed}) admits an obvious Minkowski spacetime limit 
\be\label{L2-mink}
{\cal L}^{(2)} +{\cal L}_{\rm gf} = - \frac{1}{2} (\partial_\mu a_\nu^i)^2.
\ee
The propagator is now easily obtained by going to the momentum space, and is obviously a multiple of the projector $P^{(3,1)}$ times $1/k^2$. To get all the factors right we introduce into the action a source term, and integrate out the connection
\be
S^{(2)}[a,J]=\int\frac{\rmd^4k}{(2\pi)^4}\left[-\frac{1}{2}a^{\mu i}(-k) k^2 a^{i}_\mu(k)+J^{\mu i}(-k) a^i_{\mu}(k)\right].
\ee
Integrating out the connection we get
\be
S^{(2)}[J]=\int\frac{\rmd^4k}{(2\pi)^4}\frac{1}{2}J^{\mu i}(-k)\frac{P^{(3,1)}_{\mu i|\nu j}}{k^2}J^{\nu j}(k),
\ee
where the usual $\im\epsilon$ prescription is implied. The propagator is then
\be\label{prop}
\langle a_{\mu i}(-k)a_{\nu j}(k)\rangle=\left.\frac{1}{\im}\frac{\delta}{\delta J^{\mu i}(-k)}\frac{1}{\im}\frac{\delta}{\delta J^{\nu j}(k)}e^{\im S[J]}\right|_{J=0}=\frac{P^{(3,1)}_{\mu i|\nu j}}{\im\, k^2}
\ee

\subsection{The ghost sector}

We do not need to add any ghosts for the gauge-fixing condition (\ref{diff-gf}), or if we do these ghosts will not interact with any other fields. Thus, we only need to worry about the ghosts for fixing the non-Abelian ${\rm SO}(3)$ gauge-symmetry. This is done as in the case of Yang-Mills gauge theory, with the only non-triviality being that the projector $P^{(3,1)}$ is inserted into the gauge-fixing condition (\ref{so-gf}). We thus get the following ghost sector Lagrangian
\be\label{L-ghost}
{\cal L}_{gh} = \frac{3}{2} \bar{c}^i D^\mu P^{(3,1)}_{\mu i| \nu j} \left( D^\nu c^j- \frac{\im}{\sqrt{g^{(2)}}}\epsilon^{jkl} a^{\nu k} c^l\right),
\ee
where $c^i, \bar{c}^i$ are ghosts, and we have extended the derivative to the full, non-linear covariant derivative including the connection perturbation. We have also multiplied the ghost Lagrangian by $3/2$ to have the standard ghost $1/\im k^2$ propagator. The quantity $g^{(2)}$ is a coupling constant that in particular appears in the rescaling of the connection perturbation in order to put its kinetic term in the canonical form (\ref{L2-mink}). 

In the Minkowski limit, in which all computations will be done, we can replace $D_\mu\to \partial_\mu$. The $\Sigma$-part of the projector in the kinetic term can then be dropped, as the partial derivatives commute. We then end up with the standard Yang-Mills theory ghost propagator
\be
\langle \bar{c}^i(-k) c^j(k)\rangle  = \frac{\delta^{ij}}{\im\, k^2}.
\ee

\section{Interactions}
\label{sec:inter}

Having derived the propagator we only need the Feynman rule vertex factors, as well as polarizations to be used to project the external legs of diagrams onto physical graviton scattering amplitudes. We have already gave expressions (\ref{polars}) for the later when spelling out the mode decomposition. However, we will also need the covariant versions, and these will require introducing spinors. At the same time can get sufficiently far in the analysis of interactions without using spinors. Thus, we first work out the interactions. We first compute the full de Sitter interactions, and will specialize to the interaction vertices as relevant in the Minkowski limit in the section that computes their spinor versions.

\subsection{Decomposition of $Da$}

Before we do the algebra that exhibits the structure of the interaction vertices, let us introduce a convenient representation for the Lie-algebra valued two-form $D_{[\mu} a_{\nu]}^i$. We can write
\be\label{Da}
D_{[\mu} a_{\nu]}^i = \frac{1}{4} (Da)^{ij} \Sigma_{\mu\nu}^j +\frac{1}{8} \epsilon^{ijk} (D^\rho a_\rho^j) \Sigma^k_{\mu\nu} + (\widetilde{D a^i})_{\mu\nu},
\ee
where 
\be\label{Da-sd}
(Da)^{ij} := \Sigma^{\mu\nu (i} D_\mu a_\nu^{j)}, \qquad (Da)^{ij}=(Da)^{(ij)}, \qquad {\rm Tr}(Da)=0
\ee
is the symmetric tracefree matrix that encodes the self-dual components of $D_{[\mu} a_{\nu]}^i$, and $(\widetilde{D a^i})_{\mu\nu}$ stands for the anti-self-dual part
\be\label{Da-asd}
(\widetilde{D a^i})_{\mu\nu} := (D a^i)_{\mu\nu}^{\rm asd}.
\ee
Our gauge-fixing condition (\ref{gf-cond}), together with the fact that the propagator contains the $P^{(3,1)}$ projector, implies that in all vertices and on all the lines, internal and external, the connection $a_\mu^i$ can be taken to belong to just its $S_+^3\otimes S_-$ irreducible component. Indeed, on the internal lines this projection is carried out by the propagator. On the external lines it will be performed by the polarization tensors, see below. We can thus use the gauge-fixing condition for $a_\mu^i$. Then the matrix $\Sigma^{\mu\nu i} D_\mu a_\nu^{j}$ is traceless, and we have denoted its symmetric part by $(Da)^{ij}$ and wrote the anti-symmetric part as a separate (second) term in (\ref{Da}). As we shall see below, the above components of $D_{[\mu} a_{\nu]}^i$ encode different information, and this is why it is convenient to separate the self- and anti-self-dual parts of $D_{[\mu} a_{\nu]}^i$ in the vertices. 

Let us use the above expansion of $Da$ to rewrite some terms that frequently appear in the interaction vertices. We have
\be\label{DaDa}
\frac{1}{\im} \epsilon^{\mu\nu\rho\sigma} D_\mu a_\nu^i D_\rho a_\sigma^j = \frac{1}{2} (Da)^{ik} (Da)^{kj}-\frac{1}{2} (Da)^{k(i}\epsilon^{j)kl} (D^\mu a_\mu^l) + \frac{1}{8} (\delta^{ij}\delta^{kl} - \delta^{ik}\delta^{jl}) (D^\mu a_\mu^k)(D^\nu a_\nu^l)  \\ \nonumber - 2(\widetilde{Da^i})^{\mu\nu} (\widetilde{Da^j})_{\mu\nu} 
\ee
and
\be\label{Da-eps}
\frac{1}{\im} \epsilon^{\mu\nu\rho\sigma} D_\mu a_\nu^i \epsilon^{jkl} a_\rho^k a_\sigma^l = \frac{1}{2} (Da)^{ik} (\Sigma \epsilon aa)^{kj} -\frac{1}{4} \epsilon^{ikl}(\Sigma \epsilon aa)^{kj} (D^\mu a_\mu^l)- 2(\widetilde{Da^i})^{\mu\nu} \epsilon^{jkl} a_\mu^k a_\nu^l,
\ee
where we have introduced
\be\label{Saa}
(\Sigma \epsilon aa)^{ij}:= \Sigma^{\mu\nu i} \epsilon^{jkl} a_\mu^k a_\nu^l.
\ee
We note that $(\Sigma \epsilon aa)^{ij}$ is automatically symmetric as a consequence of (\ref{gf-cond}). 

\subsection{Cubic interaction}

The cubic interaction vertex is obtained from the third order terms in the expansion of the action (\ref{variations-action-general}). Dividing the third variation by $3!$, and rescaling the variation of the connection $\delta A_\mu^i = (\im/\sqrt{g^{(2)}}) a_\mu^i$ we get the following third order Lagrangian
\be\label{L3-temp}
3\im \,M^2 (g^{(2)})^{3/2} {\cal L}^{(3)} = g^{(3)} (Da)^{ij} (Da)^{jk} (Da)^{ki}
\\ \nonumber
- \frac{3g^{(2)}}{2} \left( \frac{1}{\im}\epsilon^{\mu\nu\rho\sigma} D_\mu a_\nu^i D_\rho a_\sigma^j - M^2 (\Sigma\epsilon  aa)^{ij} \right) (Da)^{ij} - \frac{1}{\im} f(\delta) M^2 \epsilon^{\mu\nu\rho\sigma} D_\mu a_\nu^i \epsilon^{ijk} a_\rho^j a_\sigma^k.
\ee
We now use (\ref{DaDa}) and (\ref{Da-eps}) to rewrite (\ref{L3-temp}) as 
\be \label{L3}
3\im M^2 (g^{(2)})^{3/2} {\cal L}^{(3)} = \left(g^{(3)} - \frac{3g^{(2)}}{4}\right) (Da)^{ij} (Da)^{jk} (Da)^{ki} + \frac{3g^{(2)}}{16}  (Da)^{ij} (D^\mu a_\mu^i) (D^\nu a_\nu^j) \\ \nonumber
+3g^{(2)} (Da)^{ij} (\widetilde{Da^i})^{\mu\nu} (\widetilde{Da^j})_{\mu\nu}  
+ \frac{M^2}{2} (3g^{(2)}-f(\delta)) (Da)^{ij}(\Sigma \epsilon aa)^{ij} + 2M^2 f(\delta) (\widetilde{Da^i})^{\mu\nu} \epsilon^{ijk} a_\mu^j a_\nu^k.
\ee
We note that in the case of GR, see (\ref{g-GR}), the first term in the line 1 and the second term in the line 2 above are absent and we get simply
\be\label{L3-GR}
\im M_p M {\cal L}_{\rm GR}^{(3)} = (Da)^{ij} (\widetilde{Da^i})^{\mu\nu} (\widetilde{Da^j})_{\mu\nu}
+ \frac{1}{16}  (Da)^{ij} (D^\mu a_\mu^i) (D^\nu a_\nu^j)+ 2M^2 (\widetilde{Da^i})^{\mu\nu} \epsilon^{ijk} a_\mu^j a_\nu^k.
\ee
Below we shall see that only the first of these 3 terms in the cubic GR Lagrangian is important for the scattering of two gravitons of opposite helicities. Note that the terms cubic in the derivatives of the connection blow up in the limit $M\to 0$, both in the general theory case and in GR. Thus, care will have to be taken when going to this limit. Note also that the cubic interaction starts with $(\partial a)^3$ terms, and thus seems to be very different from the $(\partial h)^2 h$ cubic vertex in the metric formulation. Still, we will see that in the case of (\ref{L3-GR}) one is working with just a different description of the same GR interactions of gravitons. 

We would like to emphasize how much simpler the cubic vertex (\ref{L3-GR}) is as compared to the 13 terms one finds in the expansion of the Einstein-Hilbert Lagrangian around the Minkowski background metric, see \cite{Goroff:1985th}, formula (A.5) of the Appendix. The cubic vertex (\ref{L3-GR}) is still more complicated than the one in the case of Yang-Mills theory, but we shall see that in many cases (e.g. for purposes of computing MHV amplitudes) one effectively needs only the first term, which is of the same degree of complexity as in the Yang-Mills case. The analogy with Yang-Mills will become even more striking when we write down the spinor expression for this cubic vertex below. 

\subsection{Quartic interaction}

We now work out the quartic term. Dividing the fourth variation of the action from (\ref{variations-action-general}) by $4!$ we get
\be\nonumber
&{}& -12 M^4 (g^{(2)})^2 {\cal L}^{(4)} =  - g^{(4)} (Da)^{ij} (Da)^{ij} (Da)^{kl} (Da)^{kl} \\ \nonumber
&+& 6 g^{(3)} P_{ijkl} (Da)^{im} (Da)^{mj} \left( \frac{1}{\im}\epsilon^{\mu\nu\rho\sigma} D_\mu a_\nu^k D_\rho a_\sigma^l - M^2 (\Sigma\epsilon  aa)^{kl} \right) 
\\ \nonumber
&+&\frac{6}{\im}   g^{(2)} M^2 (Da)^{ij}  \epsilon^{\mu\nu\rho\sigma} D_\mu a_\nu^i \epsilon^{jkl} a_\rho^k a_\sigma^l + g^{(2)} (Da)^{ij}(Da)^{ij} \left( \frac{1}{\im}\epsilon^{\mu\nu\rho\sigma} D_\mu a_\nu^k D_\rho a_\sigma^k - M^2 (\Sigma\epsilon  aa)^{kk} \right) \\ \nonumber
&-& \frac{3g^{(2)}}{2} P_{ijkl} \left( \frac{1}{\im}\epsilon^{\mu\nu\rho\sigma} D_\mu a_\nu^i D_\rho a_\sigma^j - M^2 (\Sigma\epsilon  aa)^{ij} \right)\left( \frac{1}{\im}\epsilon^{\mu\nu\rho\sigma} D_\mu a_\nu^k D_\rho a_\sigma^l - M^2 (\Sigma\epsilon  aa)^{kl} \right)
\ee
For purposes of this paper we will only need the 4-vertex when evaluated completely on-shell. Thus, let us use the Lorentz gauge condition $D^\mu a_\mu^i=0$. This simplifies both (\ref{DaDa}) and (\ref{Da-eps}). Using these expansions, and collecting the terms we get
\be\nonumber
-12 M^4 (g^{(2)})^2 {\cal L}^{(4)} = \left( -g^{(4)} +\frac{1}{2} g^{(3)} + \frac{7g^{(2)}}{16} \right) (Da)^{ij} (Da)^{ij} (Da)^{kl} (Da)^{kl} \\ \label{L4}
-3(4g^{(3)} - g^{(2)}) (Da)^{ik} (Da)^{kj} (\widetilde{Da^i})^{\mu\nu} (\widetilde{Da^j})_{\mu\nu} + (4g^{(3)} - 3g^{(2)}) (Da)^{ij} (Da)^{ij} (\widetilde{Da^k})^{\mu\nu} (\widetilde{Da^k})_{\mu\nu} \\ \nonumber
- \frac{3M^2}{2}( 4g^{(3)} - 3g^{(2)}) \left( (Da)^{ik}(Da)^{kj} -\frac{1}{3} {\rm Tr}((Da)^2) \delta^{ij}\right) (\Sigma \epsilon aa)^{ij} \\ \nonumber
-\frac{3g^{(2)}}{2} P_{ijkl} \left( 2(\widetilde{Da^i})^{\mu\nu} (\widetilde{Da^j})_{\mu\nu} + M^2  (\Sigma \epsilon aa)^{ij}\right)\left( 2(\widetilde{Da^k})^{\mu\nu} (\widetilde{Da^l})_{\mu\nu}+ M^2  (\Sigma \epsilon aa)^{kl}\right)  \\ \nonumber
- 12 M^2 g^{(2)} (Da)^{ij} (\widetilde{Da^i})^{\mu\nu} \epsilon^{jkl} a_\mu^k a_\nu^l.
\ee
In the case of GR many of these terms become zero and we get a much simpler (on-shell $D^\mu a_\mu^i=0$) 4-vertex for GR:
\be\label{L4-GR}
2M_p^2 M^2 {\cal L}^{(4)}_{\rm GR} = (Da)^{ik} (Da)^{kj} (\widetilde{Da^i})^{\mu\nu} (\widetilde{Da^j})_{\mu\nu} + 2 M^2  (Da)^{ij} (\widetilde{Da^i})^{\mu\nu} \epsilon^{jkl} a_\mu^k a_\nu^l \\ \nonumber
+ P_{ijkl} \left( (\widetilde{Da^i})^{\mu\nu} (\widetilde{Da^j})_{\mu\nu} + \frac{M^2}{2}  (\Sigma \epsilon aa)^{ij}\right)\left( (\widetilde{Da^k})^{\mu\nu} (\widetilde{Da^l})_{\mu\nu}+ \frac{M^2}{2}  (\Sigma \epsilon aa)^{kl}\right).
\ee
This should be compared with a much more formidable expression in the case of the metric-based GR, see \cite{Goroff:1985th}, formula (A.6). Even with the graviton field on-shell and the background metric taken to be flat, this occupies about  half a page, as compared to just two lines in (\ref{L4-GR}). We also note that both the GR 4-vertex as well as the general vertex (\ref{L4}) start with terms $(\partial a)^4$, to be compared with just two derivatives present in the metric-based vertex $(\partial h)^2 hh$. This is part of a general pattern, and in our gauge-theoretic description the order $n$ vertex starts from $(\partial a)^n$ terms. 

\subsection{Ghost sector interactions}

For completeness, we also spell out the ghost sector interactions, even though we will not need this in the present paper. Thus, the second, interaction term in (\ref{L-ghost}), after some simple algebra involving integrating by parts, expanding the product of two $\epsilon$ tensors and using the gauge-fixing condition (\ref{diff-gf}), becomes
\be
{\cal L}_{\rm gh}^{inter} = -\frac{\im}{\sqrt{g^{(2)}}} \partial^\mu \bar{c}^i \left( \epsilon^{ijk}  \eta_{\mu\nu} + \frac{1}{2} \delta^{ij} \Sigma^k_{\mu\nu} \right) a^{\nu j} c^k,
\ee
which is different from the usual Yang-Mills ghost vertex in the fact that an extra $\Sigma$-term is present. The factors in front of this interaction vertex that come from rescaling of the connection perturbation are also unusual.

\section{Spinor technology and the helicity spinors}
\label{sec:spinor}

As is common to any modern derivation of the scattering amplitudes, the formalism of helicity states turns out to be extremely convenient. These are most efficiently described using spinors, or, as some literature calls them, twistors. The recent wave of interest into the spinor helicity methods originates in \cite{Witten:2003nn}. The method itself is, however, at least twenty years older, see e.g. \cite{De Causmaecker:1981by}, \cite{Gunion:1985vca}. We start by listing some formulas involving spinors, mainly to establish the conventions. Our notations and conventions are more similar to those in the gravity literature, but we hope that the interested particle physics reader will still be able to follow. 

\subsection{Soldering form}

The soldering form provides a map from the space of vectors to the space of rank two spinors (with two indices of opposite types). We use the conventions with a Hermitian soldering form:
$$ (\theta_{\mu}^{AA'})^* = \theta_\mu^{AA'}.$$
The metric is obtained as a square of the soldering form:
\be\label{metric}
\eta_{\mu\nu} = - \theta_{\mu A}{}^{A'} \theta_{\nu B}{}^{B'} \epsilon^{AB} \epsilon_{A'B'},
\ee
where the minus sign is dictated by our desire to work with a Hermitian soldering form, while at the same time have signature $(-,+,+,+)$. We can also rewrite this formula as
\be
\eta_{\mu\nu}= \theta^A_{\mu A'} \theta_{\nu A}{}^{A'},
\ee
so that the minus sign disappears. The contraction that appears in this formula, i.e. unprimed indices contracting bottom left to down right, and the primed indices contracting oppositely, will be referred to as {\it natural} contraction. We will sometimes use index free notation and then the natural contraction will be implied. Another useful formula is
\be
\theta^{\mu AA'} \theta_{\mu BB'} = - \epsilon_B{}^A \epsilon_{B'}{}^{A'}.
\ee

\subsection{The spinor basis}

It is very convenient to introduce in each spinor space $S_+, S_-$ a certain spinor basis. Since each space is (complex) 2-dimensional we need two basis vectors for each space. Let us denote these by 
$$o_A, \iota_A \in S_+, \qquad o^{A'}, \iota^{A'}\in S_-.$$
Note that we shall assume that the basis in the space of primed spinors is the complex conjugate of the basis in the space $S_+$:
$$ \iota^{A'}=(\iota^A)^*, \qquad o^{A'}=(o^A)^*.$$

The basis vectors are pronounced as "omicron" and "iota". Since the norm of every spinor is zero, we cannot demand that each of the basis vectors is normalised. However, we can demand that the product between the two basis vectors in each space is unity. Thus, the basis vectors satisfy the following normalisation:
$$\iota^A o_A = 1, \qquad \iota^{A'} o_{A'}=1.$$
Of course, a spinor basis in each space $S_+, S_-$ is only defined up to an ${\rm SL}(2,\C)$ rotation. Any ${\rm SL}(2,\C)$ rotated basis gives an equally good basis, and it can be seen that any two bases can be related by a (unique) ${\rm SL}(2,\C)$ rotation. 

Once a spinor basis is introduced, we have the following expansion of the spinor metric, that is the $\epsilon_{AB}$ symbol
\be
\epsilon_{AB}=o_A \iota_B -\iota_A o_B.
\ee
A similar formula is also valid for $\epsilon_{A'B'}$. 

\subsection{The soldering form in the spinor basis}

The following explicit expression for the soldering form $\theta_{\mu A}{}^{A'}$ in terms of the basis one-forms $t_\mu$ and $x_\mu, y_\mu, z_\mu$, as well as the spinor basis vectors $o^A, o^{A'}, \iota^A, \iota^{A'}$ can be obtained:
$$\theta_\mu^{AA'} = \frac{t_\mu}{\sqrt{2}} ( o^A o^{A'} + \iota^A \iota^{A'}) + \frac{z_\mu}{\sqrt{2}}  ( o^A o^{A'} - \iota^A \iota^{A'}) + \frac{x_\mu}{\sqrt{2}}  ( o^A \iota^{A'} + \iota^A o^{A'})+\frac{\im y_\mu}{\sqrt{2}}  ( o^A \iota^{A'} - \iota^A o^{A'}).$$ 
Note that the above expression is explicitly Hermitian. 

\subsection{A doubly null tetrad}

Collecting the components in front of equal spinor combinations in the above formula for the soldering form we can rewrite it as:
\be\label{theta}
\theta_\mu^{AA'} =l_\mu o^A o^{A'}+ n_\mu \iota^A \iota^{A'}+ m_\mu o^A \iota^{A'} + \bar{m}_\mu \iota^A o^{A'},
\ee
where
$$ l_\mu = \frac{t_\mu+z_\mu}{\sqrt{2}}, \qquad n_\mu = \frac{t_\mu-z_\mu}{\sqrt{2}}, \qquad m_\mu = \frac{x_\mu+\im y_\mu}{\sqrt{2}}, \qquad \bar{m}_\mu = \frac{x_\mu-\im y_\mu}{\sqrt{2}}.$$
Note that $l,n$ are real one-forms, while $\bar{m}_\mu = m_\mu^*$. The above collection of one-forms is known as a {\it doubly null tetrad}. Indeed, it is easy to see that all 4 one-forms introduced above are null, e.g. $l^\mu l_\mu=0$. The only non-zero products are
$$ l^\mu n_\mu = -1, \qquad m^\mu \bar{m}_\mu=1.$$
Thus, the Minkowski metric can be written in terms of a doubly null tetrad as
$$ \eta_{\mu\nu}= -l_\mu n_\nu - n_\mu l_\nu + m_\mu \bar{m}_\nu + \bar{m}_\mu m_\nu,$$
which can also be verified directly by substituting (\ref{theta}) into the formula (\ref{metric}) for the metric. 

\subsection{Self-dual two-forms}

The self-dual two-forms that play the central role in this article can be written down more naturally (i.e. without any reference to the time plus space decomposition of the tetrad internal index) in terms of spinors. We use the following definition:
\be
\Sigma^{AB} := \frac{1}{2} \theta^{A}{}_{A'}\wedge \theta^{BA'},
\ee
or, without the form notation
\be\label{Sigma-spinor-def}
\Sigma^{AB}_{\mu\nu} = \theta_\mu{}^{(A}_{A'} \theta_\nu{}^{B) A'},
\ee
where we used the fact that symmetrization on the unprimed spinor indices has the same effect as anti-symmetrization on the spacetime indices. 

Explicitly, in terms of the null tetrad and the spinor basis we get
\be\label{Sigma}
\Sigma^{AB} = l\wedge m \, o^A o^B + \bar{m}\wedge n \, i^A i^B + (l\wedge n - m\wedge \bar{m}) i^{(A} o^{B)}.
\ee
Using $\epsilon_{\mu\nu\rho\sigma}=24\im \, l_{[\mu} n_{\nu} m_{\rho} \bar{m}_{\sigma]}$ it can be checked that these forms are indeed self-dual $$\epsilon_{\mu\nu}{}^{\rho\sigma} \Sigma_{\rho\sigma}^{AB} = 2\im \Sigma_{\mu\nu}^{AB}.$$

Let us also give the following useful formula for the decomposition of a contraction of two soldering forms (via a primed index) in terms of the metric and the self-dual two-forms:
\be
\theta_\mu{}^A_{A'} \theta_\nu^{BA'} = -\frac{1}{2} \epsilon^{AB} g_{\mu\nu} + \Sigma_{\mu\nu}^{AB}.
\ee
This is easily checked by either contracting with $\epsilon_{AB}$, which produces minus the metric on both sides, or by symmetrizing with respect to $AB$, which reproduces (\ref{Sigma-spinor-def}). 

\subsection{${\rm SU}(2)$ spinors}

We need to introduce the notion of ${\rm SU}(2)$ spinors when we consider the Hamiltonian formulation of any fermionic theory. In our case, we need this notion to establish a relation between our polarization tensors (\ref{polars}) and some spacetime covariant expressions that we shall write down below.  

To define ${\rm SU}(2)$ spinors we need a Hermitian positive-definite form on spinors. This is a rank 2 mixed spinor $G_{A'A}: G^*_{A'A}=G_{A'A}$, such that for any spinor $\lambda^A$ we have $(\lambda^*)^{A'} \lambda^A G_{A'A}>0$. Here $(\lambda^*)^{A'}$ is the complex conjugate of $\lambda^A$. We can define the ${\rm SU}(2)$ transformations to be those ${\rm SL}(2,\C)$ ones that preserve the form $G_{A'A}$. Then $G_{A'A}$ defines an anti-linear operation $\star$ on spinors via:
\be
(\lambda^\star)_A:=G_{AA'}(\lambda^*)^{A'}.
\ee
We require that the anti-symmetric rank 2 spinor $\epsilon_{AB}$ is preserved by the $\star$-operation:
\be
(\epsilon^\star)_{AB}=\epsilon_{AB},
\ee
which implies the following normalisation condition
\be\label{G-norm}
G_{AA'} G^{A'}{}_{B}=\epsilon_{AB}.
\ee
Using the normalisation condition we find that $(\lambda^{\star\star})^A=-\lambda^A$ or
\be
\star^2=-1.
\ee
Thus, the $\star$-operation so defined is similar to a "complex structure", except for the fact that it is anti-linear:
\be
(\alpha \lambda^A +\beta \eta^A)^\star= \bar{\alpha} (\lambda^\star)^A+\bar{\beta} (\eta^\star)^A.
\ee

Now for the purpose of comparing to results of the 3+1 decomposition, we need to introduce a special Hermitian form that arises once a time vector field is chosen. We can consider the zeroth component of the soldering form
\be
\theta_0^{AA'}\equiv \theta_\mu^{AA'} \left( \frac{\partial}{\partial t}\right)^\mu = \frac{1}{\sqrt{2}} \left( o^A o^{A'} + \iota^A \iota^{A'}\right).
\ee
It is Hermitian, and so we can use a multiple of $\theta_0^{AA'}$ as $G^{AA'}$. It remains to satisfy the normalisation condition (\ref{G-norm}). This is achieved by
\be\label{herm-form}
G^{AA'} := \sqrt{2} \theta_0^{AA'}.
\ee
We then define the spatial soldering form via
\be\label{sigma-sp}
\sigma^{i\, AB}:= G^{A}{}_{A'} \theta^{i\, BA'},
\ee
which is automatically symmetric $\sigma^{i\,AB}=\sigma^{i\,(AB)}$ because its anti-symmetric part is proportional to the product of the time vector with a spatial vector, which is zero. Explicitly, in terms of the spinor basis introduced above we have
\be\label{sigma-small}
\sigma^{i\,AB}= m^i o_A o_B - \bar{m}^i \iota_A \iota_B-\frac{z^i}{\sqrt{2}}(\iota_A o_B+o_A\iota_B ).
\ee
The action of the $\star$-operation on the basis spinors is as follows:
\be\label{star}
(o^\star)^A =  - \iota^A, \qquad (\iota^\star)^A =  o^A.
\ee
It is then easy to see from (\ref{sigma-small}) that the spatial soldering form so defined is anti-Hermitian with respect to the $\star$ operation:
\be
(\sigma^{i\,\star})^{AB}=-\sigma^{i\,AB}.
\ee
It is not hard to deduce the following property of the product of two spatial soldering forms:
\be\label{sigma-ident}
\sigma^i_A{}^B\sigma^j_B{}^C=\frac{1}{2}\delta^{ij} \epsilon_A{}^C+\frac{\im}{\sqrt{2}}\epsilon^{ijk} \sigma^k_A{}^C.
\ee

\subsection{Converting to the spinor form}

We are now ready to use the spinor objects introduced above. First, let us discuss how the expressions written in ${\rm SO}(3)$ notations used so far can be converted into spinor notations. Indeed, we have so far worked with the connection perturbation being $a_\mu^i$. It is now convenient to pass to the spinor description, in which all indices of $a_\mu^i$ are converted into spinor ones. This is done with the soldering form for the spacetime index, and with Pauli matrices for the internal one. 

To fix the form of the multiple of the Pauli matrices that is relevant here, we will require that under this map the identity matrix $\delta^{ij}$ becomes the matrix $\epsilon^{(A|C|}\epsilon^{B)D}$ in spinor notation. Indeed, both matrices have trace 3. Thus, we denote the map from objects with ${\rm SO}(3)$ indices to those with pairs of unprimed spinor indices by $T^{i AB}$ and require it to have the property:
\be
\delta^{ij} T^{i AB} T^{j CD} = \epsilon^{(A|C|}\epsilon^{B)D}.
\ee
This fixes $T^{i AB}$ up to a sign. 

Now, to determine what multiple of this object appears in the relation between $\Sigma^i_{\mu\nu}$ and $\Sigma^{AB}_{\mu\nu}$, both of which have been defined before, we need to look into the algebra satisfied by them. We have  the algebra used many times in the preceding text:
\be\label{s-a-1}
\Sigma^i_\mu{}^\rho \Sigma^j_{\rho\nu} = -\delta^{ij} \eta_{\mu\nu} -\epsilon^{ijk} \Sigma^k_{\mu\nu}.
\ee
At the same time, a simple computation of the same contraction of $\Sigma^{AB}_{\mu\nu}$ gives
\be\label{s-a-2}
\Sigma^{AB}_\mu{}^\rho \Sigma^{CD}_{\rho\nu} = -\frac{1}{2} \epsilon^{(A| C |} \epsilon^{B) D} \eta_{\mu\nu} + \frac{1}{2} \epsilon^{A(C} \Sigma^{D)B}_{\mu\nu}  + \frac{1}{2} \epsilon^{B(C}\Sigma^{D)A}_{\mu\nu} .
\ee
The coefficient in front of the first term here is half that in (\ref{s-a-1}). We thus learn that there is a factor of $\sqrt{2}$ in the conversion of an ${\rm SO}(3)$ index into a pair of spinor ones:
\be\label{spin-con-1}
\frac{1}{\sqrt{2}} \Sigma^i_{\mu\nu} T^{i AB} = \Sigma^{AB}_{\mu\nu}.
\ee
To fix the sign of the quantity $T^{i\, AB}$ we just need to compare the terms containing $\Sigma$ in (\ref{s-a-1}) and (\ref{s-a-2}). Substituting (\ref{spin-con-1}) into (\ref{s-a-2}) we get
\be\label{s-a-3}
\sqrt{2} \epsilon^{ij}{}_{k} T^{i\, AB} T^{j\, CD} = - \epsilon^{A(C} T_k^{D)B} - \epsilon^{B(C} T_k^{D)A}.
\ee
We thus see that the matrices $T^{i AB}$ satisfy the following algebra:
\be
T^{i AB} T^{j}_B{}^C = -\frac{1}{2} \delta^{ij} \epsilon^{AC} + \frac{1}{\sqrt{2}}\epsilon^{ijk} T^{k AC},
\ee
which fixes them uniquely. We see that these quantities are just 
$$T^{i AB}=-\im \sigma^{i AB},$$
where $\sigma^{i AB}$ are the spatial soldering forms introduced above. Explicitly, in terms of the spinor basis, as well as a basis $m^i,\bar{m}^i,z^i$ in $\R^3$ we have:
\be\label{T}
T^{i\,AB}= -\im m^i o_A o_B + \im \bar{m}^i \iota_A \iota_B+\frac{\im}{\sqrt{2}} z^i (\iota_A o_B+o_A\iota_B ).
\ee
Note that $T^{i AB}$ is $\star$-Hermitian, i.e. $(T^{i\star})^{AB} = T^{i AB}$. 

We can now write down the conversion rule of the $\epsilon^{ijk}$ tensor. Thus, introducing
$$ \epsilon^{(AB) \ (CD) \ (EF)}: = \epsilon^{ijk} T^{i \, AB} T^{j\, CD} T^{k \, EF}$$ 
we get from (\ref{s-a-3})
\be\label{eps-conv}
\sqrt{2}\ \epsilon^{(AB) \ (CD)}{}_{(EF)}\:X^{(EF)}=-\epsilon^{A(C} X^{D)B}-\epsilon^{B(C} X^{D)A}.
\ee
This can be rewritten more conveniently as a rule for the commutator
\be
\sqrt{2} \epsilon^{(AB)}{}_{(CD)(EF)} X^{CD} Y^{EF}= X^{AE} Y_{E}{}^B+ X^{BE}Y_E{}^A,
\ee
where the spinor contraction is in a natural order. 

\subsection{Further on spinor conversion}

Let us now discuss the rules of dealing with the spacetime indices. Each such index has to be converted into a mixed type pair of spinor indices using the soldering form $\theta^{\mu AA'}$. We shall refer to the operator of the partial derivative with its spacetime index converted into a pair of spinor indices as the Dirac operator:
\be
 \partial_\mu := \theta_{\mu AA'} \partial^{AA'}.
\ee
Note that, because of our signature choice, and thus a minus sign in (\ref{metric}), we have $\partial^{AA'} = -\theta^{\mu AA'} \partial_\mu$. One has to be careful about these minus signs. 

We now come to objects that have both types of indices, spacetime and internal. The conversion of these is that we write them as the corresponding soldering forms times objects with only spinor indices. Thus, e.g. for the connection we write
\be
a_\mu^i = T^{i AB} \theta_\mu^{MM'} a_{ABMM'},
\ee
which defines what we mean by the connection with all its indices translated into the spinor ones. This choice of the normalization factor in the above formula is convenient, because as we already discussed before the Kronecker delta $\delta^{ij}$ goes under this map into the object $\epsilon^{(A}{}_C \epsilon^{B)}{}_D$, which is the identity map on the space of symmetric rank two spinors. The only unusual translation rule is 
\be\label{s-spin}
\Sigma^i_{\mu\nu} := \sqrt{2} T^{i AB} \theta_\mu^{MM'}\theta_\nu^{NN'} \Sigma^{AB}_{MM'NN'},
\ee
where
\be
\Sigma^{AB}_{MM'NN'} = \epsilon^{(A}{}_M\epsilon^{B)}{}_N \epsilon_{M'N'}.
\ee
The only reason for putting the factor of $\sqrt{2}$ in (\ref{s-spin}) is that the objects so defined have the same algebra (\ref{s-a-1}) as we are used to, and as was used on multiple occasions in deriving the form of the interaction terms in the Lagrangian. 

The final useful formula for the purposes of conversion is
\be\label{tt-as}
\theta^{MM'}_{[\mu} \theta^{NN'}_{\nu]} = \frac{1}{2} \epsilon^{NM} \Sigma^{M'N'}_{\mu\nu} + \frac{1}{2} \epsilon^{M'N'} \Sigma^{MN}_{\mu\nu},
\ee
where
\be
\Sigma^{MN}_{\mu\nu} = \theta^{(M}_{\mu M'} \theta^{N) M'}_\nu, \qquad  \Sigma^{M'N'}_{\mu\nu}= \theta^{M(M'}_\mu \theta_{\nu M}^{N')}.
\ee
Note that the natural contractions appear in these definitions, and, as a result, the anti-self-dual two-forms $\Sigma^{M'N'}_{\mu\nu}$ are {\it minus} the complex conjugates of the self-dual ones $\Sigma^{MN}_{\mu\nu}$. The formula (\ref{tt-as}) then implies
\be
V_{[\mu} U_{\nu]} = \frac{1}{2} V_{MM'} U_N{}^{M'} \Sigma^{MN}_{\mu\nu} + \frac{1}{2} V^M{}_{M'} U_{MN'} \Sigma^{M'N'}_{\mu\nu},
\ee
where again natural contractions appear. The first term here is the self-dual part, and the second is anti-self-dual part of the two-form $V_{[\mu}U_{\nu]}$.

\subsection{Momentum spinors}

Consider a massless particle of a particular 3-momentum vector $\vec{k}$. The 4-vector $k^{\mu}=(|k|,\vec{k})$ is then null. As such, it can be written as a product of two spinors $k^A k^{A'}=\theta^{\mu}_{AA'}k_\mu $. In the case of Lorentzian signature the spinors $k^A, k^{A'}$ must be complex conjugates of each other (so that the resulting null 4-vector is real). It is then clear that $k^A$ is only defined modulo a phase. Moreover, as the vector $\vec{k}$ varies, i.e. as $\vec{n}=\vec{k}/|\vec{k}|$ varies over the sphere $S^2$, there is no continuous choice of the spinor $k^A$. We make the following choice:
\be\label{kA}
k^A \equiv k^A(\vec{k}) := 2^{1/4} \sqrt{\omega_k} \left( \sin(\theta/2) e^{-\im\phi/2} \iota^A + \cos(\theta/2) e^{\im\phi/2} o^A\right),
\ee
where $o^A,\iota^A$ is a basis in the space of unprimed spinors, and $\omega_k=|k|$. Here $\theta,\phi$ are the usual coordinates on $S^2$ so that the momentum vector in the direction of the positive z-axes corresponds to $\theta=\phi=0$. We see that the corresponding spinor is $2^{1/4} \sqrt{\omega_k}o^A$. The formula (\ref{kA}) can be checked using the expression (\ref{theta}) for the soldering form.

We can now see effects of the change of the momentum vector direction. Consider, for example, what happens when the momentum direction gets reversed. This corresponds to $\theta\to \pi-\theta$ and $\phi\to \phi+\pi$. We get
\be
k^A(-\vec{k}) = \im \, 2^{1/4} \sqrt{\omega_k} \left( - \cos(\theta/2) e^{-\im\phi/2} \iota^A + \sin(\theta/2) e^{\im\phi/2} o^A\right).
\ee
We now note that 
\be
k^A(-k) = \im (k^\star)^A(k),
\ee
where the action of the $\star$-operation on the basis spinors is given in (\ref{star}). Now, using the fact that $\star^2=-1$ it is easy to see that flipping the sign of the momentum twice we get minus the original momentum spinor. In other words, $k^A$ takes values in a non-trivial spinor bundle over $S^2$.

\subsection{Helicity spinors}

The aim of this subsection is to use the rules for the $i\to (AB)$ conversion deduced above, as well as the definition (\ref{kA}) of the momentum spinors $k^A$ to write down convenient expressions for the polarization tensors (\ref{polars}) in the spinor language. 

Our polarization tensors are built from the vectors $m^i(k),\bar{m}^i(k)$, where the direction of the $z^i$ axes is chosen to be that of the momentum 3-vector $\vec{k}$. Thus, let us start by assuming that $\vec{k}$ points along the positive $z$-direction. Then from (\ref{T}) we have $T^{i AB} m^i =  \im\, \iota^A \iota^B, T^{i AB} \bar{m}^i = -\im \, o^A o^B$, and therefore
\be\label{polars-1}
m^i m^j \to - \iota^A \iota^B \iota^C \iota^D, \qquad \bar{m}^i \bar{m}^j \to - o^A o^B o^C o^D
\ee
when converted to spinor notations. We can, however, use the available freedom of gauge ${\rm SO}(3)$ rotations and consider polarization tensors (spinors) more general than those above. Indeed, we can always shift our (spatial projection of the) connection by a gauge transformation $a_{ij}\to a_{ij} + (\partial_{(i} \phi_{j)})^{tf}$, where also the tracefree part needs to be taken in order to preserve the tracelessness of the $a_{ij}$. Such a shift being pure gauge, it does not have any effect on the scattering amplitudes. So, we can freely add to both polarization tensors an object of the form $z_{(i} \phi_{j)}$, where again a tracefree part is assumed. Moreover, the vector $\phi_i$ can be different for the positive and negative helicity polarizations. Using the spinor conversion rules written above, it is not hard to see that this means that one will obtain correct scattering amplitudes when using instead of (\ref{polars-1}) the following expressions
$$
\iota^A \iota^B \iota^C \iota^D \to (\iota+ \alpha o)^{(A} (\iota+\beta o)^B (\iota+\gamma o)^C \iota^{D)}, \quad
o^A o^B o^C o^D \to (o+\alpha'\iota)^{(A} (o+\beta'\iota)^B (o+\gamma'\iota)^C o^{D)},
$$
for arbitrary coefficients $\alpha,\beta,\gamma,\alpha',\beta',\gamma'$. For reasons to become clear below, the most convenient choice is
\be\label{polars-2}
\iota^A \iota^B \iota^C \iota^D \to \frac{q^{(A} q^B q^C \iota^{D)}}{(q^E o_E)^3}, \qquad o^A o^B o^C o^D \to \frac{o^{(A} o^B o^C (p^\star)^{D)}}{(\iota^E (p^\star)_E)},
\ee
where $q^A, p^A$ are arbitrary spinors, and $\star$ is the operation on ${\rm SU}(2)$ spinors introduced above. Note that while in the choice of the first polarization spinor we have replaced as many as 3 copies of $\iota^A$ by an arbitrary reference spinor $q^A$, in the second polarization we only changed a single copy of $o^A$ to $(p^\star)^A$. The reason for this will become clear below. 

Let us now rewrite the spinor expressions for the full polarization tensors (\ref{polars}) using the spinors $k^A= 2^{1/4}\sqrt{\omega_k} o^A$ and $(k^\star)^A = - 2^{1/4}\sqrt{\omega_k} \iota^A$. We get
\be\label{hel-1}
\varepsilon^{- ABCD}(k) = M \frac{ q^{(A} q^B q^C (k^\star)^{D)}}{(q^E k_E)^3}, \qquad \varepsilon^{+ ABCD}(k) = \frac{1}{M} \frac{k^{(A} k^B k^C (p^\star)^{D)}}{(k^*)^E (p^\star)_E}.
\ee
Note that all the annoying factors of $\sqrt{2}$ in the original formulas (\ref{polars}), as well as some minus signs present in the intermediate expressions, have now cancelled. Note also that while the previous spinor expressions were only valid in a frame where the 3-momentum was pointing in the $z$-direction, the expressions (\ref{hel-1}) are valid in an arbitrary frame. 

It remains to observe that one will obtain (\ref{hel-1})  as the spin 2 parts of the spatial projections of the following mixed spinors:
\be\label{helicity-spin}
\varepsilon^{- ABCA'}(k) = M \frac{ q^A q^B q^C k^{A'}}{\ket{q}{k}^3}, \qquad \varepsilon^{+ ABCA'}(k) = \frac{1}{M} \frac{k^A k^B k^C p^{A'}}{\bra{p}{k}},
\ee
where we have introduced the usual notations for the spinor contractions
\be
\ket{\lambda}{\eta} := \lambda^A \eta_A, \qquad \bra{\lambda}{\eta} = \lambda_{A'} \eta^{A'}.
\ee
We note that the helicity spinors are normalized so that
\be
\varepsilon^{-\, ABC A'} \epsilon^{+}_{ABC A'}=1.
\ee

The expressions (\ref{helicity-spin}) are the main outcome of this heavy in conventions section. We note that these expressions could have been guessed as the only ones with the correct dimensions, as well as with the right homogeneity degree zero dependence on the reference spinors $q^A, p^{A'}$, and the right degree of homogeneity under the rescaling of the momentum spinors $k^A\to t k^A, k^{A'}\to t^{-1} k^{A'}$. Indeed, it is clear that under these rescalings (keeping the 4-momentum $k^Ak^{A'}$ unchanged) we get
\be
\varepsilon^{- ABCA'}(k)  \to t^{-4} \varepsilon^{- ABCA'}(k),\qquad \varepsilon^{+ ABCA'}(k)\to t^4 \varepsilon^{+ ABCA'}(k).
\ee
However, under any such a guess possibly important numerical factors could have been missed, and it is gratifying to see that after establishing all the conversion formulas, the helicity spinors turned out to be just the simplest expressions possible, without any complicating numerical prefactors. We note that the final spacetime covariant expressions (\ref{helicity-spin}) explain our choice (\ref{polars-2}) at the level of the spatially projected expressions. 

The only complication that remains to be discussed is the fact that the positive helicity gravitons have to be taken to be slightly massive, as we have seen in the section on the Minkowski space limit. Because of this, the meaning of the spinor $k^A$ that is used in the positive helicity spinor in (\ref{helicity-spin}) is not yet defined. To settle this, we shall represent the massive 4-vector $k^2=2M^2$ of the positive helicity gravitons as follows
\be\label{mass-shell}
k^{AA'}= k^A k^{A'} + M^2 \frac{p^A p^{A'}}{\ket{p}{k}\bra{p}{k}}.
\ee
This gives precisely the required $k^2=-k^{AA'} k_{AA'}=2M^2$. Here $p^A p^{A'}$ are a reference spinor and its complex conjugate. At this point it can be arbitrary, but it is convenient to take it to be the same as the one that appears in the positive helicity spinor in (\ref{helicity-spin}). It is now the spinor $k^A$ that appears in the decomposition (\ref{mass-shell}) is what one has to use in the positive helicity spinor in (\ref{helicity-spin}). We emphasize that only the positive helicity momentum 4-vectors should be taken to be massive, while the negative helicity does not need this complications, and the corresponding momenta 4-vectors come directly as a product of two spinors. 

\subsection{A relation to the metric helicity states}

It is instructive to see how the metric description helicity spinors can be obtained from the expressions (\ref{helicity-spin}). For this we need to recall the passage to the metric perturbation variable that was explained in great detail in \cite{Delfino:2012zy}. In that paper we have seen that the metric perturbation is obtained by applying to the connection the operator $\bar{D}$. One should also rescale by $1/M$ to keep the mass dimension correct. At the spacetime covariant level the operator $\bar{D}$ corresponds to the operation of taking the anti-self-dual part of the two-form $da^i$. In the spinor notations, this boils down to the following expression for the metric perturbation
\be\label{h-a}
h_{AB\, A'B'} \sim \frac{1}{M} \partial^E_{A'} a_{B' E AB},
\ee
where $\partial_{AA'}= - \theta^\mu_{AA'} \partial_\mu$ is the Dirac operator, and the $\sim$ sign means that we are only interested in this relation modulo numerical factors. Applying this to the helicity spinors (\ref{helicity-spin}), and ignoring the arising numerical factors, one immediately sees that the usual metric spinor helicity states get reproduced:
\be
h^-_{AA'BB'}(k) \sim \frac{q_A k_{A'} q_B k_{B'}}{\ket{q}{k}^2}, \qquad h^+_{AA'BB'}(k) \sim \frac{k_A p_{A'} k_B p_{B'}}{\bra{p}{k}^2}.
\ee
Note that for the negative helicity the metric formulation helicity spinor arises by a single $q$ spinor in the numerator of (\ref{helicity-spin}) contracting with the momentum $k$ spinor, removing one of the factors of $\ket{q}{k}$ from the denominator. There is also the cancellation of the factor of $M$ in the connection helicity spinor with $1/M$ in the passage to the metric perturbation. For the positive helicity the mechanism of obtaining the usual metric formulation helicity spinor is more subtle. Indeed, if the positive helicity graviton 4-momentum was null, then we would be contracting two momentum $k$ spinors, which would give a zero result. Instead, it is the second, mass term in (\ref{mass-shell}) that gives a non-zero contribution. The factor of $M^2$ in this second term then is nicely cancelled by the $1/M$ in the helicity spinor and the additional factor of $1/M$ in (\ref{h-a}). We therefore see that it is essential that the positive helicity graviton is kept massive till the Minkowski limit can be taken. 

Once again, the fact that the usual metric helicity states get reproduced could be taken as the sufficient reason to work with (\ref{helicity-spin}). However, we find the given above derivation of (\ref{helicity-spin}) that does not involve any reference to the metric more self-contained. 

\section{Feynman rules in the spinor form}
\label{sec:feyn}

Now that we understand how expressions can be converted into the spinor language, we can write down the derived above Feynman rules in the spinor form. Here we will also pass to the form relevant in the Minkowski limit, i.e. replace all the derivatives by the partial ones. As we shall see, there are many advantages in working with spinors, as some operations that are not easy to deal with in the ${\rm SO}(3)$ notation become elementary once one expresses them using spinors. The prime example is the projection on the $S_+^3\otimes S_-$ representation of the Lorentz group that appeared in our derivation of the propagator. In the spinor language this simply corresponds to the symmetrization on the unprimed spinor indices. We shall also see that the interaction vertices take a particularly simple form once the spinor notations are applied. 

\subsection{Propagator}

We have previously found the propagator to be given by $1/\im k^2$ times the projector $P^{(3,1)}_{\mu i|\nu j}$, given in (\ref{proj-31}), on the $S_+^3\otimes S_-$ irreducible component of objects of type $a_{\mu}^i$ with one spacetime and one internal index. To find what this projector becomes once converted into the spinor form one can multiply the spacetime indices with the soldering forms, and the internal indices with $T^{i AB}$. However, one does not need to do this computation as it is clear that this projector is simply the product of the identity operator acting on the primed index, times the operator of symmetrization of the 3 unprimed spinor indices. Thus, we can write
\be
\langle a_{EFGE'}(-k) a^{ABCA'}(k) \rangle =\frac{1}{\im k^2} 
\epsilon_E{}^{(A}\epsilon_F{}^{B}\epsilon_G{}^{C)}\epsilon_{E'}{}^{A'}.
\ee

\subsection{Pieces of the interaction vertices}

Here we develop a dictionary translating the various blocks that appear in the interaction vertices into the spinor form. As we recall from (\ref{Da}), one of the main building blocks of the vertices is the two-form $D_{[\mu} a^i_{\nu]}$, and various quantities constructed from it. We remind that from now on we replace the covariant derivative $D$ by the partial one. 

The first block to be translated is the $(Da)^{ij}$ symmetric tracefree matrix, whose definition is given in (\ref{Da-sd}). Applying the rules given above we get:
\be
(Da)^{ij} \to \sqrt{2} \partial^{(A}{}_{M'} a^{BCD)M'},
\ee
where the result is easy to understand, and the factor of $\sqrt{2}$ comes from the same factor in the translation (\ref{s-spin}) of $\Sigma^i_{\mu\nu}$. Further, we have
\be
D^\mu a_\mu^i \to \partial^{M}_{M'} a^{AB}{}_{M}{}^{M'}.
\ee
Finally, we have
\be\label{da-asd-1}
(\widetilde{Da^i})_{\mu\nu} \to  \frac{1}{2} \Sigma^{M'N'}_{\mu\nu} \partial^M{}_{M'} a^{AB}{}_{MN'}.
\ee

We also need the spinor representation of the two-form $\epsilon^{ijk} a^j_\mu a^k_\nu$. Its self-dual part encoded in the matrix (\ref{Saa}) is given by
\be
(\Sigma\epsilon aa)^{ij} \to a^{CE(A}{}_{M'} a^{B)}{}_E{}^{DM'} + a^{DE(A}{}_{M'} a^{B)}{}_E{}^{CM'} .
\ee
It is $(AB)\to (CD)$ symmetric, but it has trace given by the full contraction of the two connections. The anti-self-dual part of $\epsilon^{ijk} a^j_\mu a^k_\nu$ is given by
\be
(\epsilon^{ijk} a^j_\mu a^k_\nu)_{asd} \to \frac{1}{\sqrt{2}}\Sigma^{M'N'}_{\mu\nu} a^{EF(A}{}_{M'} a^{B)}{}_{EFN'}.
\ee

We now introduce some notations to simplify the above spinorial expressions. The idea behind this notation is that we omit pairs of naturally contracted indices. Thus, we define
\be
(\partial a)^{ABCD}:=\partial^{(A}{}_{M'}a^{BCD)M'},\quad (\partial a)^{M'N'AB}:=\partial^{C(M'}a_C{}^{ABN')},\quad (\partial a)^{AB}:=\partial^{M}_{M'} a^{AB}{}_{M}{}^{M'},
\\ \nonumber
(aa)^{ABCD}:=a^{ABM}{}_{M'}a^{CD}{}_M{}^{M'},\quad (aa)^{M'N'CD}:=a^{CD(AM'}a_{CD}{}^{B)N'}.
\ee
The blocks appearing in the interaction vertices can then be written very compactly in terms of these quantities. 

\subsection{Cubic interaction}

As we discussed in the section on the Minkowski limit, we keep only the leading order terms of all the quantities in the limit $M\to 0$. Converting all the terms in (\ref{L3}) into spinor form we get
\be \label{L3-M}
3\im M^2 (g^{(2)})^{3/2} {\cal L}^{(3)} =  2^{3/2} \left(g^{(3)} - \frac{3g^{(2)}}{4}\right) (\partial a)_{AB}{}^{CD} (\partial a)_{CD}{}^{EF} (\partial a)_{EF}{}^{AB} \\ \nonumber
+ 2^{-1/2} 3 g^{(2)} (\partial a)^{ABCD} (\partial a)_{M'N'AB} (\partial a)^{M'N'}{}_{CD}  + 2^{1/2} \frac{3g^{(2)}}{16}  (\partial a)^{ABCD} (\partial a)_{AB} (\partial a)_{CD} \\ \nonumber
- 2^{1/2} M^2 (3g^{(2)}-f(\delta)) (\partial a)^{ABCD} (aa)_{ABCD} + 2^{1/2} M^2 f(\delta) (\partial a)^{M'N' AB} (aa)_{M'N' AB},
\ee
where an additional factor of $1/2$ in the first term in the second line came from the factors of $1/2$ in (\ref{da-asd-1}), with one of them being cancelled by the factor of $2$ that appears in the contraction of two $\Sigma$'s. 

The last term in the second line in (\ref{L3-M}) is only (possibly) relevant for loop computations, for in any tree diagram at least one of the two factors of $(\partial a)^{AB}$ gets hit by an external state, which gives zero. Thus, we shall ignore this term in the present paper. Let us now consider the second term in (\ref{L3-M}) in more details, which can be seen to be the only relevant term in the $M\to 0$ limit of the GR cubic interaction.

\subsection{The parity-preserving cubic vertex}

In the case of GR the coefficients in front of the term in the first line, and the first term in the third line in (\ref{L3-M}) become zero, and we are left with the simple
\be\label{cubic-GR}
\im \mathcal{L}^{(3)}_{\rm GR}=\frac{1}{2} \frac{\kappa}{M}(\partial a)^{ABCD} (\partial a)_{M'N'AB}(\partial a)^{M'N'}{}_{CD} + \kappa M (\partial a)^{M'N' AB} (aa)_{M'N' AB},
\ee
where we have used (\ref{g-GR}). Here we introduced the usual notation
\be
\kappa^2:=32\pi G, \qquad \kappa= \frac{\sqrt{2}}{M_p}.
\ee

Below we shall see that it is the first term in (\ref{cubic-GR}) that gives most of the interesting physics, and, in particular, is the one relevant for computations of the MHV amplitudes. The second term seems to be suppressed by a factor of $M^2/E^2$, where $E$ is energy,  as compared to the first. However, due to the subtleties of taking the Minkowski limit we cannot just throw it away, and it turns out to give a contribution to some scattering amplitudes. For now, let us analyze the much more interesting first term and come back to the second piece of the GR cubic vertex later. 

First, we would like to make a pause and emphasize the similarity of the first term in (\ref{cubic-GR}) to the vertex of Yang-Mills theory, when rewritten in the spinor notations. Thus, one starts with the Yang-Mills Lagrangian in the form ${\cal L}_{\rm YM}\sim (F_{sd})^2$, where $F_{sd}$ is the self-dual part of the curvature. Applying the described above spinor conversion rules, at the linearized level this gives 
\be
{\cal L}^{(2)}_{\rm YM} \sim (\partial^{(A}_{A'} A^{B) A'})^2,
\ee
where $A_{AA'}$ is the connection, and we omitted the Lie-algebra indices. Our linearized Lagrangian (\ref{free}), when converted into the spinor form, is precisely analogous, except that the connection field in the case of gravity has two more unprimed spinor indices. Let us now look at the Yang-Mills cubic interaction. Again translating into the spinor form we get
\be\label{cubic-YM}
{\cal L}^{(3)}_{\rm YM} \sim (\partial^{(A}_{A'} A^{B) A'}) A_{AB'} A_B{}^{B'}.
\ee
Again, the analogy to the first term in (\ref{cubic-GR}) is striking. Basically, the gravitational interaction described by the first term in (\ref{cubic-GR}) is the only possible one that can be constructed following the Yang-Mills pattern, but now with more unprimed indices on the connection. Indeed, generalizing the first block $ (\partial^{(A}_{A'} A^{B) A'})$ to the case with more indices one gets $(\partial^{(A}_{A'} a^{BCD) A'})\equiv(\partial a)^{ABCD}$. We would now like to have a symmetric block involving two connections with free indices being 4 unprimed spinor indices $ABCD$. Thus, two indices must come from one connection, and the other two from the other. For reason to become clear below, we also want to have some quantities constructed out of the connections that contract only in the primed indices. This means that we have to convert one of the unprimed indices in each connection to a primed using the Dirac operator. This results precisely in (\ref{cubic-GR}). The other possible choice of having connections contracting directly, with no Dirac operators involved, i.e. $a_{AB}{}^E{}_{A'} a_{CDE}{}^{A'}$, can be easily seen not to give the desired on-shell amplitudes, see below. We thus learn that (\ref{cubic-GR}) is the only possible generalization of the Yang-Mills cubic vertex that gives the correct on-shell amplitudes. The second term in (\ref{cubic-GR}) is superficially more analogous to (\ref{cubic-YM}) than the first, as it involves just one derivative. However, we will see that it does not give the standard answers for the on-shell amplitudes, and so is not the "right" generalization.  

Let us now check that the vertex given by (\ref{cubic-GR}), when evaluated on the graviton helicity spinors gives just the required $--+$ and $++-$ amplitudes, i.e. the squares of these amplitudes for spin one particles (it is easy to check that the $---$ and $+++$ amplitudes resulting from it are zero). 

Let us first compute the $--+$ amplitude. The required helicity states are given in (\ref{helicity-spin}). We first note that the second term in (\ref{cubic-GR}) cannot contribute to this amplitude. Indeed, no matter which pair of legs of this term the two negative helicity states are inserted, there is always a contraction of the reference spinors $q^A$ and the result is zero. Thus, we only have to consider the first term in (\ref{cubic-GR}) for this helicity combination. Let us now recall that the combination $(\partial a)^{ABCD}$ only gives a non-zero result when applied to a positive helicity state. Thus, we must insert the positive helicity wave-function in this leg. The other two legs are symmetric, and we insert into them the remaining two negative helicity states. Let us denote the negative helicity momenta by $k_1, k_2$, and the positive momentum by $k_3$. We will assume that all the momenta are incoming. After applying all the derivatives present in the vertex to their corresponding states we obtain the following contraction
\be\label{3--+}
\im \frac{\kappa}{M} \frac{1}{M} 3^A 3^B 3^C 3^D M \frac{q_A q_B 1_{M'} 1_{N'}}{\ket{1}{q}^2} M \frac{q_C q_D 2^{M'} 2^{N'}}{\ket{2}{q}^2} = \im \kappa \bra{1}{2}^2 \frac{\ket{3}{q}^4}{\ket{1}{q}^2\ket{2}{q}^2},
\ee
where we have used the usual notation $k_1^A \equiv 1^A$, etc. We can now use the momentum conservation equation, which we contract with the reference spinor $q_A$ to get $\ket{1}{q} 1^{A'} + \ket{2}{q} 2^{A'} + \ket{3}{q} 3^{A'} = 0$. This immediately gives $\ket{3}{q}^2/\ket{1}{q}^2 = \bra{1}{2}^2/\bra{3}{2}^2, \ket{3}{q}^2/\ket{2}{q}^2 = \bra{1}{2}^2/\bra{3}{1}^2$, which allows us to rewrite (\ref{3--+}) as
\be\label{--+}
{\cal M}^{--+} = \im \kappa \frac{\bra{1}{2}^6}{\bra{1}{3}^2 \bra{2}{3}^2}.
\ee 
This is just the expected square of the spin one result. 

The opposite $++-$ helicity configuration is computed similarly, but now we cannot ignore the second term in (\ref{cubic-GR}). So, the computation is somewhat more involved. Let us consider the first term in (\ref{cubic-GR}) first. We note that if two positive helicity gravitons hit the two symmetric legs of this vertex, the reference spinors $p^{A'}$ will contract, and this gives zero. Thus, one of the positive helicity gravitons must necessarily be inserted into the leg $(\partial a)^{ABCD}$. This of course also comes from the fact that $(\partial a)^{ABCD}$ gives zero when applied to a negative helicity state. Thus, the negative helicity graviton must hit one of the two symmetric legs. If we choose the state inserted into $(\partial a)^{ABCD}$ to be the graviton number two, we get the following contraction
\be
- \im\frac{\kappa}{M} \frac{1}{M} 2^A 2^B 2^C 2^D M \frac{1_A 1_B p_{M'} p_{N'}}{\bra{1}{p}^2} M \frac{q_C q_D 3^{M'} 3^{N'}}{\ket{3}{q}^2} = -\im \kappa \ket{1}{2}^2 \frac{ \bra{3}{p}^2 \ket{2}{q}^2}{\ket{3}{q}^2 \bra{1}{p}^2}.
\ee
To this we must add the contribution from $1$ and $2$ gravitons exchanged, which gives the overall contribution from the first term in (\ref{cubic-GR}) as
\be\label{3-contr-1}
-\im \kappa  \ket{1}{2}^2\frac{\bra{3}{p}^2}{\ket{3}{q}^2}\left(\frac{\ket{2}{q}^2}{\bra{1}{p}^2} + \frac{\ket{1}{q}^2}{\bra{2}{p}^2} \right).
\ee
To this, we must add the contribution from the second term in (\ref{cubic-GR}). In this case, the positive helicity gravitons can only be inserted into the two symmetric legs of this vertex. This gives the following contraction
\be\label{3-contr-2}
2\im\kappa M 3^{M'} 3^E \, M \frac{q_E q^A q^B 3^{N'}}{\ket{q}{3}^3} \frac{1}{M} \frac{1^E 1^F 1_A p_{M'}}{\bra{p}{1}} \frac{1}{M} \frac{2_E 2_F 2_B p_{N'}}{\bra{p}{1}} = -2\im \kappa \ket{1}{2}^2 \frac{\bra{3}{p}^2}{\ket{3}{q}^2} \frac{\ket{1}{q}\ket{2}{q}}{\ket{3}{q}^2}.
\ee
Adding (\ref{3-contr-1}) and (\ref{3-contr-2}) we get
\be
{\cal M}^{++-}=-\im \kappa  \ket{1}{2}^2\frac{\bra{3}{p}^2}{\ket{3}{q}^2}\left(\frac{\ket{2}{q}}{\bra{1}{p}} + \frac{\ket{1}{q}}{\bra{2}{p}} \right)^2.
\ee
We would like to rewrite the answer in terms of the angle brackets only. To this aim we take the $\bra{3}{p}$ into the brackets, then use the momentum conservation and then the Schouten identity. This gives the desired
\be\label{++-usual}
{\cal M}^{++-} = -\im \kappa \frac{\ket{1}{2}^6}{\ket{1}{3}^2 \ket{2}{3}^2}.
\ee 
We thus see that the GR cubic vertex is parity-invariant, as it should be of course, in the sense that the amplitude for an opposite configuration of helicities is given by the complex conjugate of the original amplitude. Note, however, that to obtain this result both terms in (\ref{cubic-GR}) were used in an essential way. This can be viewed as a justification for the presence of the second term in (\ref{cubic-GR}), while it is the first term only that is important for many calculations, such as e.g. that of MHV graviton amplitudes. 

Let us now discuss the corresponding Feynman rules. When we write the vertex factor corresponding to the first term in (\ref{cubic-GR}) we obtain 3 terms, since the first instance of $(\partial a)$ corresponding to the self-dual part of the two-form $da$ can be applied to any of the 3 legs of the vertex. We will not write the vertex factor as we will never need it in this paper, and because its explicit form containing many $\epsilon$ symbols and symmetrizations is not very illuminating. Instead, we shall draw a picture of the contractions involved in the vertex for just one of the terms, when momenta $k_{1,2}$ that we assume are incoming are on the lines corresponding to $(\partial a)_{asd}$, and the momentum $k_{12}=k_1+k_2$ is on the line corresponding to $(\partial a)_{sd}$. In this picture the red lines correspond to primed, and black to unprimed spinor indices. The symbols of momenta in circles stand for the factors of $k^{AA'}$. This part of the full vertex is given by
\be\label{V-GR}
\frac{\im \kappa}{M} \quad \lower0.6in\hbox{\includegraphics[width=1.2in]{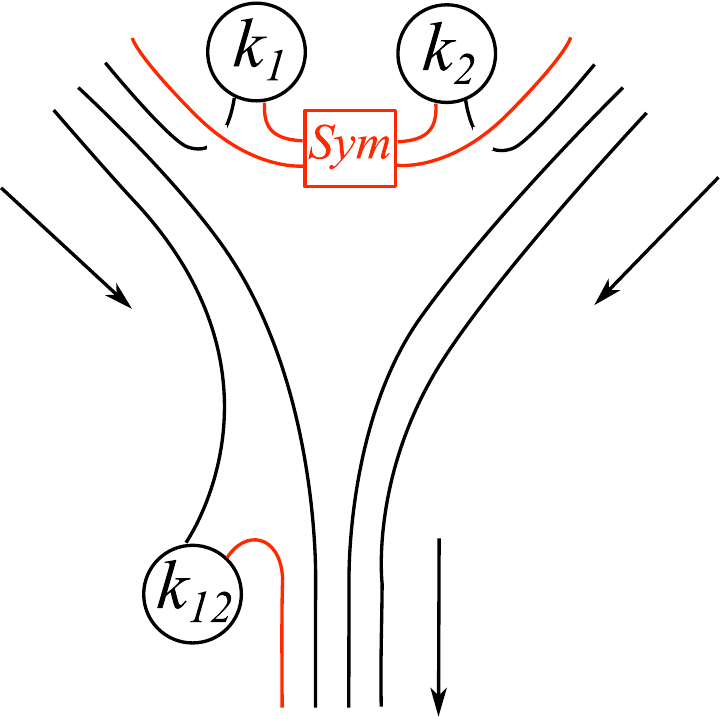}}
\ee
where one also has to symmetrize over the two external legs that can get contracted to $k_{12}$. As we shall see later, for the purpose of computing the most interesting (graviton-graviton or more generally MHV amplitudes) this will be the only surviving contribution to the full vertex. 

Let us now briefly discuss the case of a general theory. In this case the vertex of interest in the tree level computations has several pieces. The most "interesting" part is still essentially the most interesting part of the GR vertex, i.e. the first term in (\ref{cubic-GR}), but with the different prefactor 
\be\label{V-1}
\frac{\im \sqrt{2}}{M^2 \sqrt{g^{(2)}}}  \quad  \lower0.6in\hbox{\includegraphics[width=1.2in]{3vertexASD.pdf}}
\ee
We can therefore expect that the Newton constant measuring the strength of interactions of gravitons in the general theory is
\be\label{Mp}
\frac{1}{16\pi G} = M^2 g^{(2)},
\ee
which is essentially the coupling constant $g^{(2)}$, expressed in the units of $M$, the only dimensionful parameter present in the theory. This expectation will be confirmed below when we compute the parity-preserving graviton scattering amplitude. 

Let us also give a pictorial representation for the second term in (\ref{cubic-GR}). We have
\be\label{V-strange}
- 2 \im \kappa M \quad \lower0.6in\hbox{\includegraphics[width=1.3in]{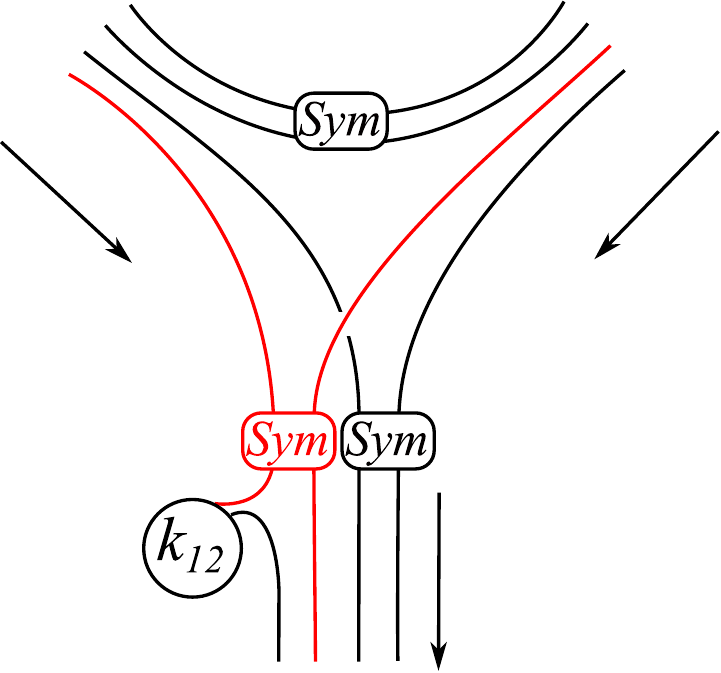}} 
\ee
where the factor of two in front comes from the two ways that the symmetric legs can be applied. The prefactor here is as is relevant for the case of GR. Let us also give a general expression for this vertex. From (\ref{L3-M}) we get the following graphical representation 
\be\label{V-3}
- \frac{\im}{3} \kappa^3 M^3 f(\delta) \quad \lower0.6in\hbox{\includegraphics[width=1.3in]{3vertex1derASD.pdf}} 
\ee
Here we have already summed over the two ways that the symmetric legs can be applied.

\subsection{The parity-violating cubic vertex}

For a general theory there are two more vertices. One comes from the first term in (\ref{L3-M}). Assuming again the convention that the two momenta are incoming and one is outgoing, and taking into account the symmetry factor of $3!$, the vertex can be graphically represented as
\be\label{V-2}
\frac{\im \sqrt{2} (4g^{(3)}-3g^{(2)})}{M^2 (g^{(2)})^{3/2}} \quad  \lower0.6in\hbox{\includegraphics[width=1.2in]{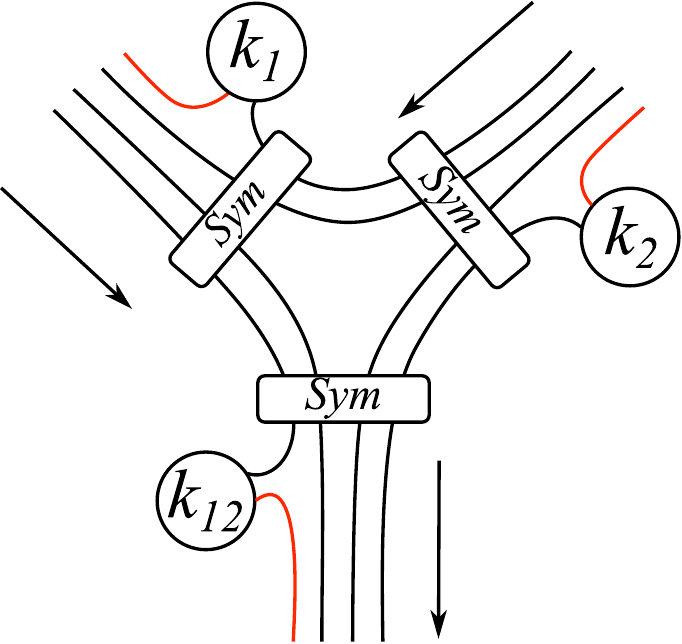}}
\ee

To see what kind of interaction this generates, let us evaluate this vertex on the graviton polarization spinors. It is immediately clear that it only produces a non-zero result when all the helicities are positive, because the negative helicity inserted into the combination $(\partial a)^{ABCD}$ gives a zero result. After applying all the derivatives to the external states, we get for the amplitude coming from this vertex
\be
{\cal M}^{+++} = \frac{\sqrt{2} (4g^{(3)}-3g^{(2)})}{\im M^5 (g^{(2)})^{3/2}}  \ket{1}{2}^2 \ket{2}{3}^2 \ket{3}{1}^2.
\ee
Using the already known to us fact (\ref{Mp}) that the Planck mass $M_p^2=1/16\pi G$ equals $M^2 g^{(2)}$ we can express the quantity $g^{(2)}$ here in terms of the Planck mass. Also, to understand the behaviour of this amplitude in the $M\to 0$ limit we specialize to a family of theories considered in the Appendix, which are guaranteed to have only Planckian modifications of GR. For this family the difference $4g^{(3)}-3g^{(2)}= -(27/4)\beta^2 M^2/M_p^2$, and the amplitude becomes
\be\label{3-par-viol}
{\cal M}^{+++} = \im \frac{27\kappa^5\beta^2}{16} \ket{1}{2}^2 \ket{2}{3}^2 \ket{3}{1}^2.
\ee
Here $\beta$ is a parameter controlling deviations from GR, see (\ref{mod-action}). 

We note that the amplitude of the type (\ref{3-par-viol}) can arise in a gravity theory with the $(Riemann)^3$ term in the Lagrangian, see e.g. \cite{ArkaniHamed:2008gz}, discussion following formula (57). However, a theory with this term in the Lagrangian would be parity-preserving, and thus would also have a non-zero ${\cal M}^{---}$ amplitude. We stress that in our theories this is not the case, with only one chiral half of this amplitude being present, and the amplitude ${\cal M}^{---}$ continuing to vanish as in GR. We thus learn that the higher order in curvature correction that is present in a general member of our family of theories is not $(Riemann)^3$ but rather $(Weyl_{sd})^3$, in other words the cube of the self-dual part of Weyl curvature. No anti-self-dual part cubed is present, and this is why the 3 negative helicity amplitude continues to be zero. Below we shall discuss implications of this fact in more details. 

\subsection{Another first derivative vertex}

We finally consider the last remaining piece of a general cubic vertex. It comes from the first term in the last line of (\ref{L3-M}). Graphically, it can be represented as
\be\label{V-4}
\im \sqrt{2} \, \frac{3g^{(2)}-f(\delta)}{3(g^{(2)})^{3/2}} \quad  \lower0.6in\hbox{\includegraphics[width=1.2in]{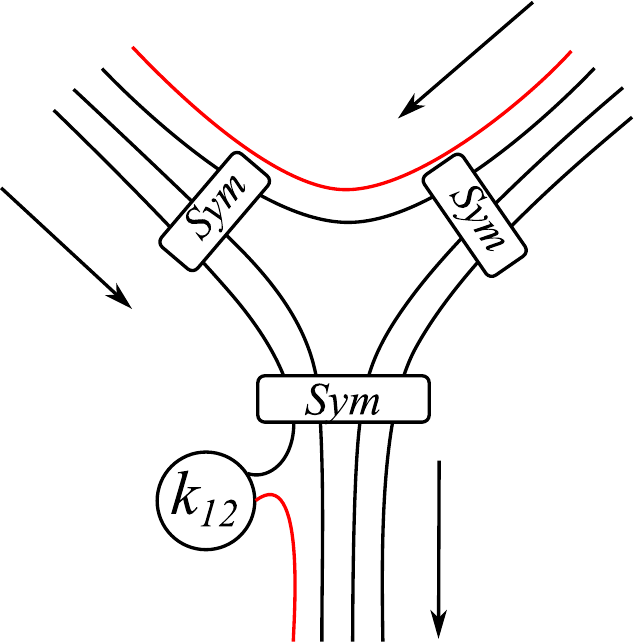}}
\ee

As before, let us understand what kind of interaction this vertex represents by evaluating it on-shell. It is easy to see that there can be at most a single negative helicity graviton. Indeed, any pair of legs in this vertex is connected by a black line. Thus, if there are two negative helicity states, there will be a contraction of the reference $q^A$ spinors, which results in a null amplitude. Thus, the only possible amplitude is $++-$. We also see that one of the plus helicities must be inserted into the bottom leg. This is because the derivative present in the bottom leg would give zero when applied to a negative helicity, and also because if two positive helicity states are inserted from the top, there would be a contraction of the reference spinors $p^{A'}$ along the red lines. For this configuration of helicities, however, there is no need to compute the result, as it can be seen to go to zero as $M\to 0$. Indeed, we know that $g^{(2)} = M_p^2/M^2$. From the consideration of the Appendix we know, see (\ref{g2-f}), that the difference $3g^{(2)}-f(\delta)=(27/4)\alpha\beta^2 $ goes to zero as $M^2/M_p^2$. This means that the prefactor in this vertex goes as $M^5/M_p^5$, which is too fast to give any surviving contribution when contracted with the $1/M$ factor coming from the helicity states. So, there is no surviving in the $M\to 0$ limit amplitude that this vertex produces. It turns out that this vertex does not contribute to the graviton-graviton amplitudes either, so we can safely ignore it for the rest of this paper.

\subsection{Quartic interaction}

As for the cubic vertex, we now take the full expression (\ref{L4}), replace all the derivatives by the partial ones, and convert everything into spinors. Our life is considerably simplified by an observation that in order to contribute to the graviton-graviton amplitudes the 4-vertex needs to have at least 4 derivatives. This will be demonstrated below. Taking this observation into account, we can ignore the lower derivative parts of the vertex (\ref{L4}). The 4-derivative part that we keep, converted into the spinor form, reads
\be\nonumber
-4! M^4 (g^{(2)})^2 {\cal L}^{(4)} =  \left(- 8 g^{(4)} +4 g^{(3)} + (7/2) g^{(2)}\right) (\partial a)_{AB}{}^{CD} (\partial a)_{CD}{}^{AB}  (\partial a)_{EF}{}^{MN} (\partial a)_{MN}{}^{EF} \\ \label{L4-M}
-24(4g^{(3)} - g^{(2)}) (\partial a)^{ABEF} (\partial a)_{EF}{}^{CD} (\partial a)_{M'N' AB} (\partial a)^{M'N'}{}_{CD} 
\\ \nonumber
+8(4g^{(3)} - 3g^{(2)}) (\partial a)^{ABCD} (\partial a)_{ABCD} (\partial a)_{M'N'}{}^{EF} (\partial a)^{M'N'}{}_{EF} 
\\ \nonumber
-12 g^{(2)} (\partial a)_{M'N'}{}^{(AB} (\partial a)^{CD)M'N'} (\partial a)_{E'F'AB} (\partial a)^{E'F'}{}_{CD}  \, .
 \ee
We emphasize that this is an on-shell vertex, with all legs satisfying $\partial^\mu a_\mu^i=0$ condition, and that only the 4-derivative part was kept. Thus, some terms potentially relevant for loop computations have not been written down. Only two of these terms survive in the GR case, and we get
\be\label{L4-GR-M}
 {\cal L}^{(4)}_{\rm GR} =\frac{\kappa^2}{M^2} (\partial a)^{ABEF} (\partial a)_{EF}{}^{CD} (\partial a)_{M'N' AB} (\partial a)^{M'N'}{}_{CD} 
\\ \nonumber
+\frac{\kappa^2}{4M^2} (\partial a)_{M'N'}{}^{(AB} (\partial a)^{CD)M'N'} (\partial a)_{E'F'AB} (\partial a)^{E'F'}{}_{CD}  \, .
 \ee

We will only write down the vertex factors corresponding to the first term in (\ref{L4-M}), the reason being that the other terms cannot contribute to the graviton-graviton scattering (for a particular convenient choice of the reference spinors). This will become clear in the next section. Thus, in particular the GR 4-vertex present in (\ref{L4-GR-M}) is not at relevant as far as the graviton-graviton scattering is concerned, and we have written it only for reference.  

For the vertex factor we will use the rule that all 4 momenta are incoming. Taking into account the symmetry factors, the associated vertex is 
\be\label{4vert-sd-2}
\frac{\im ( 8g^{(4)} - 4g^{(3)} -(7/2) g^{(2)} )}{3 M^4 (g^{(2)})^2}   \quad  \lower0.6in\hbox{\includegraphics[width=1.2in]{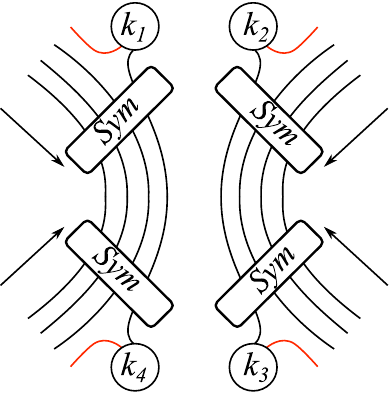}} \, ,
\ee
where one needs to the sum over possible permutations. There are 3 terms that are numbered e.g. by what $k_1$ gets connected to. 

Just as a reference, we will also give pictures of the index contractions present in the other 3 terms appearing in (\ref{L4-M}). As we have said, these other terms will not contribute to any computations in this paper, but the pictures below will be instrumental in seeing this. We give them in the order that they appear in (\ref{L4-M}), without any associated prefactors.
\be\label{4vert-asd}
\quad  \lower0.6in\hbox{\includegraphics[width=1.2in]{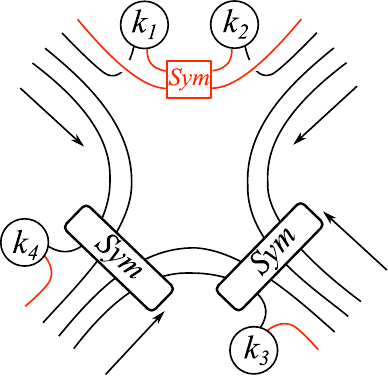}}\quad  \lower0.6in\hbox{\includegraphics[width=1.2in]{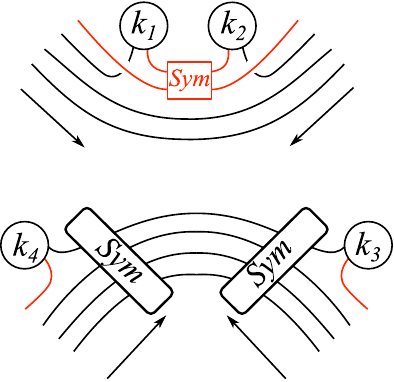}}\quad  \lower0.6in\hbox{\includegraphics[width=1.2in]{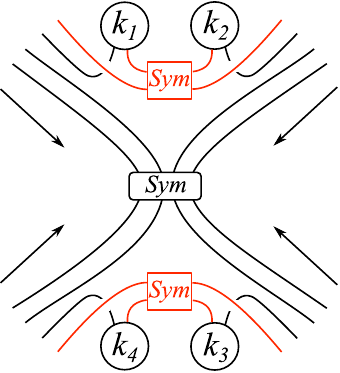}}
\ee
The first and the last of these are the ones that appear in (\ref{L4-GR-M}).

\section{Graviton-graviton scattering}
\label{sec:scatt}

This is the last section of the present paper, where we put to use all the technology that we developed. As is the case with the metric-based GR, even the simplest $--++$ amplitude is somewhat easier to compute as just the first non-trivial case of a more general MHV amplitude with all but two positive helicity gravitons. For such amplitudes one can write the so-called BCFW recursion relation \cite{Britto:2005fq}, and then the $--++$ amplitude is given by a sum of two terms, both involving just the known $--+$ amplitudes (\ref{--+}). However, here we avoid developing the technology of recursion relations, postponing it to the next paper. Instead, we compute all amplitudes of interest by directly evaluating the relevant Feynman diagrams, of course using the spinor helicity technology that we already have developed. We will see that the $--++$ amplitude receives contributions from only tri-valent graphs with vertices (\ref{V-GR}), and there are two diagrams to compute. The degree of complexity of this computation is very similar to the analogous textbook computation in Yang-Mills theory, see e.g. \cite{Srednicki:2007qs}. We first do the computation of the $--++$ amplitude, and then consider the $-+++$ and $++++$ amplitudes (which are only non-zero in our more general parity-violating theories). 

\subsection{The all negative and all negative one positive helicity amplitudes}

The amplitudes with at most one positive helicity graviton can be shown to vanish in full generality, just by a simple count of the number of derivatives present in any given diagram. 

Let us first consider the all minus case, with $n$ external legs. In this case the helicity states (\ref{helicity-spin}) each carry 3 copies of the negative helicity reference spinor $q^A$. This spinor can then be chosen to be the same for all the gravitons. Thus, we have $3n$ copies of $q^A$ in the diagram, and we have to contract them all with some momenta that appear as a result of evaluating in the derivatives present in the vertices. It is then easy to see that there are not enough derivatives to avoid contracting the $q^A$'s between themselves. Indeed, the largest possible number of derivatives is in a diagram with 3-valent vertices only. There are $n-2$ such vertices, and thus at most $3(n-2)$ derivatives present (as the largest power of the derivative in each vertex is 3). Since each derivative can only eat one copy of $q^A$, there is not enough derivatives for these amplitudes to be non-zero. 

The argument with all minus one plus proceeds similarly, but in this case one chooses all the reference spinors of the negative helicity gravitons to be the momentum spinor $k^A$ of the positive helicity graviton. Then there are again $3n$ instances of $k^A$ and only at most $3(n-2)$ derivatives, which is not enough to avoid the $k^A$'s contracting. 

\subsection{The $--++$ amplitude}

This is the only non-vanishing amplitude in the case of GR, as we shall also explicitly see using our methods. 

Let us first verify that the 4-vertex containing diagrams cannot contribute to this amplitude. This is an argument of the same type that we already gave to show that the all minus and all minus one plus amplitudes are zero. Indeed, we now have 6 copies of $q^A$ coming from the two negative helicity states, and we can choose them to be equal to the momentum spinor $k_3^A$ of one of the positive helicity graviton. We thus get 9 copies of $k_3^A$ that need to contract with something else than themselves. Let us count other object available. We have 3 copies of the other positive helicity momentum spinor $k_4^A$ coming from its helicity spinor (\ref{helicity-spin}). We thus need at least 6 derivatives to absorb the remaining 6 copies of $k_3^A$. These can only come from tri-valent diagrams, as the other diagrams contain less derivatives. Incidentally, the same argument shows that only the 3-valent graphs contribute to any MHV amplitude (which in our convention is the amplitude with at most two plus helicity gravitons). 

It remains to see that only the vertices (\ref{V-1}) can contribute to this amplitude. As we just saw, we need to have at least 6 derivatives, so the only other possible vertex is (\ref{V-2}). We cannot use this vertex twice, because it can only take the positive helicity states. Thus, let us assume that just one of the vertices used is (\ref{V-2}). Since this vertex can only take the positive helicity gravitons, both available positive helicity states must go in it. The other vertex is then (\ref{V-1}), with two negative helicity states inserted in it. It is then a simple verification to see that in the internal edge of the diagram we will have $q^A q^B q^C$ coming from the negative helicities contracting with $k_3^{(A} k_3^{B} k_4^{C)}$ or with $k_3^{(A} k_4^{B} k_4^{C)}$ coming from the positive helicities. Since we have chosen $q^A$ to be the momentum spinor of one of the positive helicity gravitons, this contraction is zero. Again, precisely the same argument works for a general MHV amplitude, and establishes that only the vertex (\ref{V-1}) is relevant in MHV computations. This means that after the identification (\ref{Mp}) is made, all MHV amplitudes for a general member of our family of theories are the same as in GR. In particular, the $--++$ amplitude defining the Newton's constant is the same, and so in the following considerations we can assume the form (\ref{V-GR}) of the relevant vertex. 

Thus, we only need to consider the 3-valent graphs with vertices (\ref{V-1}). There are 3 such graphs ($s,t,u$ channels), and each vertex factor has 3 terms. Thus, there are in principle 27 terms to consider. However, most of them are zero. 

A very convenient way to organize the computation is to consider the possible ways of putting two negative helicities into the vertex (\ref{V-GR}). It is clear that they must go into the two legs in which the vertex is symmetric. Indeed, any pair of external legs containing the $(\partial a)^{ABCD}$ leg (this is the bottom leg in the pictorial representation) is connected by a black line. Thus, the insertion of a pair of negative helicity states in any other way but from the top gives a zero result. Recalling that the top legs came from the anti-self-dual part of the two-form $da^i$, we shall refer to them as the ASD legs. Similarly, the bottom leg will be referred to as SD. 

We can now recall our choice $q^A=k_3^A$. It means that for many purposes the positive helicity graviton of momentum $k_3$ behaves like a negative heliclity one. In particular, if we are to put this positive helicity graviton into the vertex (\ref{V-GR}) together with some negative helicity graviton, it is easy to see that both necessarily must go into the ASD legs. Indeed, if the positive helicity goes into the bottom leg in the picture (\ref{V-GR}) it is easy to see that at least one of the $q^A$'s from the negative helicity graviton will contract with $k_3^A$ along the black lines. Thus, in case of the graviton of momentum $k_3$ and some negative helicity graviton, both must go into the upper, ASD legs in the figure (\ref{V-GR}). 

Let us now compute the result of such insertion of two negative or one negative one positive of momentum $k_3$ helicity states into the vertex (\ref{V-GR}). For two negative states that we label by $1$ and $2$, we get the following quantity
\be\label{--a}
\im M\kappa \, q^A q^B q^C  (k_1+k_2)^{DD'} q_D \frac{\bra{1}{2}^2}{\ket{1}{q}^2\ket{2}{q}^2} .
\ee
Similarly, when the states $1$ and $3$ are put together into this vertex we get
\be\label{-+a}
- \im M\kappa \, q^A q^B q^C  (k_1+k_3)^{DD'} q_D \frac{\bra{1}{p}^2}{\ket{1}{q}^2\bra{3}{p}^2} ,
\ee
where an extra minus is from $\partial^E_{M'} a_{EABN'}$ where we get $k_1^E q_E=\ket{1}{q}$, but in the denominator of these helicity states we have $\ket{q}{1}^3$, and this different order of the contraction produces an extra minus sign. Two of such minus signs have cancelled each other in (\ref{--a}). We remind the reader that in (\ref{-+a}) we have used $k_3^A=q^A$. 

The important point about the results (\ref{--a}) and (\ref{-+a}) is that they can now be put together with some other state via a 3-vertex (\ref{V-GR}) only in a single way, again from the upper two legs in the picture. The reason for this is that both of them, as the original helicity states, contain 3 copies of $q^A$, thus, the argument we gave above about the only possible way to couple such states applies. This means that the vertex (\ref{V-GR}) works only "in one direction" coupling the negative helicities or the positive helicity graviton of momentum $k_3$ by taking them in its ASD legs. As an aside remark, we note that this immediately implies that even in a general MHV amplitude with two plus helicity gravitons, the vertex (\ref{V-2}) cannot be used. Indeed, it could only be used to couple the two positive helicity gravitons to all negative gravitons already coupled in some way with (\ref{V-1}). However, the one leg off-shell current with any number of negative helicity gravitons is necessarily proportional to $q^A q^B q^C$, and this will vanish when contracted with what comes from the vertex (\ref{V-2}), as we have already discussed above. In the next paper we will put these facts to work and compute the one leg off-shell all negative helicity current. For now we continue our consideration of the $--++$ amplitude.

The above picture of the vertex (\ref{V-GR}) working as a coupler of states just in one direction gives 3 possibilities to consider. One can either couple first the two negative helicity gravitons $1$ and $2$, and then couple the result to $3$, or first couple $1$ and $3$, and then couple to $2$, or first couple $2$ to $3$ and then to $1$. 
\be\nonumber
\includegraphics[height=1in]{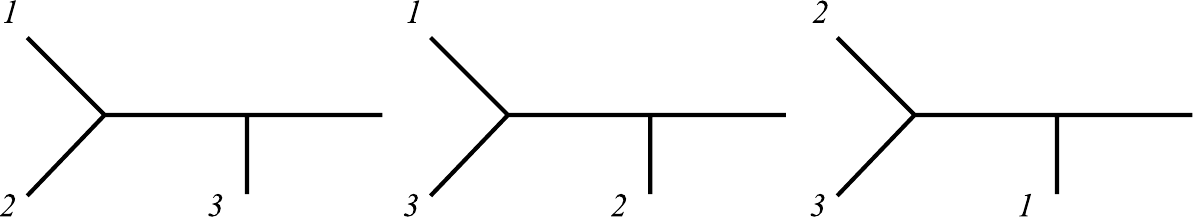}
\ee
In all these cases one couples to the positive helicity graviton $4$ at the very end. These are of course just the 3 different $s,t,u$ channels, but we now have just 3 terms to consider instead of 27. 

The final simplification comes from the availability to choose the positive helicity reference spinor $p^{A'}$ conveniently. We recall that we have not yet made any choice of this in the analysis so far, so we are now free to make the most convenient choice. We choose $p^{A'}=2^{A'}$, which eliminates the possibility (\ref{-+a}) to couple $3$ directly to $2$. This leaves just the first two terms in the above picture to consider. 

Let us compute the diagram when $1$ and $2$ get coupled first, and then couple to $3$. We have computed the result of coupling $1$ and $2$ in (\ref{--a}). We then multiply this result by the propagator $1/\im k^2$, where $k^2=k^A{}_{A'} k_A{}^{A'}$ computes, using the fact that both $k_1$ and $k_2$ are null, to just $k_{12}^2=2\ket{1}{2}\bra{1}{2}$. We now couple together (\ref{--a}) with the propagator added at the end with the positive helicity state $3$. After applying the derivatives present in the vertex we get the following contraction
\be\nonumber
(\im)^2 \kappa^2 \, q_A q_B (k_1+k_2)^{E}{}_{M'}  q_E  (k_1+k_2)^{F}{}_{N'} q_F (k_1+k_2+k_3)^D{}_{D'} \frac{\bra{1}{2}^2}{\ket{1}{q}^2\ket{2}{q}^2} \frac{1}{2\im \ket{1}{2}\bra{1}{2}} M \frac{q_{C} q_{D} p^{M'} p^{N'}}{\bra{3}{p}^2}
\\ \nonumber
= 2\im M \left(\frac{\kappa}{2}\right)^2 q_A q_B q_C (k_1+k_2+k_3)^D{}_{D'} q_D \frac{\bra{1}{2}}{\ket{1}{2}} \frac{\langle q| k_1+k_2| p]^2}{\ket{1}{q}^2\ket{2}{q}^2\bra{3}{p}^2}.
\ee
We finally insert into this result the last remaining positive helicity $(1/M) 4^A 4^B 4^C p^{D'}/\bra{p}{4}$. We can also use the momentum conservation to replace $k_1+k_2+k_3$ by $-k_4$ (all momenta are incoming). Substituting also $q^A=3^A$ and $p^{A'}=2^{A'}$ we get, overall
\be\label{amp-1}
2\im \left(\frac{\kappa}{2}\right)^2  \frac{\bra{1}{2}^3}{\ket{1}{2}} \frac{\ket{3}{4}^4}{\ket{2}{3}^2\bra{2}{3}^2}.
\ee

We now compute the other non-vanishing diagram, where $1$ gets first connected to $3$, and then the result connects to $2$. Adding to (\ref{-+a}) the propagator, taking the helicity state for $2$ and applying all the derivatives gives the following contraction
\be\nonumber
(\im)^2 \kappa^2 \, q_A q_B (k_1+k_3)^{E}{}_{M'}  q_E  (k_1+k_3)^{F}{}_{N'} q_F (k_1+k_2+k_3)^D{}_{D'} \frac{\bra{1}{p}^2}{\ket{1}{q}^2\bra{3}{p}^2} \frac{1}{2\im \ket{1}{3}\bra{1}{3}} M \frac{q_{C} q_{D} p^{M'} p^{N'}}{\ket{2}{q}^2}
\\ \nonumber
= 2\im M \left(\frac{\kappa}{2}\right)^2 q_A q_B q_C (k_1+k_2+k_3)^D{}_{D'} q_D \frac{\bra{1}{p}^2}{\ket{1}{3}\bra{1}{3}} \frac{\langle q| k_1+k_3| p]^2}{\ket{1}{q}^2\ket{2}{q}^2\bra{3}{p}^2}.
\ee
We now put in this the last remaining state $4$, and use the values of $q^A, p^{A'}$ to get
\be\label{amp-2}
2\im \left(\frac{\kappa}{2}\right)^2  \frac{\bra{1}{2}^4}{\ket{1}{3}\bra{1}{3}} \frac{\ket{3}{4}^4}{\ket{2}{3}^2\bra{2}{3}^2}.
\ee
Adding (\ref{amp-1}) and (\ref{amp-2}) and using the momentum conservation we get
\be
- 2\im \left(\frac{\kappa}{2}\right)^2 \frac{\ket{3}{4}^4}{\ket{2}{3}^2\bra{2}{3}^2}  \frac{\bra{1}{2}^3 \ket{1}{4} \bra{1}{4}}{\ket{1}{2} \ket{1}{3} \bra{1}{3}}.
\ee
We now convert as many square bracket contractions into the round ones using the momentum conservation identities, e.g. $\bra{1}{4}/\bra{1}{3}=-\ket{2}{3}/\ket{2}{4}$. We finally get
\be\label{--++1}
{\cal M}^{--++} = 2\im \left(\frac{\kappa}{2}\right)^2 \ket{3}{4}^6 \frac{1}{\ket{1}{3}\ket{1}{4}\ket{2}{3}\ket{2}{4}} \frac{\bra{1}{2}}{\ket{1}{2}},
\ee
which is the usual GR result, see below. 

To rewrite (\ref{--++1}) in a more recognizable form, we evaluate the spinor contractions present in the center of mass frame, and rewrite everything in terms of the Mandelstam variables. The relevant contractions are given in (\ref{contr}). We get
\be\label{--++}
{\cal M}^{--++} = \im \left(\frac{\kappa}{2}\right)^2 \frac{s^3}{tu},
\ee
which is the form one find this result in e.g. \cite{Bern:2002kj}, formula (17), modulo the factor of $\im$ that most likely has to do with different conventions, or in \cite{Grisaru:1975bx}, formula (40) with the coefficient $c$ given after formula (42). Our factors in (\ref{--++}) precisely match those in \cite{Grisaru:1975bx}. 

We emphasize that the result (\ref{--++}) holds for a general member of our family of theories. It is a particular case of a more general result that all MHV amplitudes (i.e. amplitudes with just two positive helicity gravitons) are the same for all members of our family, provided one identifies the Newton's constant as in (\ref{Mp}). The generality of the result (\ref{--++}) is important, because it immediately tells us important information about the types of higher derivative terms that are present in the Lagrangian of our theories if one interprets them as metric theories. Indeed, for a general counterterm corrected Einstein-Hilbert Lagrangian one expects (\ref{--++}) to get modified, see e.g. formula (61) of \cite{ArkaniHamed:2008gz}. The fact that this does not happen for our class of theories tells us that no such $R^4$ type modifications are present. Once again it illustrates a very tightly constrained character of the modifications present in our different from GR gravitational theories. 

\subsection{The $-+++$ amplitude}

We now compute the first example of an amplitude that is zero in GR, but non-zero in a general parity-violating theory (the $+++$ amplitude encountered in the previous section is zero once the momentum is taken to be real and the momentum conservation is imposed).

The first check that we need to do is that both for this, and the $++++$ amplitude only the 4-vertices with at least 4 derivatives can contribute, and so it is sufficient to restrict one's attention to (\ref{L4-M}). Let us first run the argument for the all plus case. Here we have 4 reference spinors $q^{A'}$, and these need to be contracted with something else than themselves. Therefore, one must have at least 4 derivatives. The same argument also works in the case $-+++$, if one chooses the positive helicity reference spinors to be equal to the momentum spinor for the negative helicity state. 

Let us now check that there cannot be any contribution to this amplitude from the 4-valent vertices. Since we have one negative helicity, only the vertices in (\ref{4vert-asd}) could contribute, with the negative helicity inserted into one of the ASD legs, i.e. where the contraction of the derivative with the connection is of the form $\partial^E_{M'} a_{EABN'}$. However, such a leg gets necessarily contracted with another ASD leg, where we will have a positive helicity graviton. As usual, we can choose the reference momenta of all positive helicity gravitons to be the same and equal to the momenta of the negative one. Let the negative helicity graviton be of momentum $k_1$, then we choose $p^{A'}=1^{A'}$. It is now easy to see that the primed index contraction will give that of $p^{A'}$ with itself, and so these 4-valent graph diagrams cannot contribute. Another way to phrase this argument is to note that there is always a red line connecting at least one pair of vertices in (\ref{4vert-asd}). Such red lines give a contraction of the reference spinors $p^{A'}$, or of these reference spinors with the momentum spinors $1^{A'}$, and this gives a zero result. 

It remains to compute contributions from the 3-valent graphs. There appears to be many possibilities, as we have 4 different types of 4-valent vertices to consider. Indeed, because we need just 4 derivatives to be present, it appears that even the one-derivative vertices can give some contribution. However, as usual, most of the possibilities give a zero result. In particular, the one-derivative vertices cannot contribute. Let us check this first. 

The way to check which diagrams can contribute is to follow the red lines. With our choice $P^{A'}=1^{A'}$ we know that we cannot put together states in a pair of legs that is connected by a red line. E.g., in the GR vertex (\ref{V-1}) we can only put the states one in an upper leg, one in a lower, but not two in the upper ASD legs, because this would result in a zero spinor contraction. If we do put two states into this vertex in the only allowed way, we will get the reference spinor $p^{A'}$ on the free red line, later to be contracted into some other 3-valent vertex. The same argument applies to (\ref{V-strange}) and (\ref{V-3}) where we again see that in order to avoid the contractions of the positive helicity reference spinors we need to use a very specific insertion. Thus, in the case of (\ref{V-strange}) we must necessarily insert the external states from the top, with again a reference spinor $p^{A'}$ appearing on the red line of the free leg (bottom in this case). With (\ref{V-3}) we can only insert the external states one into a top leg, one into a bottom, as these are not connected by a red line. Again, there is a $p^{A'}$ spinor appearing on the free leg. This immediately tells us that the vertices (\ref{V-1}), (\ref{V-strange}) and (\ref{V-3}) cannot be paired, i.e. any diagram involving only vertices of these types is zero (for this choice of helicities). This means that one of the vertices must be (\ref{V-2}). In particular, this argument explains why this amplitude is zero in GR (in our formulation). 

The easiest way to see that the diagrams involving vertices (\ref{V-strange}), (\ref{V-2}) and (\ref{V-3}), (\ref{V-2}) cannot contribute is to work out the $M$-dependence. The prefactor in (\ref{V-2}) goes as $M^3/M_p^5$. The helicity states give $M$ from the negative helicity times $1/M^3$ from the positive. Overall, this leaves a factor of $M$ to first positive power. Thus, in order for these diagrams to give a surviving in $M\to 0$ contribution the other 3-valent vertex must have a negative power of $M$ in front. However, this is not the case. The vertex (\ref{V-strange}) goes as $M/M_p$, and the vertex (\ref{V-3}) goes as $M^5/M_p^5$. Thus, no non-zero Minkowski limit contributions are produced in this case. 

The only possibly non-vanishing in the $M\to 0$ limit diagram is therefore one involving GR vertex (\ref{V-1}) and the vertex (\ref{V-2}). We must necessarily insert the negative graviton into the vertex (\ref{V-1}), in one of the upper legs. Also, some positive helicity graviton should go into the bottom leg of this vertex. Let us compute the corresponding contraction, including the propagator at the end. We take the positive graviton to be the one of momentum $k_2$, and get
\be
-\frac{\im\kappa}{M} M \frac{q_E q_F 1^{M'} 1^{N'}}{\ket{1}{q}^2} \frac{1}{M} 2^E 2^F 2^{(A} 2^B (k_1+k_2)^{C)}{}_{N'} \frac{1}{2\im \ket{1}{2}\bra{1}{2}} \\ \nonumber
= \frac{\kappa}{2M^2} \frac{\ket{2}{q}^2}{\ket{1}{2}\ket{1}{q}^2} 2^A 2^B 2^C 1^{M'}.
\ee
The free indices here are $ABCM'$. We have to contract this object with one obtained by combining the gravitons $3$ and $4$ via the vertex (\ref{V-2}). This gives
\be
\frac{\im \sqrt{2} (4g^{(3)}-3g^{(2)})}{M^2 (g^{(2)})^{3/2}} \frac{1}{M^2} 3^E 3^F 4_E 4_F 3_{(A} 3_B 4_C 4_{D)} (k_3+k_4)^D{}_{M'}.
\ee
The symmetrization here, together with its contraction with $k_3+k_4$ can be written as
\be
3_{(A} 3_B 4_C 4_{D)} (k_3+k_4)^D{}_{M'} = \frac{1}{2}\left( 3_D 3_{(A} 4_B 4_{C)} + 4_D 4_{(A} 3_B 3_{C)}\right) (k_3+k_4)^D{}_{M'} \\ \nonumber
= \frac{1}{2}\left( \ket{4}{3} 3_{(A} 4_B 4_{C)} 4_{M'}  + \ket{3}{4} 4_{(A} 3_B 3_{C)} 3_{M'} \right) .
\ee
Overall, after contracting the above two quantities, we get for this contribution to the amplitude
\be\label{amp-3}
\frac{\im (4g^{(3)}-3g^{(2)})}{2M^2 M_p^4} \frac{\ket{2}{q}^2\ket{3}{4}^3\ket{2}{3}\ket{2}{4}}{\ket{1}{2}\ket{1}{q}^2}\left( \ket{2}{4} \bra{1}{4} -\ket{2}{3} \bra{1}{3}\right).
\ee 
This is $34$ symmetric, as it should be. Note that we have replaced $\kappa$ with $\sqrt{2}/M_p$ and also used the fact that $M^2g^{(2)}=M_p^2$. We should now add contributions from the 2 more diagrams in which $1$ first gets connected to $3$ and then to $1$ and $4$, and another one where $1$ first gets connected to $4$ and then to the rest. What one gets can be checked to be $q$-independent, as it should be of course. So, we shall make a choice and set $q^A=4^A$, so that there is just one more contribution to consider. It reads
\be\label{amp-4}
\frac{\im (4g^{(3)}-3g^{(2)})}{2M^2 M_p^4} \frac{\ket{3}{q}^2\ket{2}{4}^3\ket{3}{2}\ket{3}{4}}{\ket{1}{3}\ket{1}{q}^2}\left( \ket{3}{4} \bra{1}{4} -\ket{3}{2} \bra{1}{2}\right).
\ee 
We now set $q^A=4^A$ and add the above two quantities. We can also use the momentum conservation to note that the quantities in brackets in both (\ref{amp-3}) and (\ref{amp-4}) are equal, so that we can keep only the first one of them in each case, and double the result. We thus get
\be
\frac{\im (4g^{(3)}-3g^{(2)})}{M^2 M_p^4} \bra{1}{4} \frac{\ket{2}{4}^3\ket{3}{4}^3 \ket{2}{3}}{\ket{1}{4}^2} \left( \frac{\ket{2}{4}}{\ket{1}{2}} - \frac{\ket{3}{4}}{\ket{1}{3}}\right). 
\ee
Using the Schouten identity this finally gives
\be\label{-+++1}
{\cal M}^{-+++} = \im \frac{(4g^{(3)}-3g^{(2)})}{M^2 M_p^4}  \frac{\bra{1}{4}}{\ket{1}{2}\ket{1}{3}\ket{1}{4}} \ket{2}{4}^3\ket{3}{4}^3 \ket{2}{3}^2.
\ee
Using the momentum conservation this can be seen to be $234$ symmetric. 

It is instructive to rewrite the result (\ref{-+++1}) using the Mandelstam variables. The relevant spinor contractions in the center of mass frame are given in the Appendix, see (\ref{contr}). One gets
\be\label{-+++M}
{\cal M}^{-+++} = \frac{(4g^{(3)}-3g^{(2)})}{8\im M^2 M_p^4}  stu.
\ee
Specializing to the one-parameter family of theories guaranteed to contain only Planckian modifications of gravity, we use (\ref{diff-g-1}) and get
\be\label{-+++}
{\cal M}^{-+++} = \im \frac{27\beta^2}{32 M_p^6}  stu.
\ee
Here $\beta$ is a parameter that controls the strength of deviations from GR, see (\ref{mod-action}).  We see that at high energies this amplitude goes as $E^6/M_p^6$. 

\subsection{The $++++$ amplitude}

We now study the final graviton-graviton amplitude, involving 4 incoming gravitons of positive helicity. This amplitude vanishes if there are just the GR vertices, but is non-vanishing in general, as we shall now compute. 

Many of the argument of the previous section apply, and we can see that the 4-vertices in (\ref{4vert-asd}) do not contribute. We also know that from the 3-valent diagrams, only ones involving the vertex (\ref{V-2}) can contribute. We will analyze the 3-valent diagrams containing just one copy of the vertex (\ref{V-2}) in the Appendix. Let us now consider the diagram containing two vertices of the type (\ref{V-2}). We know that the prefactor in this vertex goes as $M^3/M_p^5$. Given two such prefactors, and taking these together with the factor of $1/M^4$ coming from the positive helicity states, we see that the result goes as $M^2$, and so does not survive in the Minkowski limit. 

We thus have to consider only the contributions from the 4-valent graphs with the vertex (\ref{4vert-sd-2}), as well as 3-valent graphs involving just one copy of the vertex (\ref{V-2}). The later are worked out in the Appendix. On the other hand, the computation of the 4-valent diagram is very easy, because after the derivatives get applied to the external states, one is just left with $(1/M) k^A k^B k^C k^D$ quantities, where $k^A$ is the corresponding momentum spinor, to contract as dictated by the black lines in figure (\ref{4vert-sd-2}). The result is
\be\nonumber
\frac{\im ( 8 g^{(4)} - 4g^{(3)} -(7/2) g^{(2)} )}{3 M^8 (g^{(2)})^2}  ( \ket{1}{3}^4 \ket{2}{4}^4+ \ket{1}{2}^4\ket{3}{4}^4 + \ket{1}{4}^4\ket{2}{3}^4).
\ee

We can rewrite the above results more compactly by going into the center of mass frame and introducing the Mandelstam variables (\ref{mandel}). Using the results of spinor contractions given in (\ref{contr}) we get 
\be\label{++++M}
{\cal M}^{++++}_{\rm 4-vert} = \frac{\im(2 g^{(4)} - g^{(3)} -(7/8) g^{(2)})}{12  M^4  M_p^4 }  (s^4+t^4+u^4).
\ee
For theories having only Planckian modifications, such as the family described in the Appendix, we have (\ref{diff-g-2}), and 
\be
{\cal M}^{++++}_{\rm 4-vert} = \im \frac{27\beta^3 }{48  M_p^8 }  (s^4+t^4+u^4).
\ee
Adding this to the contribution (\ref{app-3-valent}) from the 3-valent diagrams from the Appendix we get the following answer for this amplitude
\be\label{++++}
{\cal M}^{++++}= \im \frac{135\beta^2}{16M_p^6} stu+ \im \frac{27\beta^3 }{48  M_p^8 }  (s^4+t^4+u^4).
\ee
At sub-Planckian energies the first term dominates and the amplitude goes as $E^6/M_p^6$. 

\section{Discussion}

The main outcome of this work is a derivation of the Feynman rules, as well as of the prescription for computing the Minkowski space graviton scattering amplitudes. For the convenience of the reader these are listed in the Appendix. 

Our main physical result is the computation of the graviton-graviton scattering amplitudes. We recall that in GR there is just one non-vanishing such amplitude (in the convention that all particles are e.g. incoming), and this is the amplitude with two gravitons of one and two of the opposite helicity. Using the terminology of maximally helicity violating (MHV) amplitudes (in our conventions these are amplitudes with just two positive helicities), the GR graviton-graviton amplitudes is the simplest MHV amplitude. Our computation confirms the GR result, which is completely expected since we work with just a different, but equivalent at the tree level formulation of general relativity. 

However, some aspects of this computation deserve to be emphasized. As we already stressed in the Introduction, the gauge-theoretic formulation of GR is quite different from the metric-based one, with one of the most significant distinctive features being that the graviton conformal mode does not propagate even off-shell. The fact that such a reformulation is possible does not come as a surprise. Indeed, in the formulation due to Bern, see e.g. \cite{Bern:2002kj}, one introduces an additional scalar field, and then does a field redefinition of the metric variable in such a way that the conformal mode of the graviton completely disappears (in favor of the new scalar field). Then, for the purposes of computing the tree-level scattering amplitudes the scalar field can be forgotten about completely, because it is not sourced on the external lines, and thus cannot propagate on the internal lines either. What happens in our formulation is similar, except that the statement that the conformal mode does not propagate is true in even greater generality, as it extends to loops. Thus, the gauge-theoretic formulation can be expected to start to differ from the metric-based GR when one computes quantum effects. And at the tree level, our formulation automatically produces the simple 3-vertex (\ref{cubic-GR}) which is at the heart of the simplicity of the present formalism. While many simplifications of the metric-based GR cubic vertex are possible by a clever choice of gauge-fixing as in e.g. \cite{vandeVen:1991gw}, or by the trick of introducing an additional scalar field as in \cite{Bern:2002kj}, our formulation produces a simple and practical for calculations expression for this vertex automatically, without too much thought. 

We have also stressed a very strong analogy between the linearized Lagrangian, as well as the cubic interaction in our gauge-theoretic formulation of GR and Yang-Mills. As we already discussed after introducing the cubic interaction in (\ref{cubic-GR}), both the linearized Lagrangian and the cubic interaction in gauge-theoretic gravity are more or less direct generalizations of the corresponding objects in Yang-Mills theory. The generalization in question is the passage from an object with one unprimed one primed spinor index $A_{MM'}$, which is the Yang-Mills spin 1 field with its color index suppressed, to an object $A_{ABMM'}$ with 3 unprimed indices and one primed, which is the ${\rm SU}(2)$ gauge field that we used for our description of gravitons. 

We now come to the discussion of even more intriguing aspects of the present formalism. One of the most dramatic outcomes of the studied here (and related) reformulations of GR is that general relativity is {\it not the only theory of interacting massless spin 2 particles}. To understand how our theories manage to avoid the GR uniqueness, let us recall how the uniqueness statement is established. There are many proofs, but for our purposes it will be convenient to concentrate on the proof from unitarity. The proof starts by dimensional reasoning. It is first shown that the amplitudes for 3 on-shell gravitons (with analytically continued to complex values momenta) are completely fixed by their behaviour with respect to the little group. All helicity combinations are in general possible, but only the spinor contractions appearing in ${\cal M}^{--+}$ and ${\cal M}^{-++}$ amplitudes have small enough dimensions (energy squared) to result from a vertex without too many derivatives. Indeed, in the usual formalism they are produced by the $(1/M_p) h^2 \partial^2 h$ cubic vertex of GR. To give rise to non-zero amplitudes ${\cal M}^{---}$ or ${\cal M}^{+++}$ one would need a vertex of the form $(1/M_p^5) (\partial^2 h)^3$. This vertex, if present in the Lagrangian, would give a theory with field equations containing more than two derivatives of the basic field. This is undesirable, as generically higher derivatives are known to lead to instabilities already in the classical theory. This fixes the only cubic amplitudes to be ${\cal M}^{--+}$ and ${\cal M}^{-++}$, and then the only possible quartic amplitude is shown to be ${\cal M}^{--++}$. Its part that has singularities in the $s,t,u$ plane is shown to be completely fixed by the tree-level unitarity. Non-singular contributions to the ${\cal M}^{--++}$ amplitude are also possible, but they can only come from terms in the Lagrangian that have too many derivatives. This shows uniqueness of GR as the only interacting theory of massless spin 2 particles without too many derivatives in the Lagrangian. 

When we reformulate GR as a theory of connections we are using a different spin 2 representation of the Lorentz group to describe gravitons. Namely, it is the connection field $a^{ABCA'}$ that is used instead of $h^{ABA'B'}$ in the metric GR. The natural helicity spinors to be used are then $q^A q^B q^C k^{A'}/\ket{k}{q}^3$ for one of the polarisations (in our conventions negative), and $k^A k^B k^C q^{A'}/ [kq]$, where $q^A, q^{A'}$ are some reference spinors. However, these helicity spinors are dimensionful, unlike the natural constructs in metric GR. Thus, one needs to multiply them by a dimensionful parameter to produce the polarisations with correct mass dimension zero. This has the effect that the helicity spinors are now given by (\ref{helicity-spin}). In particular, the minus polarisation contains a factor of $M$, and the plus polarisation contains a factor of $1/M$, where in our framework $M$ is the square root of the cosmological constant (our reformulation only makes sense with a non-zero scalar curvature). 

This form of the polarisations immediately shows that the connection formulation vertex that gives rise to e.g. ${\cal M}^{--+}$ amplitude must contain one more derivative as compared to the metric formalism. Indeed, we have $M^2$ from the negative polarisations and $1/M$ from the positive helicity graviton. Thus, to produce the same containing $1/M_p$ final result we need a cubic vertex with the pre factor $1/MM_p$, which thus contains three, not two derivatives as in the metric GR. Thus, the familiar GR vertex $(1/M_p) h^2 \partial^2 h$ is now cast into the form $(1/MM_p) (\partial a)^3$.  Note that in spite of the fact that there is one more derivative in the vertex as compared to GR, the derivatives only appear as a power of the first derivative of the field. This means that the field equations produced from this contain still just second derivatives (as well as non-linearities in the first derivative). 

The final point of this discussion is that, unlike in the metric formalism, it is now possible to have the ${\cal M}^{+++}$ amplitude {\it without increasing the order of field equations}. Indeed, there is $1/M^3$ coming from the polarisation spinors. The spinor contractions present in the ${\cal M}^{+++}$ amplitude have dimensions of $M^6$, and so this amplitude is of the form $1/M^5$ times the spinor contractions. Taking into account that we have a factor of $1/M^3$ coming from the polarisations, we see that the vertex that produces a non-zero such amplitude is still of the form that we have already encountered, namely $1/M^2 (\partial a)^3$. More precisely, it comes out to be of the form $M^3/M_p^5 (\partial a)^3$ so that there are no factors of $M$ left in the final result for the ${\cal M}^{+++}$ amplitude. But what is important for us is that the connection formulation made it possible for the amplitudes ${\cal M}^{--+}$ as well as ${\cal M}^{+++}$ to come from schematically the same cubic interaction vertex $1/M^2 (\partial a)^3$. Of course, the tensorial contractions present in the vertices that lead to ${\cal M}^{--+}$ and ${\cal M}^{+++}$ are different, but what is important for us is that they both contain just the third power of the first derivative of the connection. The amplitude ${\cal M}^{-++}$ also works out correctly, but with a bit more work, see the main text. 

Note also that the amplitude ${\cal M}^{---}$ does not fit this framework, because of too many factors of $M$ coming from the polarisations. Indeed, one gets $M^3$ from there, but the answer one wants to get must contain a factor of $1/M^5$. Thus, there must be $1/M^8$ in the vertex, which corresponds to $(1/M^8) (\partial^3 a)^3$. There are clearly too many derivatives here to have second order field equations, and this is why this vertex is excluded. 

To summarise, the reformulation of the theory of interacting gravitons in terms of a different representation of the Lorentz group puts the amplitudes ${\cal M}^{--+}, {\cal M}^{-++}$ and ${\cal M}^{+++}$ all on the same footing, in the sense that they are all produced by a vertex of the type $(1/M^2) (\partial a)^3$. There is no conflict with second order in derivatives nature of the field equations, and this is why there is more than one theory of interacting gravitons with just second order equations of motion {\it provided one uses connection, not metric to describe gravitons}. We have seen that the essence of why this works lies in the fact that the two helicities are treated asymmetrically, with a factor of $M$ in one (negative) helicity and $1/M$ in the positive one. It is this factor of $1/M$ in the positive helicity that allows for a new cubic amplitude ${\cal M}^{+++}$, and in fact for a new amplitude ${\cal M}^{+\ldots +}$ at any number of gravitons, without increasing the order of field equations. This explains why there is an infinite-parametric family of theories of interacting gravitons {\it without higher derivatives in the field equations}, provided one uses the connection formalism. 

Let us contrast this story with what is possible in the metric formalism. In the metric formulation it is possible to have amplitudes that are absent in GR, e.g. ${\cal M}^{+++}$ but these come from terms in the Lagrangian that contain higher derivatives. The developed gauge theoretic formalism allows for these amplitudes without increasing the order of field equations. However, the resulting theories are necessarily chiral (parity-violating), as only the mostly plus amplitudes can get produced by the above mechanism. 

As the new interactions possible in the gauge-theoretic framework are parity-violating, it appears likely that {\it if} the condition of parity-invariance is added, then GR may indeed be the unique theory of interacting gravitons. We have not attempted to prove any such statement here, but since all the deviations from GR that we have found are in the direction of parity violation, it is possible that such GR uniqueness statement can be shown to hold. However, there is no physical reason to restrict one's attention to only parity-invariant theories. Indeed, we know that Nature does violate parity invariance (in the Standard Model) in the strongest possible way. It is gratifying to see that gravitational theories where parity invariance is not built in from the start are also possible. 

One interesting property of the whole class of gravitational theories that we are studying is that they all admit an unambiguous definition of the Newton's constant. Indeed, in our story Newton's constant can be defined as the strength of interaction of two gravitons, and can be read off from the amplitude for two gravitons of some helicities to undergo a scattering process that does not change their helicities. In all our theories this amplitude has the same dependence on the Mandelstam variables $s,t,u$ as in GR, namely ${\cal M}^{--++}\sim s^4/stu$. The dimensionful coefficient in front of this amplitude comes out as $(2 M^2 g^{(2)})^{-1}$, where $M^2$ is the dimensionful parameter setting the radius of curvature of the background on which we work ($M^2\sim\Lambda$), and $g^{(2)}$ is the first coupling constant that characterises a theory from our class. Recall that $g^{(2)}$ is read off (\ref{f3}) from the matrix of second derivatives of the defining function. We can then define the Newton's constant so that the amplitude has exactly the same form as in GR. This fixes $(16\pi G)^{-1} = M^2 g^{(2)}$. We stress that this definition makes sense for an arbitrary member of our infinite-parametric family of theories. It is clear that this definition of the Newton's constant only makes sense because our theories do not contain the most general imaginable higher derivative modification of GR. Indeed, with terms of the form $R^4$ in the metric formalism it would be possible to have contributions to the ${\cal M}^{--++}$ amplitude that have a different functional dependence on the $s,t,u$ variables. In this case the Newton's constant (the coefficient in front of the $s^4/stu$ part of the amplitude) could not be simply read off from a single measurement of the graviton scattering. As it will be clear from the next paper in this series, the fact that the ${\cal M}^{--++}$ amplitude is the same as in GR generalises to a stronger statement that all MHV amplitudes for our infinite-dimensional class of theories are the same as in GR. 

Let us now discuss the obtained parity-violating amplitudes in more details. The first such amplitude (\ref{3-par-viol}) we encountered was for 3 gravitons (with complexified momenta), all of positive helicity. At the same time, the $---$ amplitude continues to be zero, as in GR. This tells us that, at least to the cubic order in perturbation theory, we can think about our parity-violating gravitational theories as General Relativity corrected by a term $(1/M_p^2)(Weyl^+)^3$, where $Weyl^+$ is the self-dual part of the Weyl curvature tensor. Indeed, this term, if added to the GR Lagrangian, would not give any contribution to the amplitudes containing at least one negative helicity graviton, and would produce precisely an amplitude with the structure of (\ref{3-par-viol}) for the $+++$ helicity configuration. This term is of the same type as the famous Goroff-Sagnotti \cite{Goroff:1985th} counterterm $(Weyl)^3$, except that now it is only the self-dual part of curvature tensor that is present and the Lagrangian is not invariant under parity. Moreover, this Lagrangian, in Lorentzian signature, is complex, in view of the fact that all parity-odd terms in it are purely imaginary. This raises immediate concerns about the unitarity of these theories. 

In this paper we have little to say about this issue, except that there are known examples where a perfectly unitary theory may be rewritten in a way that makes unitarity non-manifest. One such example, of particular relevance here, was considered in e.g. \cite{TorresGomez:2012sr}, where a Dirac Lagrangian for spin 1/2 particles was rewritten in terms of 2-component spinors, and then spinors of one of the helicities, say primed, were integrated out. This results in a {\it complex} second-order in derivatives Lagrangian for spin 1/2 particles, which makes the hermiticity (and thus unitarity) of the original first-order description completely hidden. But in spite of being complex, the new second-order Lagrangian continues to describe unitary dynamics. Our hope is that something similar is at play with our parity-violating deformations of GR and that, in spite of the Lagrangian being complex, the dynamics is still unitary. One encouraging fact is that, as it can be checked, the 2-to-2 graviton scattering amplitudes we have obtained satisfy tree-level unitarity requirements, in the sense that these amplitudes have singularities in the $s,t,u$ plane with residues given by products of smaller tree-level amplitudes. We will not attempt any investigation of the unitarity issues in this paper, leaving this aside for future work.   

Let us discuss what 2-to-2 graviton parity-violating amplitudes tell us. These are amplitudes that do not vanish even for real momenta. First, there is a set of amplitudes where parity is flipped just by one unit. Applying the crossing symmetry and converting two gravitons from incoming to outgoing states, we can draw these parity-violating processes as
\be\nonumber
 \lower0.6in\hbox{\includegraphics[height=1.2in]{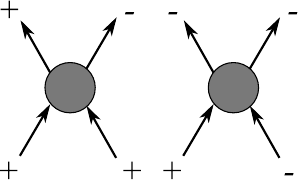}} 
\ee
In both cases one of the positive helicity gravitons is converted into a negative helicity one. We have seen (\ref{-+++}) that this amplitude goes as $stu/M_p^6$, where $s,t$ and $u$ are the usual Mandelstam variables. It is thus of equal importance as the usual parity-preserving amplitude $(1/M_p^2) s^3/tu$ at Planck energies, which is just an illustration of the fact that a generic member of our family of gravity theories is very different from GR at high energies. If one could extrapolate beyond the Planck barrier (which in reality one cannot because the perturbation theory breaks down), one could say that at higher energies the parity-violating processes scaling as $E^6$ become even more important than the parity-preserving ones scaling as $E^2$.

Another set of the processes is when the helicity is flipped by two units. This can be drawn as
\be\nonumber
 \lower0.6in\hbox{\includegraphics[height=1.2in]{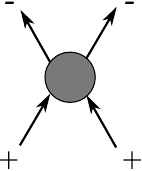}} 
 \ee
Two positive helicity gravitons are converted by this process into two negative helicity ones, with the amplitude computed in (\ref{++++}) having one contribution going as $stu/M_p^6$ and another one proportional to $(s^4+t^4+u^4)/M_p^8$. Again, at Planck energies these processes become equally important to the parity-preserving ones. 

One is then led to admittedly speculative, but thought-provoking picture of the dynamics of gravity at high energies, as predicted by our theories. Indeed, the parity-violating processes only go in one direction (the theories are not T-invariant). It is then clear that eventually all gravitons will get converted into gravitons of a single helicity (negative in our conventions). Whether this indeed has anything to do with what happens at the Planck scale remains to be seen, but it is clear that this picture is a valid, and to some extent unexpected outcome of our gauge-theoretic approach.

To conclude the discussion of the parity-violating amplitudes let us make one more comment on how our theories manage to avoid the proof of GR uniqueness, see also remarks above. There is a proof given in \cite{Grisaru:1975bx} that works directly with the scattering amplitudes. One point about the work \cite{Grisaru:1975bx} that makes it not general enough is the assumption of the parity symmetry. This is clearly violated by our theories. The absence of parity is also the reason why in our case it is possible to get non-zero answers for parity-violating processes, while in \cite{Grisaru:1975bx} parity can be easily shown to lead to vanishing of the $2,2;-2,-2$ amplitude, in the conventions of these authors. The vanishing of another amplitude $2,2;2,-2$ is shown in \cite{Grisaru:1975bx} by a more subtle argument, which does not relate to parity in a direct way. The argument is that the kinematical singularities of this amplitude are known, and then the non-singular piece must behave (for dimensional reasons) as such a high power of momentum that this is impossible in a theory with at most two derivatives in a vertex. The way this argument is circumvented is instructive. While our theories lead to second-order in derivatives field equations, they have arbitrary number of derivatives at vertices when expanded, in the sense that the $n$-th order vertex can contain as many as $n$ derivative, but always with just a single derivative acting on each leg of the vertex. This is the reason why the argument in \cite{Grisaru:1975bx} for the case of the $2,2;2,-2$ amplitude is not applicable to our theories, and indeed, we see a non-vanishing single helicity flip amplitude (\ref{-+++}). 

Let us now come to a discussion of open issues. One unsatisfactory aspect of our approach is our inability to control the form of the defining function of the theory, e.g. as a function of the energy scale. Thus, in contrast to the metric approach where the Einstein-Hilbert action is clearly the most dominant term at low energies, we do not at present understand why at low energies the defining function of our theory should be chosen the way it is for GR. From our results we can only see that the scattering amplitudes for parity violating processes go as much higher power of energy than is the case for GR amplitudes. This does indicate that at low energies these processes can be neglected, which to some extent explains why the parity-preserving GR is the low energy theory. But it would be nice to convert this qualitative argument into a quantitative, and this is missing at the moment. 

Another, related aspect is our lack of understanding of the renormalizability properties of our class of theories. The question is whether any counterterms that are not already present in the Lagrangian need to be added in the process of renormalization. If, as originally conjectured in \cite{Krasnov:2006du}, it happens that this class of theories is closed under the renormalization, one could compute the associated RG flow, and then answer the question of how the defining function flows with energy in a quantitative way. In the absence of any such result, we could only choose the defining function by hand, which is what we did in obtaining the above stated amplitude results. In particular, our conclusion that the parity-violating processes are of importance only at Planck energies was based on a particular choice of the defining function worked out in Appendix D. For this choice, one is guaranteed to have only Planckian modifications of gravity. However, if the defining function does not approach the GR one at low energies as in this simple case, then the energy scales where the parity-violating processes become important are different. In particular, because the cosmological constant $\Lambda=M^2$ gets involved, see (\ref{-+++M}), (\ref{++++M}), the associated energy scales become much lower than $M_p$, which is a potentially very interesting effect. Thus, it is worth stressing that a general member of our family of theories is very different from GR in its properties. Not only it is parity-violating, but more importantly, for a generic choice of the couplings $g^{(2)}, g^{(3)}, \ldots$ the theory becomes strongly coupled at a much lower energy scale than $M_p$, even if one makes $M_p\gg M$ by making the coupling $g^{(2)}$ large, see (\ref{Mp}). This makes GR a very special point in the family of theories that we studied, and makes the question of why it is GR that is seen at low energies very interesting.  

One other important outcome of this work is a better understanding of the modification of GR that is provided by a generic member of our family of theories. We have already discussed the parity-violating aspect of this modification, as well as the potential difficulties with unitarity. Another aspect so far less emphasized is the fact that the graviton-graviton parity-preserving amplitude is unmodified at all, and is the same for all members of our family. An even stronger statement is true: All MHV amplitudes, defined as amplitudes with just two positive helicity gravitons, are exactly the same as in GR. We have sketched an argument to this effect in the main text, and will study these amplitudes in more details in the next paper. Let us concentrate on the graviton-graviton case. A general metric theory with higher powers of curvature present in the Lagrangian will modify the graviton-graviton amplitude, basically due to the fact that it introduces new (higher derivative) vertices. The fact that there is no such modification taking place for all our theories tells us a lot about them. It is possible to establish several general statements for second-derivative gravity theories. In particular, it can be shown that the graviton-graviton parity-preserving amplitude cannot be anything else except what it is in GR. Thus, it is not too surprising that this particular amplitude does not get modified. But it is worth emphasizing that this signals once more that the theories considered are {\it not} the most general gravity theories, the latter tending to introduce additional propagating degrees of freedom and, in particular, modify the graviton-graviton parity preserving amplitude. 

Another very interesting aspect of the modification provided by our theories is that (at tree level) they continue to give zero results for the mostly negative helicity amplitudes, such as amplitudes with all gravitons of negative helicity, or just with a single positive helicity particle. This is of course the case for GR, but it continues to hold for all members of the class we have studied. This property can in turn be shown to be related to the property of MHV amplitudes being unmodified. And this property can again be used to characterize the type of corrections present in our theories as compared to GR. Indeed, a general metric Lagrangian quantum corrected by all curvature invariants will typically render both all minus and all plus amplitudes non-zero. For example, even the simplest term Riemann curvature cubed (the Goroff-Sagnotti term) that needs to be added as a counterterm at two loops will give rise to non-zero answers for both the all negative and the all positive graviton-graviton amplitude. The fact that this does not happen in our case tells us that the type of counterterm corrections that have been included in our theories on top of what is present in GR is very special. Indeed, as we have already discussed, this means that only the self-dual part of the Weyl curvature cubed is present in our Lagrangian, rather than the full Weyl cubed. It may thus seem that our class of theories has no chance of being closed under the renormalization, as containing not the full Goroff-Sagnotti term, but rather only its chiral half. While this is a legitimate worry, we also recall that our theories are expected to differ from GR as far as quantum corrections are concerned (being equivalent only on-shell). Thus, it is possible that their two-loop behavior is different, and that only the self-dual part of Weyl cubed arises as a counterterm. More work is needed to see if this is the case. However, it is interesting that results in the present paper suggest yet another clear test of how the closedness under the renormalization can be probed (by performing the two-loop computation \cite{Goroff:1985th} for GR but in the gauge-theoretic language).

We close by stating once more our amazement at how efficient and powerful the presented gauge-theoretic formulation of gravity is for practical computations. We did not expect this amount of simplicity when we embarked on the present study. It also comes to us as a surprise that the present family of gravity theories can be understood in rather complete generality, and many statements that are true for all members of the family are possible. In the next paper from this series we will see even more instances of this, in that it will be possible to characterize all tree level scattering amplitudes, not just MHV, to certain extent. The studied here reformulation of gravity as a diffeomorphism invariant gauge theory thus continues to be a source of fascination. 

\section*{Acknowledgements} KK was supported by an ERC Starting Grant 277570-DIGT, and partially by a fellowship from the Alexander von Humboldt foundation, Germany. CS was partially supported by the ERC Grant 277570-DIGT. 

\section*{Appendix A: Center of mass frame momentum spinors and Mandelstam variables}

The center of mass frame expressions for our momentum spinors can be obtained from (\ref{kA}). We take the particles $1,2$ to be moving in the direction of the $\vec{z}$ axes, positive and negative respectively. This gives
\be
1^A = 2^{1/4} \sqrt{\omega_k} o^A, \qquad 2^A = \im 2^{1/4} \sqrt{\omega_k} \iota^A.
\ee
The particles $3,4$ we take to be the scattered ones, moving at an angle $\theta$ to the $\vec{z}$ axes. For simplicity we put $\phi=0$. We get
\be
3^A = 2^{1/4}\sqrt{\omega_k} \left( \sin(\theta/2) \iota^A + \cos(\theta/2) o^A\right), \qquad 
4^A = \im 2^{1/4}\sqrt{\omega_k} \left( \sin(\theta/2) o^A - \cos(\theta/2) \iota^A\right).
\ee
The non-zero contractions are
\be\label{contr}
\ket{1}{2} = -\im \sqrt{2}\omega_k, \quad \ket{1}{3} = - \sqrt{2}\omega_k \sin(\theta/2), \quad \ket{1}{4} = \im \sqrt{2}\omega_k \cos(\theta/2), \\ \nonumber
\ket{2}{3} = \im \sqrt{2}\omega_k \cos(\theta/2), \quad \ket{2}{4} = - \sqrt{2}\omega_k \sin(\theta/2), \quad \ket{3}{4} = \im \sqrt{2}\omega_k.
\ee
It is also customary to introduce the following Mandelstam variables
\be\label{mandel}
s = -4\omega_k^2, \quad t= 4\omega_k^2 \sin^2(\theta/2), \quad u=4\omega_k^2 \cos^2(\theta/2).
\ee

\section*{Appendix B: Feynman rules}

In this Appendix we collect all the Feynman rules derived in the main text, directly in spinor notations that are most convenient for practical computations. 

In spinor notations the vertices involve only the factors of momentum spinors $k^{AA'}$ on the legs of the vertex, as well as factors of spinor metrics $\epsilon^{AB}, \epsilon^{A'B'}$. Writing down the corresponding expressions can quickly lead to horribly looking formulas. For this reason it is much more efficient to draw the vertices, indicating the factors of $\epsilon^{AB}, \epsilon^{A'B'}$ by lines. The only drawback of this procedure is that it is not easy to keep track of the signs (remember that raising-lowering a pair of spinor indices induces a minus sign). We have not tried to develop any convention for these signs, just going back to the corresponding term in the Lagrangian and seeing how the indices contract when there is a question. But it is possible that a more systematic sign convention can be developed. Here, in view of the fact that only simple computations are done, our rules are sufficient. 

With these remarks in mind, let us state the rules of the game. First, we only state here the rules of computing the Minkowski space amplitudes. The way these are obtained as a limit of more general de Sitter graviton amplitudes is explained in the main text. The field that propagates in our theory is an ${\rm SU}(2)$ connection, but after all the gauge-fixings and translation into the spinor notations, this is a field $a_{ABCC'}$ with 4 spinor indices, 3 unprimed and one primed. It is moreover symmetric in its 3 unprimed spinor indices, thus forming an object that takes values in an irreducible representation of the Lorentz group. Thus, only $4\times 2=8$ components of the field propagate, as compared to $10$ in the usual metirc treatment. As in any textbook example, the scattering amplitudes are obtained from the field (connection) correlation functions by certain reduction formulas. These are most practical in the momentum space representation, where the correlation function is that of the Fourier coefficients of the field operator. The scattering amplitude then reads
\be\label{LSZ-app}
\langle k_- \ldots  k_+ \ldots  | p_- \ldots p_+ \ldots \rangle  = (-1)^{m+m'} (2\pi)^4 \delta^4\left(\sum k - \sum p\right)  
\\ \nonumber
\, \langle T \varepsilon^+(k_-) \cdot a(k_-) \ldots \varepsilon^-(k_+) \cdot a(k_+)\ldots  \varepsilon^-(p_-) \cdot a(p_-) \ldots  \varepsilon^+ (p_+) a(p_+)\rangle_{\rm amp},
\ee
where we denoted the contractions of the spinor indices involved by a dot, in other words $\varepsilon(k) \cdot a(k) \equiv \varepsilon_{ABCC'}(k) a^{ABCC'}(k)$, and $k_-, k_+$ are the momenta of the outgoing negative and positive helicity gravitons, $p_-, p_+$ are those of the incoming particles, and $m,m'$ are the numbers of positive helicity outgoing and incoming gravitons respectively. As usual, the momentum space amplitude is amputated from its external line propagators, and there is a factor of the total momentum conservation. 

The quantities $\varepsilon^\pm$ are the graviton helicity states that read
\be\label{helicity-app}
\varepsilon^{-}_{ABCA'}(k) = M \frac{ q_A q_B q_C k_{A'}}{\ket{q}{k}^3}, \qquad \varepsilon^{+}_{ABCA'}(k) = \frac{1}{M} \frac{k_A k_B k_C p_{A'}}{\bra{p}{k}},
\ee
where $M$ is the mass scale of the background de Sitter space, and $q^A, p^{A'}$ are the negative and positive helicity reference spinors, which we denote (for convenience) by different letters.

We have the usual statement of the crossing symmetry, which is that one can change an outgoing state into an incoming one, if one flips the direction of the arrow on the corresponding external leg, and flips the helicity. In addition, there is a factor of minus sign for any such flip, but this is a result of our convention choices. Because of the crossing symmetry, one can assume all particles to be e.g. incoming, which is what we do. 

Finally, the rule is that all positive helicity (incoming) particles are taken to be slightly massive, with the mass related to the mass scale $M$ in (\ref{helicity-app}). The meaning of $k_A$ in this formula is then explained by the following decomposition of the 4-momentum $k^{AA'}$
\be\label{mass-shell-app}
k^{AA'}= k^A k^{A'} + M^2 \frac{p^A p^{A'}}{\ket{p}{k}\bra{p}{k}}.
\ee
The convention is that the reference spinors $p^A, p^{A'}$ in this formula are the same as what is used in the positive helicity spinor in (\ref{helicity-app}). The negative helicity particles are all massless. 

The propagator of the theory is best represented as a drawing, consisting of a set of lines contracting the indices, and a black box denoting the symmetrization of the unprimed spinor indices. Black lines represent unprimed indices, while the red line is for the single primed index. The propagator then reads
\be\nonumber
\frac{1}{\im k^2} \quad \lower0.07in\hbox{\includegraphics[height=0.3in]{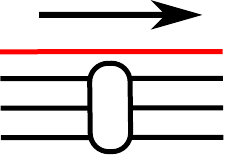}} 
\ee

In the same graphic notation, the 3-valent vertex that is relevant for the computation of the MHV amplitudes reads
\be\label{V-GR-app}
\frac{\im \kappa}{M} \quad \lower0.6in\hbox{\includegraphics[width=1.2in]{3vertexASD.pdf}} \, ,
\ee
where $\kappa^2=32\pi G$. For GR this is the only relevant vertex (at tree level, for MHV amplitudes). The general member of our family of theories contains an additional 3-vertex, which reads
\be\label{V-2-app}
\im \kappa^3 M (2g^{(3)}-(3/2)g^{(2)}) \quad  \lower0.6in\hbox{\includegraphics[width=1.2in]{3vertexSD.pdf}} \, 
\ee
where $(2g^{(3)}-(3/2)g^{(2)})$ is a certain combination of the coupling constants that vanishes in the case of GR. For the family of theories considered in the next section, this combination goes as $M^2/M_p^2$, see (\ref{diff-g-1}). 

The other vertex we will need in this paper is the single-derivative one present already in GR. It is not important for computations of the MHV amplitudes, but contributes to some other amplitudes, e.g. the $++-$ amplitude (for complex momenta) and to the $++++$ amplitude. Represented pictorially, it reads
\be\label{V-3-app}
-\frac{\im}{3} \kappa^3 M^3 f(\delta) \quad \lower0.6in\hbox{\includegraphics[width=1.3in]{3vertex1derASD.pdf}} \,
\ee
Drawings for the 4-valent vertices can be found in the main text. 

\section*{Appendix C: Theory with modifications at the Planck Scale}

In this section we consider a simple one-parameter family of deformations of GR. This family has the advantage that the modifications one introduces are guaranteed to be only those relevant at the Planck scale. In contrast, as we have seen in the main text, a generic defining function would give rise to strong coupling phenomena at much lower energy scales. 

As in \cite{Krasnov:2011up}, consider the following Lagrangian
\be\label{mod-action}
S[B,A,\Psi] = \frac{1}{8\pi \im G}\int \left[B^i\wedge F^i (A)- \frac{1}{2} \left( \Psi^{ij} - \frac{\Lambda}{3}\d^{ij}+ \frac{\b}{2M_p^2}  \Tr{\Psi^2} \d^{ij} \right) B^i\wedge B^j\right].
\ee
When $\beta=0$ this is just the Lagrangian of GR. One obtains a pure connection Lagrangian for GR in the form (\ref{action}) by integrating out first the two-form field $B^i$, and then the Lagrange multiplier field $\Psi^{ij}$. On-shell, the Lagrange multiplier field $\Psi^{ij}$ receives the meaning of the self-dual part of the Weyl curvature. When $\beta\not=0$ we have added to the Lagrangian a term that becomes important when $(Weyl)/M_p^2$ is of order unity. Thus, (\ref{mod-action}) is guaranteed to produce only Planckian modifications of GR (for $\beta$ of the order unity).

Integrating out the two-form field in (\ref{mod-action}) one gets
\be
S[A,\Psi] = \frac{1}{16\pi \im G}\int\left( \Psi^{ij} - \frac{\Lambda}{3}\d^{ij}+ \frac{\beta}{2M_p^2}  \Tr{\Psi^2} \d^{ij} \right)^{-1} F^i\wedge F^j.
\ee
To prepare this functional for minimizing with respect to $\Psi$ let us rescale $\tilde{\Psi}=\Psi/M^2$, where as before $M^2=\Lambda/3$. We can then rewrite the above action as
\be\label{S-A-Psi}
S[A,\tilde{\Psi}] = \frac{\im}{\alpha}\int\left( \delta^{ij} - \tilde{\Psi}^{ij} - \frac{\alpha\b}{2}  \Tr{\tilde{\Psi}^2} \d^{ij} \right)^{-1} F^i\wedge F^j.
\ee
The new Lagrange multiplier field $\tilde{\Psi}^{ij}$ is dimensionless, and 
\be
\alpha:= \frac{M^2}{M_p^2},
\ee
where as usual $M_p^2=1/16\pi G$. This way of writing the functional shows that deviations from GR are parameterized by $\beta$ always in the combination $\alpha\beta$. 

We can now extremize (\ref{S-A-Psi}) by passing to eigenvalues of both the $F^i\wedge F^j$ and $\tilde{\Psi}^{ij}$ matrices, as in \cite{Krasnov:2011pp}. Denoting the eigenvalues of $F^i\wedge F^j$ by $\lambda_{1,2,3}$, i.e. $F^i\wedge F^j = {\rm diag}(\lambda_1,\lambda_2,\lambda_3) \,d^4x$, we obtain the following expression for the extremum of (\ref{S-A-Psi}) as a function of the eigenvalues 
\be
S[A] = \im \int f[\lambda] \, d^4 x,
\ee
where 
\be\label{f-exp}
f[\lambda] = \frac{ (\sum_i \sqrt{\lambda_i})^2 }{3\a} \left( 1 + \frac{3\a \b}{2}\right.
\\ 
\nn -\frac{9(\a \b)^2}{4(\sum_i \sqrt{\lambda_i})^3}
\bigg(4 \lambda _1^{3/2}-3 \lambda _1 \sqrt{\lambda _2}-3 \sqrt{\lambda _1} \lambda _2+4 \lambda _2^{3/2}-3 \lambda _1 \sqrt{\lambda _3}
\\
\nn +6 \sqrt{\lambda _1} \sqrt{\lambda _2} \sqrt{\lambda _3}-3 \lambda _2 \sqrt{\lambda _3}-3 \sqrt{\lambda _1} \lambda _3-3 \sqrt{\lambda _2} \lambda _3+4 \lambda _3^{3/2}\bigg)
\\ 
 - \frac{27}{8}\frac{\a^3  \b^3 }{\left(\sqrt{\lambda _1}+\sqrt{\lambda _2}+\sqrt{\lambda _3}\right){}^4} \left(16 \lambda _1^2-17 \lambda _1^{3/2} \sqrt{\lambda _2}+6 \lambda _1 \lambda _2-17 \sqrt{\lambda _1} \lambda _2^{3/2} \right.
\\
\nn+16 \lambda _2^2-17 \lambda _1^{3/2} \sqrt{\lambda _3}+12 \lambda _1 \sqrt{\lambda _2} \sqrt{\lambda _3}
 +12 \sqrt{\lambda _1} \lambda _2 \sqrt{\lambda _3}-17 \lambda _2^{3/2} \sqrt{\lambda _3}
\\ \left. \left.
\nn +6 \lambda _1 \lambda _3+12 \sqrt{\lambda _1} \sqrt{\lambda _2} \lambda _3+6 \lambda _2 \lambda _3-17 \sqrt{\lambda _1} \lambda _3^{3/2}-17 \sqrt{\lambda _2} \lambda _3^{3/2}+16 \lambda _3^2\right)
+ O( (\a \b)^4 ) \right).
\ee

This expression is sufficient to compute various coupling constants that appear in (\ref{f2})-(\ref{f4}). Thus, from (\ref{f2}) we have
\be
 - \frac{g^{(2)}}{3}= \frac{\partial^2 f}{\partial \lambda_1\partial \lambda_1}\bigg|_{\lambda_i = 1}.
\ee
This gives
\be
g^{(2)} = \frac{1}{ \alpha } \left(1+\frac{3 \alpha  \b}{2}+\frac{9 \alpha ^2 \b^2}{4}+\frac{27 \alpha ^3 \b^3}{8}+  O[\a \b]^4\right).
\ee
Similarly, from (\ref{f3}) we have
\be
\frac{\partial^3 f}{\partial \lambda_1\partial \lambda_1 \partial \lambda_1}\bigg|_{\lambda_i = 1} =  \, \frac{2g^{(3)}}{9}  + \frac{g^{(2)}}{3},
\ee
which gives
\be
g^{(3)} = \frac{3}{4\a}   \left( 1 +    \frac{3}{2} \a \b  - \frac{27}{4}\a^3 \b^3
+O[\alpha\b]^4\right).
\ee
We can now see how the coefficient in front of the vertex (\ref{V-2-app}) scales with $M$. We have
\be\label{diff-g-1}
g^{(3)} -\frac{3}{4}g^{(2)} =-\frac{27}{16} \a \b^2 + O(  \a^2 \b^3 ),
\ee   
and therefore this goes as $M^2/M_p^2$. 

To compute $g^{(4)}$ in a similar fashion, we will need to write down the terms denoted by dots in (\ref{f4*}). We have
\be
\nn \frac{\partial^4 f}{\partial X^{ij} \partial X^{kl}\partial X^{mn}\partial X^{pq}}=
 \ -g^{(4)} \sum_{Perms_3}\frac{1}{3} P^{ij|ab}P^{kl|bc}P^{mn|cd}P^{pq|da}\\
  - \frac{8}{3}g^{(3)} \sum_{Perms_4}  \frac{1}{4} \delta^{ij}P^{pq|ab}P^{kl|cb}P^{mn|ac}  
-\frac{2}{3}g^{(2)}  \sum_{Perms_6} \frac{1}{6} P^{pq|kl}\delta^{ij} \delta^{mn} .
\ee
Then the fourth derivative with respect to one of the eigenvalues gives
\be
\frac{\partial^4 f}{\partial \lambda_1 \partial \lambda_1\partial \lambda_1 \partial \lambda_1}\bigg|_{\lambda_i = 1} = -\frac{2 g^{(4)} }{9}-\frac{16  g^{(3)} }{27}-\frac{4  g^{(2)} }{9}, 
\ee
which in turn gives
\be
g^{(4)} = \frac{104+156  \alpha\b + 126 \alpha^2\b^2  +297  \alpha^3\b^3}{128 \alpha } +O(\alpha^3\beta^4).
\ee
The coefficient appearing in the 4-vertex and in the $++++$ amplitude then reads
\be\label{diff-g-2}
2g^{(4)}-g^{(3)}-\frac{7}{8}g^{(2)}  = \frac{27}{4} \a^2 \b^3 +O(\alpha^3\beta^4).
\ee
Note that this goes as a higher power of $M$ than the similar difference (\ref{diff-g-1}). 

Finally, from (\ref{f-exp}) we see that
\be
f(\delta)=\frac{3}{\alpha}\left( 1+ \frac{3\alpha\beta}{2} + O((\alpha\beta)^3)\right).
\ee
This implies that
\be\label{g2-f}
3g^{(2)}-f(\delta)= \frac{27\alpha\beta^2}{4} + O(\alpha^2\beta^3).
\ee

\section*{Appendix D: Contributions to the $++++$ amplitude from the 3-valent graphs}

The purpose of this section is to compute contributions to the all plus amplitude from two different 3-valent diagrams. The computation is more involved than any of those that we have done before, and so we decided to put it in the Appendix. 

There are two types of diagrams to consider. In one, we couple the MHV-relevant vertex (\ref{V-GR-app}) to the vertex (\ref{V-2-app}), and in the other it is the 1-derivative vertex (\ref{V-3-app}) that is similarly coupled to (\ref{V-2-app}). Let us consider the 3-derivative vertex diagram first. We do the calculation in several steps. First, we will compute the diagram where the states $1,2$ are inserted into the 3-derivative vertex, and the states $3,4$ are inserted into (\ref{V-2-app}). Then we perform the sum over permutations. 

First, consider the 3-derivative vertex with the states $1,2$ inserted. We necessarily have to insert one of them into the SD leg, and another into an ASD one. Since the vertex is not symmetric in this leg, we will need to add the permutation of $1$ and $2$. So, let $1$ go into the ASD leg, and $2$ into the SD one. We change the orientation of the SD leg, so that the momentum on all the lines is now incoming, and add a minus sign resulting from this change. There is another minus sign coming from the fact that the SD leg insertion gives for $k^{(A}{}_{A'} a^{BCD)A'}$
\be
\frac{1}{M} 2^A 2_{A'} \frac{2^B 2^C 2^D p^{A'}}{\bra{p}{2}} = - \frac{1}{M} 2^A 2^B 2^C 2^D.
\ee
Insertion into the ASD leg gives, taking into account the fact that the positive helicity states are massive
\be
M \frac{1_A 1_B p_{M'} p_{N'}}{\bra{1}{p}^2}.
\ee
Combining everything, and adding a factor for the final momentum we get
\be\label{app4-1}
\frac{\im\kappa}{M} \frac{\ket{1}{2}^2}{\bra{1}{p}^2} \left( 2^{(A} 2^B 1^{C)} \bra{1}{p} + 2^A 2^B 2^C \bra{2}{p}\right) p^{A'}.
\ee
This will have to be multiplied by the propagator $1/2\im\ket{1}{2}\bra{1}{2}$, and then connected to the other vertex. However, before we do this, let us add all other contributions coming from connecting $1,2$ via the 3-derivative vertex. It is clear that we also need to add to (\ref{app4-1}) the same quantity with $1$ and $2$ exchanged. This gives
\be\label{app4-2}
\frac{\kappa}{2M}\frac{\ket{1}{2}}{\bra{1}{2}}\left( 1^A 1^B 1^C \frac{\bra{1}{p}}{\bra{2}{p}^2} +2^A 2^B 2^C \frac{\bra{2}{p}}{\bra{1}{p}^2} + 1^{(A} 1^B 2^{C)} \frac{1}{\bra{2}{p}} + 2^{(A} 2^B 1^{C)} \frac{1}{\bra{1}{p}} \right) p^{A'},
\ee
where we also multiplied by the propagator.

We now connect this to the result of insertion of states $3,4$ into the vertex (\ref{V-2-app}). Here we choose to represent the momentum on the internal line of the diagram, outgoing from the vertex (\ref{V-2-app}), as $-(1+2)^{AA'}$. We get for this vertex
\be\label{app4-4}
- \frac{\im\sqrt{2}(4g^{(3)}-3g^{(2)})}{M M_p^3} \ket{3}{4}^2 3_{(A} 3_B 4_C 4_{D)} (1^D 1_{A'} + 2^D 2_{A'}).
\ee
Contracting (\ref{app4-2}) and (\ref{app4-4}) we get, after simplifying the prefactor 
\be\nonumber
\frac{(4g^{(3)}-3g^{(2)})}{\im M^2 M_p^4} \frac{\ket{1}{2}\ket{3}{4}^2}{\bra{1}{2}\bra{1}{p}^2\bra{2}{p}^2} \left( 1^A 1^B  1^C 1^D \bra{1}{p}^4 + 2^A 2^B 2^C 2^D \bra{2}{p}^4 \right. \\ \nonumber
\left. + 2\cdot  2^A 1^B 1^C 1^D  \bra{1}{p}^3\bra{2}{p} + 2\cdot 1^A 2^B 2^C 2^D \bra{1}{p}\bra{2}{p}^3 + 2 \cdot 1^A 1^B 2^C 2^D \bra{1}{p}^2 \bra{2}{p}^2 \right)  3_{(A} 3_B 4_C 4_{D)} .
\ee
To perform the contraction, we rewrite the symmetrization $3_{(A} 3_B 4_C 4_{D)}$ in several different ways, depending on the symmetries of the expression that it gets contracted to. For the contracting object that is $BCD$ symmetric we can write
\be\label{app-sym1}
3_{(A} 3_B 4_C 4_{D)} = \frac{1}{2}\left( 3_A 3_{(B} 4_C 4_{D)} + 4_A 4_{(B} 3_C 3_{D)}\right).
\ee
For the contracting object that is $AB$ and $CD$ symmetric, as well as symmetric under the exchange of pairs $AB$ and $CD$ we can write 
\be\label{app-sym2}
3_{(A} 3_B 4_C 4_{D)} = \frac{1}{6} \left( 3_A 3_B 4_C 4_D +4_A 4_B 3_C 3_D+ 4\cdot 3_{(A} 4_{B)} 3_{(C} 4_{D)}\right).
\ee
Performing the contractions we get the following amplitude
\be\label{app4-amp1}
\frac{(4g^{(3)}-3g^{(2)})}{\im M^2 M_p^4} \frac{\ket{1}{2}\ket{3}{4}^2}{\bra{1}{2}} \Big[ \ket{1}{3}^2\ket{1}{4}^2 \frac{\bra{1}{p}^2}{\bra{2}{p}^2} + \ket{2}{3}^2\ket{2}{4}^2 \frac{\bra{2}{p}^2}{\bra{1}{p}^2}  \\ \nonumber 
+ \left( \ket{2}{3}\ket{1}{3}\ket{1}{4}^2 + \ket{2}{4}\ket{1}{4}\ket{1}{3}^2\right) \frac{\bra{1}{p}}{\bra{2}{p}}+ \left( \ket{1}{3}\ket{2}{3}\ket{2}{4}^2 + \ket{1}{4}\ket{2}{4}\ket{2}{3}^2\right) \frac{\bra{2}{p}}{\bra{1}{p}}  \\ \nonumber 
+\frac{1}{3} \left( \ket{2}{3}^2\ket{1}{4}^2 + \ket{1}{3}^2\ket{2}{4}^2 + 4 \ket{1}{3}\ket{1}{4}\ket{2}{3}\ket{2}{4}\right) \Big].
\ee

The first four terms in (\ref{app4-amp1}) are $p$-dependent, while the ones in the last line are not. We expect the result to be $p$-independent, but this should only be true after all the permutations are added. Thus, the above result is $12$ and $34$ symmetric (by construction), but is not symmetric under the permutation of $12$ with $34$. This means that we have to consider overall 6 different permutations, and add them all up. We first need to show that the result of adding all these permutations is $p$-independent. To this end, let us consider all the terms that will have $\bra{1}{p}^2$ in the denominator. These are
\be\label{app4-5}
\frac{\ket{2}{3}^2\ket{2}{4}^2\ket{3}{4}^2}{\bra{1}{p}^2} \left( \bra{2}{p}^2 \frac{\ket{1}{2}}{\bra{1}{2}} + \bra{3}{p}^2 \frac{\ket{1}{3}}{\bra{1}{3}}+ \bra{4}{p}^2 \frac{\ket{1}{4}}{\bra{1}{4}} \right).
\ee
We can now use the Schouten identity to rewrite $\bra{2}{p}^2$ in terms of the square brackets $\bra{3}{p}$ and $\bra{4}{p}$. Thus, we replace
\be
\ket{1}{2}^2\bra{1}{2}^2 = \ket{1}{3}^2 \bra{3}{p}^2 + \ket{1}{4}^4 \bra{4}{p}^2 + 2\ket{1}{3}\ket{1}{4} \bra{3}{p}\bra{4}{p}.
\ee
Then, after some simplifications with the use of the Schouten identity we get for (\ref{app4-5})
\be\label{app4-6}
- \frac{\ket{2}{3}^2\ket{2}{4}^2\ket{3}{4}^2 \ket{1}{3}\ket{1}{4} \bra{3}{4}^2}{ \ket{1}{2}\bra{1}{2}\bra{1}{3}\bra{1}{4}},
\ee
where $\bra{1}{p}^2$ cancelled out. We can now rewrite this result in a bit more convenient form, by eliminating as many square brackets as possible using the momentum conservation. This gives for (\ref{app4-6})
\be\label{app4-8}
\ket{1}{2}\ket{1}{3}\ket{1}{4}\ket{2}{3}\ket{2}{4}\ket{3}{4} \frac{\ket{3}{4}}{\bra{1}{2}}.
\ee
The fact that this is $2,3,4$ symmetric follows easily from the momentum conservation. It can then be checked that the other terms in the sum of permutations of (\ref{app4-amp1}) containing squares of $p$-dependent square brackets in the denominator are all equal to (\ref{app4-8}). Thus, running the same argument for e.g. the terms proportional to $1/\bra{2}{p}^2$ gives precisely the same result as (\ref{app4-8}). This means that the sum of permutations of the first two terms in (\ref{app4-7}) is 4 times (\ref{app4-8}). 

Let us now consider permutations of terms containing just a single power of $p$-dependent square bracket. Let us concentrate on the terms proportional to $1/\bra{1}{p}$. These give rise to the sum
\be
\frac{\ket{2}{3}\ket{2}{4}\ket{3}{4}}{\bra{1}{p}} \left( \left( \ket{1}{3}\ket{2}{4}+\ket{1}{4}\ket{2}{3}\right) \bra{2}{p} \frac{\ket{1}{2}\ket{3}{4}}{\bra{1}{2}} \right. \\ \nonumber \left. - \left( \ket{1}{2}\ket{3}{4}-\ket{1}{4}\ket{2}{3}\right) \bra{3}{p} \frac{\ket{1}{3}\ket{2}{4}}{\bra{1}{3}}-\left( \ket{1}{2}\ket{3}{4}+\ket{1}{3}\ket{2}{4}\right) \bra{4}{p} \frac{\ket{1}{4}\ket{2}{3}}{\bra{1}{4}}  \right).
\ee
Using the momentum conservation to convert the determinants inside the brackets into multiples of $\bra{1}{2}$ we get
\be\label{app4-7}
\frac{\ket{2}{3}\ket{2}{4}\ket{3}{4}^2}{\bra{1}{p}\bra{1}{2}} \left( \left( \ket{1}{3}\ket{2}{4}+\ket{1}{4}\ket{2}{3}\right) \bra{2}{p} \ket{1}{2} \right. \\ \nonumber \left. + \left( \ket{1}{2}\ket{3}{4}-\ket{1}{4}\ket{2}{3}\right) \bra{3}{p} \ket{1}{3}-\left( \ket{1}{2}\ket{3}{4}+\ket{1}{3}\ket{2}{4}\right) \bra{4}{p} \ket{1}{4} \right).
\ee
We can now again use the momentum conservation, now in the form $\ket{1}{2}\bra{2}{p}=-\ket{1}{3}\bra{3}{p}-\ket{1}{4}\bra{4}{p}$. After some simple algebra involving the Schouten identity the expression in brackets becomes $-3\ket{1}{3}\ket{1}{4}( \ket{2}{3}\bra{3}{p}+\ket{2}{4}\bra{4}{p}) = - 3\ket{1}{2}\ket{1}{3}\ket{1}{4}\bra{1}{p}$, where in the last equality we used the momentum conservation. Overall, (\ref{app4-7}) becomes
\be
-3 \ket{1}{2}\ket{1}{3}\ket{1}{4}\ket{2}{3}\ket{2}{4}\ket{3}{4} \frac{\ket{3}{4}}{\bra{1}{2}}.
\ee
As previously the case with (\ref{app4-8}), there are in total a multiple of 4 of these, coming from applying the same analysis to different $p$-dependent denominators, e.g. to terms proportional to $1/\bra{2}{p}$, etc. 

It remains to consider terms in (\ref{app4-amp1}) that are $p$-independent, together with their permutations. For this, using Schouten identity it is convenient to rewrite the term in the last line of (\ref{app4-amp1}) as
\be\label{app4-9}
2 \ket{1}{3}\ket{1}{4}\ket{2}{3}\ket{2}{4} + \frac{1}{3} \ket{1}{2}^2\ket{3}{4}^2.
\ee
Then the first of these two terms, when multiplied by the prefactor in (\ref{app4-8}), is already of the form (\ref{app4-8}). Summing over 6 different permutations we thus get a multiple of 12 of (\ref{app4-8}) from the first term in (\ref{app4-9}). For the permutations of the second term in (\ref{app4-9}) we have
\be\label{app4-10}
\frac{2}{3}\left( \frac{\ket{1}{2}^4 \ket{3}{4}^4}{\bra{1}{2}\bra{1}{2}} + \frac{\ket{1}{3}^4 \ket{2}{4}^4}{\bra{1}{3}\bra{1}{3}}+\frac{\ket{1}{4}^4 \ket{2}{3}^4}{\bra{1}{4}\bra{1}{4}}\right).
\ee
Converting the square brackets in the denominators into multiples of $\bra{1}{2}$ using the momentum conservation, and then using the Schouten identity we get for (\ref{app4-10}) $- 2$ times (\ref{app4-8}). Thus, the sum of all permutations of (\ref{app4-amp1}) is equal $(1-3)4+12-2= 2$ multiples of  (\ref{app4-8}). Together with the prefactor, this gives the following result for this part of the amplitude
\be\label{app4-res1}
\frac{2(4g^{(3)}-3g^{(2)})}{\im M^2 M_p^4} \ket{1}{2}\ket{1}{3}\ket{1}{4}\ket{2}{3}\ket{2}{4}\ket{3}{4} \frac{\ket{3}{4}}{\bra{1}{2}}.
\ee

We now compute the other diagram, where the 1-derivative vertex (\ref{V-3-app}) is connected to (\ref{V-2-app}).  As before, we first consider contributions from connecting $1,2$ into the 1-derivative vertex  and $3,4$ into the vertex (\ref{V-2-app}), and then sum over the permutations. For the 1-derivative vertex, both states are now inserted into the legs with no derivatives in them. We also need to change the orientation of the ASD leg, so that all the momenta are incoming, with a resulting extra minus sign. Overall, we get
\be
 \frac{\im}{3} \kappa^3 M f(\delta) \frac{\ket{1}{2}^2}{\bra{1}{p}\bra{2}{p}} 1^{(A} 2^B \left( 1^{C)} \bra{1}{p} + 2^{C)} \bra{2}{p}\right) p^{A'}.
\ee
We now multiply this by the propagator, and connect to the other vertex (\ref{app4-4}). We get, after simplifying the prefactor
\be
\frac{2f(\delta) (4g^{(3)}-3g^{(2)})}{3\im M_p^6} \frac{\ket{1}{2}\ket{3}{4}^2}{\bra{1}{2}} 1^A 2^B \left(   1^C 1^D \frac{\bra{1}{p}}{\bra{2}{p}} +  2^C 2^D \frac{\bra{2}{p}}{\bra{1}{p}} + 2\cdot 1^C 2^D \right) 3_{(A} 3_B 4_C 4_{D)} .
\ee
We perform the contraction using (\ref{app-sym1}), (\ref{app-sym2}). We get for this contribution to the amplitude
\be
\frac{2f(\delta) (4g^{(3)}-3g^{(2)})}{3\im M_p^6} \frac{\ket{1}{2}\ket{3}{4}^2}{\bra{1}{2}} \Big[ \frac{1}{2} \left( \ket{2}{3}\ket{1}{3}\ket{1}{4}^2 + \ket{2}{4}\ket{1}{4}\ket{1}{3}^2\right) \frac{\bra{1}{p}}{\bra{2}{p}}\\ \nonumber + \frac{1}{2} \left( \ket{1}{3}\ket{2}{3}\ket{2}{4}^2 + \ket{1}{4}\ket{2}{4}\ket{2}{3}^2\right) \frac{\bra{2}{p}}{\bra{1}{p}}   
+2 \ket{1}{3}\ket{1}{4}\ket{2}{3}\ket{2}{4}+ \frac{1}{3}\ket{1}{2}^2\ket{3}{4}^2 \Big],
\ee
where we have used (\ref{app4-9}). We have already performed the sums over permutations required here, and so we can immediately write down the result. It is given by the prefactor, times a multiple $-(3/2)\cdot 4+ 12-2=4$ of (\ref{app4-8}), in other words this part of the amplitude equals
\be\label{app4-res2}
 \frac{8f(\delta) (4g^{(3)}-3g^{(2)})}{3\im M_p^6} \ket{1}{2}\ket{1}{3}\ket{1}{4}\ket{2}{3}\ket{2}{4}\ket{3}{4} \frac{\ket{3}{4}}{\bra{1}{2}}.
\ee

We now have to add the two results (\ref{app4-res1}) and (\ref{app4-res2}), and compute the $M\to 0$ limit. Since $4g^{(3)}-3g^{(2)}= - (27/4) \beta^2 M^2/M_p^2$ plus higher order in $M$ terms, and $f(\delta)=3M_P^2/M^2$ plus order unity terms, we only need to keep these leading orders. Thus, we get, overall, our sample one-parameter family of theories with Planckian modifications
\be\label{app-3-valent}
{\cal M}^{++++}_{\rm 3-vert} = \im \frac{135\beta^2}{2M_p^6} \ket{1}{2}\ket{1}{3}\ket{1}{4}\ket{2}{3}\ket{2}{4}\ket{3}{4} \frac{\ket{3}{4}}{\bra{1}{2}} = \im \frac{135\beta^2}{16M_p^6} stu.
\ee

\end{document}